\pdfoutput=1

\documentclass[11pt,twoside,a4paper,cmspaper,final]{cms-tdr}
\def\svnVersion{167927}\def\cmsCernNo{2012-220}\def\cmsCernDate{\today}\def\cmsMessage{Submitted to Physics Letters B}

\begin{document}\cmsNoteHeader{HIG-12-028}

\hyphenation{had-ron-i-za-tion}
\hyphenation{cal-or-i-me-ter}
\hyphenation{de-vices}

\RCS$Revision: 143127 $
\RCS$HeadURL: svn+ssh://svn.cern.ch/reps/tdr2/papers/HIG-12-028/trunk/HIG-12-028.tex $
\RCS$Id: HIG-12-028.tex 143127 2012-08-09 16:50:46Z alverson $
\newlength\cmsFigWidth
\ifthenelse{\boolean{cms@external}}{\setlength\cmsFigWidth{0.95\columnwidth}}{\setlength\cmsFigWidth{0.4\textwidth}}
\newlength\cmsFigWideWidth
\ifthenelse{\boolean{cms@external}}{\setlength\cmsFigWideWidth{0.95\columnwidth}}{\setlength\cmsFigWideWidth{0.6\textwidth}}
\ifthenelse{\boolean{cms@external}}{\providecommand{\cmsLeft}{top}}{\providecommand{\cmsLeft}{left}}
\ifthenelse{\boolean{cms@external}}{\providecommand{\cmsRight}{bottom}}{\providecommand{\cmsRight}{right}}
\ifthenelse{\boolean{cms@external}}{\providecommand{\color}[1]{\relax}}{}
\providecommand{\doi}{\texttt{doi:}\begingroup \urlstyle{tt}\Url}
\cmsNoteHeader{HIG-12-028} 
\title{Observation of a new boson at a mass of 125\GeV with the CMS experiment at the LHC}

\date{\today}

\abstract{
Results are presented from searches for the standard model Higgs boson
in proton-proton collisions at $\sqrt{s}=7$~and~8\TeV
in the Compact Muon Solenoid experiment at the LHC, using data samples corresponding
to integrated luminosities of
up to 5.1\fbinv at 7\TeV and 5.3\fbinv at 8\TeV.
The search is performed in five decay modes:
$\Pgg\Pgg$, $\cPZ\cPZ$, $\PWp\PWm$, $\Pgt^+\Pgt^-$, and $\bbbar$.
An excess of events is observed above the expected background,
with a local significance of 5.0 standard deviations,
at a mass near 125\GeV, signalling the production of a new particle.
The expected significance for a standard model Higgs boson of that mass is 5.8 standard deviations.
The excess is most significant in the two decay modes with
the best mass resolution, $\Pgg\Pgg$ and $\cPZ\cPZ$; a
fit to these signals gives a mass of
$125.3\pm 0.4\,(\text{stat.})\pm 0.5\,(\text{syst.})\GeV$.
The decay to two photons indicates that the new particle is a boson with spin different from one.
}

\hypersetup{%
pdfauthor={CMS Collaboration},%
pdftitle={Observation of a new boson at a mass of 125 GeV with the CMS experiment at the LHC},%
pdfsubject={CMS},%
pdfkeywords={CMS, physics, Higgs}}

\maketitle 

\newcommand{\ObsNFL}{{\color{black}121.5}} 
\newcommand{\ObsNFH}{{\color{black}128}}   

\newcommand{\MaxLocalZ}{{\color{black}5.0}}  
\newcommand{\Zgamgam}{{\color{black}4.1}}
\newcommand{\Zfourlepton}{{\color{black}3.2}}
\newcommand{\ZhighRes}{{\color{black}5.0}}
\newcommand{\Zww}{{\color{black}1.5}}
\newcommand{\Zbosonic}{{\color{black}5.1}}

\newcommand{\MaxLocalZeight}{{\color{black}3.8}}                   

\newcommand{\MaxLocalZseven}{3.2}  

\newcommand{\MaxZmass}{{\color{black}125}}
\newcommand{\Nup}{{\color{black}6}}
\newcommand{\CLEE}{{\color{black}0.56}}

\newcommand{\GlobalZsmall}{4.6} 
\newcommand{\GlobalZmedium}{4.5} 
\newcommand{\GlobalZfull}{{\color{red}4.0}}

\newcommand{\MASS}{\ensuremath{125.3 \pm 0.4\, (\text{stat.}) \pm 0.5\, (\text{syst.})}}
\newcommand{\MASSH}{{\color{black}$125.1 \pm 0.5$}}

\newcommand{\MUHAT}{{\color{black}$0.87 \pm 0.23$}} 

\newcommand{\mH}{\ensuremath{m_{\PH}}}
\newcommand{\CLs}{\ensuremath{\mathrm{CL_s}\xspace}}

\newcommand{\ptgg}{\ensuremath{p_{\mathrm{T}}^{\gamma\gamma}}\xspace}
\newcommand{\mgg}{\ensuremath{m_{\gamma\gamma}}\xspace}

\newcommand\ptV {\ensuremath{\PT(\mathrm{V})}}
\newcommand\ZmmH  {\ensuremath{\cPZ(\Pgm\Pgm)\PH}}
\newcommand\Zudscg   {\ensuremath{\cPZ+\cPqu\cPqd\cPqs\cPqc\Pg}}
\newcommand\ZeeH  {\ensuremath{\cPZ(\Pe\Pe)\PH}}
\newcommand\ZnnH  {\ensuremath{\cPZ(\cPgn\cPgn)\PH}}
\newcommand\WmnH  {\ensuremath{\PW(\Pgm\cPgn)\PH}}
\newcommand\WenH  {\ensuremath{\PW(\Pe\cPgn)\PH}}
\newcommand\Wudscg   {\ensuremath{\PW+\cPqu\cPqd\cPqs\cPqc\Pg}}
\newcommand\Wbb   {\ensuremath{\PW\bbbar}}
\newcommand\Zbb   {\ensuremath{\cPZ\bbbar}}

\newcommand{\dyll}{\ensuremath{\cPZ/\GAMMA^*{\to \ell^+\ell^-}}}
\newcommand{\wgamma}{\ensuremath{\W\GAMMA}}

\section{Introduction}\label{sec:Intro}

The standard model (SM) of elementary particles provides a remarkably
accurate description of results from many accelerator and non-accelerator based experiments. The
SM comprises quarks and leptons as the building blocks of matter,
and describes their interactions through the exchange of force carriers: the photon for electromagnetic
interactions, the $\PW$ and $\cPZ$ bosons for weak interactions, and the gluons for strong interactions.
The electromagnetic and weak interactions are unified in
the electroweak theory. Although the predictions of the
SM have been extensively confirmed, the question of how the $\PW$ and $\cPZ$ gauge bosons acquire
mass whilst the photon remains massless
is still open.

Nearly fifty years ago it was
proposed~\cite{Englert:1964et,Higgs:1964ia,Higgs:1964pj,Guralnik:1964eu,Higgs:1966ev,Kibble:1967sv}
that spontaneous symmetry breaking in gauge theories could be achieved
through the introduction of a scalar field. Applying
this mechanism to the
electroweak theory~\cite{Glashow:1961tr,Weinberg:1967tq,sm_salam} through a complex scalar doublet field
leads to the
generation of the $\PW$ and $\cPZ$ masses, and to the prediction of the existence of the SM
Higgs boson (\PH). The scalar field also gives mass to the fundamental
fermions through the Yukawa interaction. The mass
$\mH$ of the SM Higgs boson is not predicted by
theory.
However, general considerations~\cite{Cornwall:1973tb,Cornwall:1974km,LlewellynSmith:1973ey,Lee:1977eg}
suggest that $\mH$ should be smaller than ${\sim}1\TeV$, while precision electroweak
measurements imply that $\mH < 152$\GeV at 95\% confidence level (CL)~\cite{EWKlimits}. Over the
past twenty years, direct searches for the Higgs boson have been carried out at the LEP collider,
leading to a lower bound of $\mH > 114.4$\GeV at 95\% CL~\cite{LEPlimits}, and at the Tevatron
proton-antiproton collider, excluding the mass range 162--166\GeV at 95\% CL~\cite{TEVHIGGS_2010} and detecting an excess of events, recently reported in~\cite{CDF:Hbb,CDFD0:HbbCombined,D0CombHbb}, in the range 120--135\GeV.

The discovery or exclusion of the SM Higgs boson is one of the primary scientific goals of the Large Hadron Collider (LHC)~\cite{lhc}.
Previous direct searches at the LHC were based on data from proton-proton collisions corresponding to
an integrated luminosity of 5\fbinv collected at a centre-of-mass
energy $\sqrt{s}=7$\TeV.
The CMS experiment excluded
at 95\% CL a range of masses from 127 to 600\GeV~\cite{Chatrchyan:2012tx}.
The ATLAS experiment excluded at 95\% CL the ranges 111.4--116.6,
119.4--122.1 and 129.2--541\GeV~\cite{ATLAScombJul2012_7TeV}.
Within the remaining allowed mass region, an excess of events near
125\GeV was reported by both experiments.
In 2012 the proton-proton centre-of-mass energy was increased to 8\TeV and by the end of June
an additional integrated luminosity of more than 5\fbinv had been
recorded by each of these experiments,
thereby enhancing significantly the sensitivity of the search for the Higgs boson.

This paper reports the results of a search for the SM Higgs boson using samples collected by
the CMS experiment, comprising data recorded at $\sqrt{s}=7$ and 8\TeV. The search is
performed in five decay modes,
$\PH\to\Pgg\Pgg$, $\cPZ\cPZ$, $\PWp\PWm$, $\Pgt^+\Pgt^-$, and
$\bbbar$, in the low-mass range from 110 up to 160\GeV.
In this mass range the Higgs boson production cross section is predicted to have values between 23
(29) and 10 (14)\unit{pb} at $\sqrt{s}=7$ (8)\TeV~\cite{LHCHiggsCrossSectionWorkingGroup:2011ti}.
The natural width of the SM Higgs boson over the same
range is less than 100\MeV and the width of any observed peak would be entirely dominated by instrumental mass
resolution.
In what follows, $\ell$ stands for electrons or muons, $\PH\to\PWp\PWm$ is denoted as $\PH\to\PW\PW$, $\PH \to \Pgt^+\Pgt^-$ as
$\PH \to \Pgt\Pgt$, and $\PH \to \bbbar$ as $\PH \to \cPqb\cPqb$.
For the final states $\cPZ\cPZ$ and $\PW\PW$ in the low-mass region, one or more of the $\cPZ$ or $\PW$
bosons is off mass shell.

With respect to the published
analyses~\cite{Chatrchyan:2012tw,Chatrchyan:2012dg,Chatrchyan:2012ty,Chatrchyan:2012vp,Chatrchyan:2012ww},
most analyses have been re-optimized,
incorporating improvements in reconstruction performance and event
selection, and mitigating the more challenging conditions due to
the higher LHC intensities in 2012.
The new analyses presented herein, of 8\TeV samples, and of 7\TeV
samples featuring modified event selection criteria, were performed in a ``blind'' way:
the algorithms and selection procedures were formally approved and
fixed before the results from data in the signal region were examined.
In the previously published analyses similar but less formal procedures were followed.

Within the context of this search for the SM Higgs boson, we report the observation of an excess of
events above the expected background, consistent with the production
of a new particle with
mass near 125\GeV. The observed local significance is 5.0 standard deviations ($\sigma$), compared
with an expected significance of 5.8\,$\sigma$. The evidence is
strongest in the two final states with the best mass
resolution, namely $\PH\to\Pgg\Pgg$ with a significance of 4.1\,$\sigma$ and $\PH\to\cPZ\cPZ$ (with the
$\cPZ$ bosons decaying to electrons or muons) with a significance of 3.2\,$\sigma$. The decay to two photons indicates that the new particle is a boson with spin different from one.

\section{The CMS experiment}\label{sec:Apparatus}

The possibility of detection of the SM Higgs boson played a crucial role in the conceptual design of the
CMS experiment as a benchmark to test the performance of the
detector~\cite{Pimia:1990zy,DellaNegra:1992hp,Ellis:1994sq}.
Since the SM Higgs boson
mass is not predicted by theory and its production cross section and natural width vary widely
over the allowed mass range, a search was envisaged over a large range of masses and in
diverse decay modes: pairs of photons, Z bosons, W bosons, $\Pgt$ leptons, and $\cPqb$
quarks. Planning in view of the analysis of all these channels ensured a detector capable of
observing a Higgs boson over a broad mass range
and able to detect most potential signals of new physics.

The central feature of the CMS apparatus~\cite{Chatrchyan:2008zzk}
is a superconducting solenoid of 6\unit{m}
internal diameter, which provides a magnetic field of 3.8\unit{T}. Within the field volume are a
silicon pixel and strip tracker, a lead tungstate crystal electromagnetic calorimeter (ECAL),
and a brass/scintillator hadron calorimeter (HCAL). Muons are measured in gas-ionization
detectors embedded in the steel flux-return yoke. Extensive forward calorimeters complement
the coverage provided by the barrel and endcap detectors.

Charged particles are tracked within the pseudorapidity range
$|\eta|<2.5$, where
\ifthenelse{\boolean{cms@external}}{}{\linebreak[4]}$\eta=-\ln[\tan(\theta/2)]$,
and $\theta$ is the
polar angle measured from the positive $z$ axis (along the anticlockwise beam direction).
The silicon pixel
tracker comprises 66~million $100\times150\mum^2$ pixels, arranged in three barrel
layers and two disks at each end. The silicon strip tracker, organized in ten barrel
layers and twelve disks at each end, comprises 9.3 million strips with pitch between 80 and
180\mum, with a total silicon surface area of $198\unit{m}^2$. The tracker has a
track-finding efficiency larger than 99\% for muons with transverse momentum
$\pt$ greater than 1\GeV and a transverse momentum resolution
between 1.5 and 2.5\% for charged tracks of $\pt \sim 100$\GeV in the
central region ($|\eta| <$ 1.5).
Measurements of the impact parameters of charged tracks and secondary vertices are used to
identify jets that are likely to contain the hadronisation and decay
products of $\cPqb$ quarks (``$\cPqb$ jets'').
A b-jet tagging efficiency of more than 50\% is achieved with a
rejection factor for light-quark jets of ${\sim}200$, as measured in $\ttbar$ events in data~\cite{CMS-PAS-BTV-11-004}.
The dimuon mass resolution at the $\Upsilon$ mass, dominated by
instrumental effects, is measured to be 0.6\% in the barrel region~\cite{PhysRevD.83.112004}, consistent with the design
goal.

The ECAL is a fine-grained hermetic calorimeter consisting of 75\,848
lead tungstate crystals, arranged in a quasi-projective geometry and distributed in a
barrel region ($|\eta| < 1.48$) and two endcaps that extend up to $|\eta| = 3.0$.
The front-face cross section of the crystals is $22\times 22\mm^2$ in the barrel region
and $28.6\times 28.6\mm^2$ in the endcaps.
Preshower detectors consisting of two planes of silicon sensors interleaved with a total of
three radiation lengths of lead absorber are located in front of the endcaps. Electromagnetic showers
are very narrow in lead tungstate (Moli\`ere radius of 21\mm), helping in particle
identification and in the implementation of isolation criteria.
In the central barrel region the energy resolution of electrons that
do not radiate substantially in the tracker material indicates that
the resolution of unconverted photons is consistent with design goals.
For such photons the diphoton mass resolution is
1.1\GeV at a mass of 125\GeV.

The HCAL barrel and endcaps are sampling calorimeters consisting of brass and
scintillator plates, covering $|\eta| < 3.0$. Their thickness varies from 7 to 11
interaction lengths, depending on $\eta$; a scintillator ``tail catcher'' placed outside
the coil of the solenoid, just in front of the innermost muon detector, extends the instrumented thickness to more
than 10 interaction lengths everywhere.  Iron forward calorimeters with quartz fibers, read
out by photomultipliers, extend the calorimeter coverage up to
$|\eta| = 5.0$.

Muons are measured in the range $|\eta| < 2.4$, with detection planes based on
three technologies: drift tubes ($|\eta| <$ 1.2), cathode strip chambers
($0.9 < |\eta| < 2.4$), and resistive plate chambers ($|\eta| < 1.6$).
The first two technologies provide a precise position measurement and
trigger whilst the third provides precise timing
information as well as a second and independent trigger.
The muon system consists of four stations in the barrel and endcaps, designed to ensure robust
triggering and detection of muons
over a large angular range. In the barrel region each muon station consists of twelve
drift-tube layers, except for the outermost station, which has eight layers. In the endcaps, each
muon station consists of six detection planes. The precision of the $r$-$\phi$
measurement is 100\mum in the drift tubes and varies from 60 to
140\mum in the cathode strip chambers.

The CMS trigger and data acquisition systems ensure that potentially
interesting events are recorded with high efficiency. The first level (L1)
trigger, comprising the calorimeter, muon, and global trigger
processors, uses
coarse-granularity information to select the most interesting events in less than 4\mus. The
detector data are pipelined to ensure negligible deadtime up to a L1
rate of 100\unit{kHz}.
After L1 triggering, data are transferred from the readout electronics of all subdetectors,
through the readout network, to the high-level-trigger processor farm, which
operates offline-quality reconstruction algorithms to decrease the event rate to around
0.5\unit{kHz}, before data storage.

The CMS experiment employs a highly distributed computing infrastructure, with a
primary Tier-0 centre at CERN, supplemented by seven Tier-1, more than 50 Tier-2, and many
Tier-3 centres at national laboratories and universities throughout
the world.
The CMS software running on this high-perfor\-mance
computing system executes numerous tasks, including the reconstruction
and analysis of the collected data, as well as the generation and
detailed detector simulation of Monte Carlo (MC) event samples.

\section{Event reconstruction}\label{sec:EventDesc}

The CMS ``particle-flow'' event description algorithm~\cite{CMS-PAS-PFT-09-001,CMS-PAS-PFT-10-001}
is used to reconstruct and identify each single particle with an
optimized combination of all subdetector information.
In this process, the identification of the particle (photon,
electron, muon, charged hadron, neutral hadron) plays an important
role in the determination of the particle momentum. The reconstructed particles are henceforth
referred to as objects.

Jets are reconstructed by clustering the particle-flow objects with
the anti-k$_\mathrm{T}$ algorithm~\cite{Cacciari:2008gp} using a
distance parameter of 0.5.
Additional selection criteria are applied to each event to remove
spurious features originating from isolated noise patterns in
certain HCAL regions, and from anomalous signals caused by
particles depositing energy in the
silicon avalanche photodiodes used in the ECAL barrel region.
The average number of {pp} interactions per LHC bunch crossing
is estimated to be about 9 and 19 in the 7\TeV (2011) and
8\TeV (2012) data sets, respectively.
Energy from overlapping pp interactions (``pileup"), and from the underlying
event, is subtracted using the \textsc{FastJet}
technique~\cite{Cacciari:2007fd,Cacciari:2008gn,Cacciari:2011ma},
which is based on the calculation of the $\eta$-dependent transverse momentum
density, evaluated on an event-by-event basis.

The jet momentum is determined as the vector sum of all particle
momenta in the jet.
Jet energy corrections
are derived from simulation studies and from
in situ measurements using the energy balance of dijet and
$\cPZ/\Pgg$+jet events~\cite{CMS-JME-10-011}.
These corrections are between 5\% and 10\% of the
true momentum over the entire \pt spectrum and detector acceptance.
The jet momentum resolution achieved is
$\sigma(\pt)/\pt = 85\%/\sqrt{\pt/\GeVns} \oplus
4\%$ for central jets.
A selection is applied to separate jets originating in the primary
interaction from those due to energy deposits associated
with pileup. The discrimination is based on the differences in the jet
shapes, in the relative multiplicity of charged and neutral components,
and in the fraction of transverse momentum carried by
the hardest components. Within the tracker acceptance the jet tracks are
also required to be consistent with originating at the primary vertex.

The missing transverse energy vector is taken as the negative
vector sum of all particle transverse momenta, and its magnitude is
referred to as $\ETmiss$.
The typical missing transverse energy resolution is around
$0.5 \sqrt{\Sigma\ET}\GeV$~\cite{CMS-JME-10-009}, where $\Sigma\ET$ is the scalar sum of all
particle transverse momenta in \GeVns.

The energy deposited in the ECAL is clustered both with general
clustering algorithms~\cite{CMS-PAS-PFT-10-003} and with algorithms
that constrain the clusters in $\eta$ and $\phi$ to the shapes
expected from electrons and photons with high $\pt$~\cite{CMS-PAS-EGM-10-004}.
These specialised algorithms are used to cluster electromagnetic showers without any
hypothesis regarding whether the particle originating from the interaction point was
a photon or an electron; doing this for electrons from
$\cPZ\to\Pe\Pe$ events provides a measurement of the photon trigger, reconstruction,
and identification efficiencies, as well as of the photon energy scale and resolution.
The width of the reconstructed $\cPZ$ resonance is used to quantify the
performance of the ECAL, using decays to two electrons whose energies are measured
using the ECAL alone, with only their directions being determined
from the tracks.
In the 7\TeV data set, the dielectron mass resolution at the $\cPZ$ boson
mass is 1.56\GeV in the barrel and 2.57\GeV in the endcaps, while in
the 8\TeV sample, reconstructed with preliminary calibration
constants, the corresponding values are 1.61 and 3.75\GeV.
For electrons, the reconstruction combines the clusters in the ECAL and the
trajectory in the silicon tracker~\cite{Baffioni:2006cd}.
Trajectories in the tracker volume are reconstructed using a
model of electron energy loss and fitted with a Gaussian sum filter~\cite{Adam2005}.
The electron momentum is determined from the combination of ECAL and tracker
measurements.

Muon candidates are reconstructed with two algorithms, one in
which the tracks in the silicon detector are matched to segments in
the muon chambers, and another in which a combined
fit is performed to the signals found in both the silicon tracker
and muon systems~\cite{CMS-PAS-PFT-10-003}.
The efficiency to reconstruct a muon of $\pt > 5$\GeV is larger than 95\%,
while the probability to misidentify a hadron as a muon is below 0.1\%.
For $\pt > 200$\GeV the precision of the momentum measurement improves
when the silicon tracker signals are complemented with the information from the
muon chambers.

Selection based on isolation of lepton and photon objects is used extensively.
A requirement is placed on the scalar sum of the transverse momenta of the particles reconstructed within a
distance $\DR_\text{max}$ of the object, sometimes normalised to the $\pt$ of the object. The distance $\DR$ is defined as $\DR = \sqrt{(\Delta\eta)^2+(\Delta\phi)^2}$, where
$\Delta\eta$ and $\Delta\phi$ are the pseudorapidity and azimuthal angle
differences between the particle direction and the object direction.
Typically  $\DR_\text{max}$ is chosen to be 0.3 or 0.4.

The measurement of the integrated luminosity in CMS is based on a
pixel cluster counting method, which exploits the large number of silicon
pixels, and hence their low occupancy in a {pp} collision~\cite{CMS-PAS-SMP-12-008}.
The cross section normalisation is derived from van der Meer
scans~\cite{vanderMeer:296752}.
The uncertainties in the luminosity measurements are 2.2\% and 4.4\% for the 7\TeV
and 8\TeV data sets, respectively.

\section{Searches for the standard model Higgs boson}\label{sec:Strategy}

Initial phenomenological discussions of Higgs boson production and
decay can be found in Refs.~\cite{Ellis:1975ap,Georgi:1977gs,Glashow:1978ab,Cahn:1986zv,Gunion:1987ke,Rainwater:1997dg,Rainwater:1998kj,Rainwater:1999sd}.
Four main mechanisms are predicted for Higgs boson production in pp
collisions:
the gluon-gluon fusion mechanism, which
has the largest cross section,
followed in turn by vector-boson fusion (VBF),
associated $\PW\PH$ and $\cPZ\PH$ production (VH),
and production in association with top quarks ($\ttbar\PH$).
The cross sections for the individual production mechanisms
and the decay branching fractions, together with their uncertainties,
have been computed following
Refs.~\cite{
      Djouadi:1991tka,
      Dawson:1990zj, Spira:1995rr,
      Harlander:2002wh,Anastasiou:2002yz,Ravindran:2003um,
      Catani:2003zt,
      Aglietti:2004nj, Degrassi:2004mx,
      Actis:2008ug,
      Anastasiou:2008tj,deFlorian:2009hc,Baglio:2010ae,deFlorian:2012yg,
      Bozzi:2005wk,deFlorian:2011xf,
      Passarino:2010qk,
      Stewart:2011cf,
      Djouadi:1997yw,hdecay2,
      Bredenstein:2006rh,Bredenstein:2006ha,
      Actis:2008ts,
      Denner:2011mq,
      Ciccolini:2007jr,Ciccolini:2007ec, Figy:2003nv,
      Arnold:2008rz,
      Bolzoni:2010xr,
      Han:1991ia,
      Brein:2003wg,
      Ciccolini:2003jy,
      Hamberg:1990np,
      Denner:2011rn,Ferrera:2011bk,
      Beenakker:2001rj,Beenakker:2002nc,
      Dawson:2002tg,Dawson:2003zu,
      Botje:2011sn,Alekhin:2011sk,Lai:2010vv,Martin:2009iq,Ball:2011mu,
      Baglio:2010ae,Anastasiou:2012hx}
and are compiled in
Refs.~\cite{LHCHiggsCrossSectionWorkingGroup:2011ti,Dittmaier:2012vm}.

The particular set of sensitive
decay modes of the SM Higgs boson depends strongly on $\mH$.
The results presented in this paper are based on the five most sensitive
decay modes in the low-mass region:
$\PH\to\Pgg\Pgg$;
$\PH\to\cPZ\cPZ$ followed by $\cPZ\cPZ$ decays to $4\ell$;
$\PH\to\PW\PW$ followed by decays to $2\ell 2\cPgn$;
$\PH\to\Pgt\Pgt$ followed by at least one leptonic $\Pgt$ decay;
and $\PH\to\cPqb\cPqb$ followed by $\cPqb$-quark fragmentation into jets.
This list is presented in Table~\ref{tab:chans} and comprises the full
set of decay modes and subchannels, or categories, for which both the 7~and
8\TeV data sets have been analysed.
Other lower sensitivity subchannels
($\ttbar\PH$, $\PH\to\cPqb\cPqb$; $\PW/\cPZ\PH$, $\PH\to\Pgt\Pgt$;
 $\PW/\cPZ\PH$, $\PH\to\PW\PW\to 2\ell 2\nu$; $\PH\to\cPZ\cPZ\to 2\ell 2\cPq$)
have also been studied, so far only in the 7\TeV data, and are not included
here.  Adding these analyses in the combination results in an
improvement of 0.1$\,\sigma$ in the overall expected local significance at
$\mH = 125$\GeV.

\begin{table*}[htbp]
  \begin{center}
    \topcaption{Summary of the subchannels, or categories, used in the
      analysis of each decay mode.}
    \label{tab:chans}
    \begin{tabular}{l c c c c c} \hline
Decay & Production & No. of & $\mH$ range & \multicolumn{2}{c}{Int.\ Lum.\ (\!\fbinv)} \\
mode  & tagging & subchannels & (\GeVns) & 7\TeV & 8\TeV \\ \hline\hline
\multirow{2}{*}{$\Pgg\Pgg$} & untagged & 4 & \multirow{2}{*}{110--150} & \multirow{2}{*}{5.1} & \multirow{2}{*}{5.3} \\
                                         & dijet (VBF) & 1 or 2 & & & \\ \hline
$\cPZ\cPZ$ & untagged & 3 & 110--160 & 5.1 & 5.3 \\ \hline
\multirow{2}{*}{$\PW\PW$} & untagged & 4 & \multirow{2}{*}{110--160} & \multirow{2}{*}{4.9} & \multirow{2}{*}{5.1} \\
                                           & dijet (VBF) & 1 or 2 & & & \\ \hline
\multirow{2}{*}{$\Pgt\Pgt$} & untagged & 16 & \multirow{2}{*}{110--145} & \multirow{2}{*}{4.9} & \multirow{2}{*}{5.1} \\
                                          & dijet (VBF) & 4 & & & \\ \hline
bb & lepton, $\ETmiss$ (VH) & 10 & 110--135 & 5.0 & 5.1 \\ \hline
    \end{tabular}
  \end{center}
\end{table*}

For a given value of $\mH$, the search sensitivity depends on
the production cross section, the decay branching fraction into the chosen final state,
the signal selection efficiency, the mass resolution,
and the level of background from identical or similar final-state topologies.

Samples of MC events used to represent signal and background
are fully simulated using $\GEANT$4~\cite{Agostinelli:2002hh}.
The simulations include pileup interactions matching the
distribution of the number of such interactions observed in data.
The description of the Higgs boson signal is obtained from MC
simulation using, for most of the decay modes and production processes,
the next-to-leading-order (NLO) matrix-element generator
\POWHEG~\cite{powheg1,powheg2}, interfaced with \PYTHIA 6.4~\cite{Sjostrand:2006za}.
For the dominant gluon-gluon fusion process, the
transverse momentum spectrum of the Higgs boson in the 7\TeV MC samples
is reweighted to the
next-to-next-to-leading-logarithmic (NNLL) + NLO distribution computed
with \textsc{h}q\textsc{t}~\cite{HqT1,Bozzi:2005wk,deFlorian:2011xf} and
\textsc{FeHiPro}~\cite{FeHiPro1,FeHiPro2},
except in the $\PH\to\cPZ\cPZ$ analysis, where the effect is marginal.
The agreement of the $\pt$ spectrum in the simulation at 8\TeV with
the NNLL + NLO distribution makes reweighting unnecessary.
The improved agreement is due to a modification in the \POWHEG setup
recommended in Ref.~\cite{Dittmaier:2012vm}.
The simulation of associated-production signal samples uses \PYTHIA
and all signal samples for $\PH\to\cPqb\cPqb$ are made using
\POWHEG interfaced to \textsc{herwig++}~\cite{Gieseke:2006ga}.
Samples used for background studies are generated with \PYTHIA,
\POWHEG, and \MADGRAPH~\cite{Alwall:2007st}, and the normalisations are obtained from the best
available NNLO or NLO calculations.
The uncertainty in the signal cross section related to
the choice of parton distribution functions is determined with the PDF4LHC
prescription~\cite{Botje:2011sn,Alekhin:2011sk,Lai:2010vv,Martin:2009iq,Ball:2011mu}.

The overall statistical methodology~\cite{LHC-HCG-Report} used in this paper was developed
by the CMS and ATLAS Collaborations in the context of the LHC Higgs
Combination Group.
A more concise summary of CMS usage in
the search for a Higgs boson is given in Ref.~\cite{Chatrchyan:2012tx}.
The modified frequentist criterion $\CLs$~\cite{Junk:1999kv,Read1}
is used for the calculation of exclusion limits.
Systematic uncertainties are incorporated as nuisance parameters and
are treated according to the frequentist paradigm.
The combination of searches
requires simultaneous analysis of the data selected by all individual analyses,
accounting for all statistical and systematic uncertainties and their
correlations. The probability for a background
fluctuation to be at least as large as the observed maximum excess is termed the local $p$-value, and that
for an excess
\textit{anywhere} in a specified mass
range the global $p$-value.
This probability can be evaluated
by generating sets of simulated data incorporating all correlations between analyses optimized for
different Higgs boson masses.
The global $p$-value (for the specified region)
is greater than the local $p$-value, and this fact is often referred to as
the look-elsewhere effect (LEE)~\cite{LEE}.
Both the local and global $p$-values can be expressed as
a corresponding number of standard deviations
using the one-sided Gaussian tail convention. The magnitude of a possible Higgs boson signal is characterised by the
production cross section times the relevant branching fractions, relative to
the SM expectation, denoted $\sigma/\sigma_\mathrm{SM}$ and referred to as the signal
strength.
The results presented in this paper are obtained using asymptotic formulae~\cite{Cowan:2010st},
including updates recently introduced in the \textsc{RooStats}
package~\cite{RooStats}.

Figure~\ref{fig:exp-pvalues}
shows the expected local $p$-values in the mass range 110--145\GeV for the five
decay modes reported here.
The expected significance of a SM Higgs boson signal at $\mH = 125$\GeV
when the five decay modes are combined is 5.6\,$\sigma$.
The highest sensitivity in this mass range is
achieved in the $\cPZ\cPZ$, $\Pgg\Pgg$, and $\PW\PW$ channels.
Because of the excellent mass resolution (1--2\GeV) achieved in the $\Pgg\Pgg$ and
$\cPZ\cPZ$ channels,
they play a special role in the low-mass region,
where the natural width of the SM Higgs boson is predicted to be a few \!\MeV.
The expected signature in these channels is therefore
a narrow resonance above background, with a width consistent with the detector resolution.

\begin{figure}[htbp]
  \begin{center}
    \includegraphics[width=\cmsFigWideWidth]{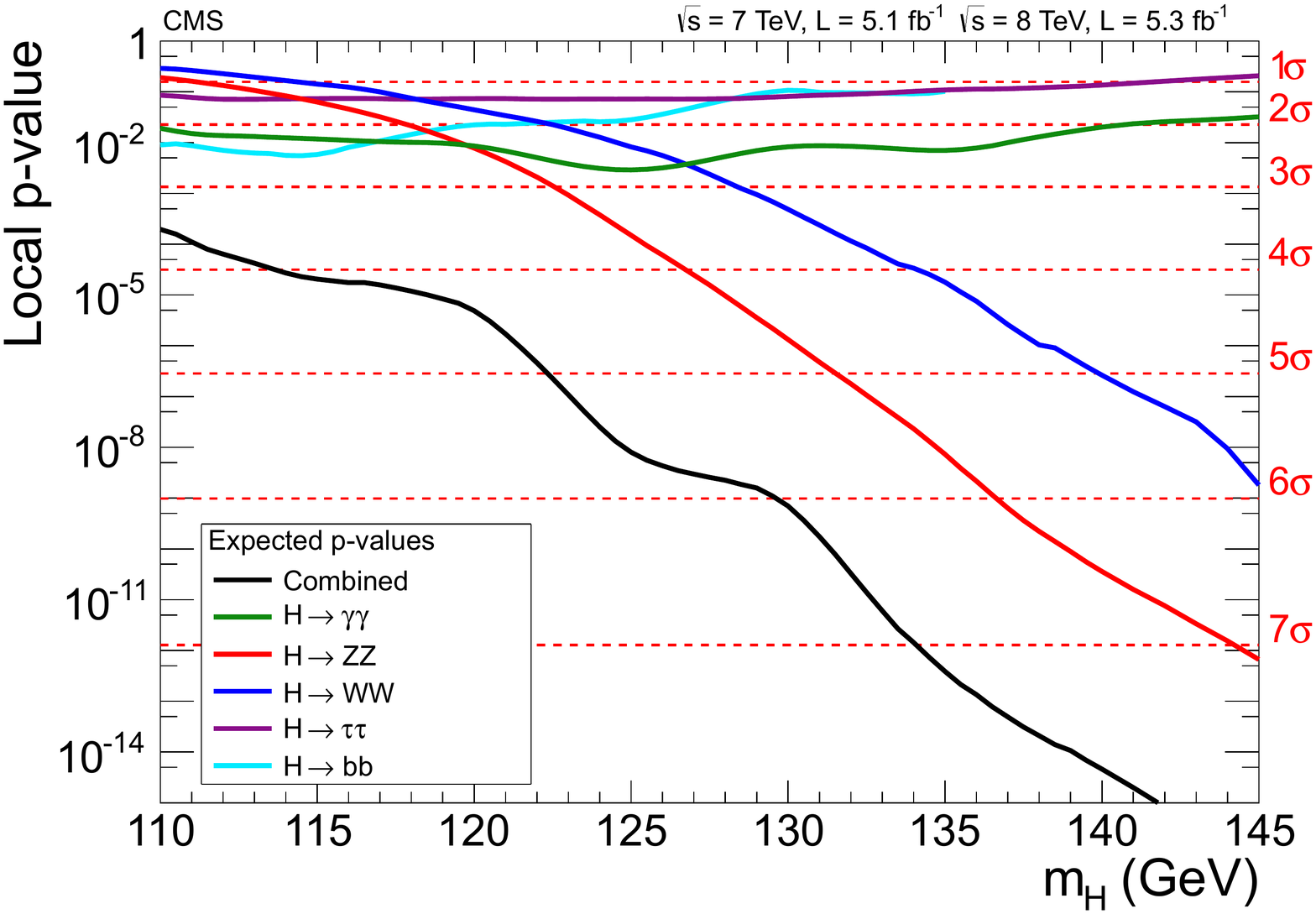}
    \caption{Expected local $p$-values for a SM Higgs boson as a function of
      $\mH$, for the decay modes $\Pgg\Pgg$, $\cPZ\cPZ$, $\PW\PW$, $\Pgt\Pgt$, and $\cPqb\cPqb$ and their combination.
}
    \label{fig:exp-pvalues}
  \end{center}
\end{figure}

\section{Decay modes with high mass resolution}

\subsection{\texorpdfstring{$\PH\to\Pgg\Pgg$}{H to gamma gamma}\label{sec:Hgg}}

In the $\PH\to\Pgg\Pgg$ analysis a search is made for a narrow peak in the diphoton
invariant mass distribution in the range 110--150\GeV,
on a large irreducible background from QCD production of two photons.
There is also a reducible background where one or more
of the reconstructed photon candidates originate from
misidentification of jet fragments.
Early detailed studies indicated this to be one of the most promising channels
in the search for a SM Higgs boson in the low-mass range~\cite{Seez1990a}.

To enhance the sensitivity of the analysis, candidate diphoton events are separated into mutually
exclusive categories of different expected signal-to-background ratios,
based on the properties of the reconstructed photons and on
the presence of two jets satisfying criteria aimed at selecting events
in which a Higgs boson is produced through the VBF process.
The analysis uses multivariate techniques
for the selection and classification of the events.
As an independent cross-check, an analysis is also performed
that is almost identical to the one described in Ref.~\cite{Chatrchyan:2012tw},
using simpler criteria based on the properties of the reconstructed photons to
select and classify events.
The multivariate analysis achieves 15\% higher sensitivity than the
cross-check analysis.

The reconstructed primary vertex that most probably corresponds
to the interaction vertex of the diphoton candidate
is identified using the kinematic properties
of the tracks associated with that vertex and their correlation with
the diphoton kinematics.
In addition, if either of the photons converts and the tracks from the conversion are
reconstructed and identified, the direction of the converted photon
contributes to the identification of the hard-scattering vertex.
More details can be found in Ref.~\cite{Chatrchyan:2012tw}.

The event selection requires two photon candidates satisfying
$\pt$ requirements and ``loose'' photon identification criteria.
These photons must be reconstructed
within the fiducial region, $|\eta|~<~2.5$, excluding the
barrel-endcap transition region, $1.44 < |\eta| < 1.57$.
A $\pt$ threshold of $\mgg/3$ ($\mgg/4$) is applied to
the photon leading (subleading) in $\pt$, where $\mgg$ is the
diphoton invariant mass.
Scaling the $\pt$ thresholds in this way avoids distortion of the
shape of the $\mgg$ distribution.
In the case of events passing the dijet selection, the requirement
on the leading photon is increased to $\mgg/2$,
further reducing background with negligible loss of signal.

Jet selection criteria are applied to the two
jets of largest $\pt$ in the event within $|\eta|< 4.7$.
The jet selection requirements are optimized
using simulated VBF signal and diphoton background events.
The $\pt$ thresholds for the two jets are 30 and 20\GeV, and their $\eta$
separation is required to be greater than 3.5.
The dijet invariant mass is required to be greater than 350 and 250\GeV
for the 7 and 8\TeV data sets, respectively.
The lower dijet invariant mass requirement for the 8\TeV data set
reflects the fact that
for the analysis of that data set, the dijet event category is
divided into two to increase the search sensitivity.
This division creates a second
``tight'' dijet-tagged category in which the dijet invariant
mass must be
greater than 500\GeV and both jets must have
$\pt> 30$\GeV.
Two additional selection criteria, relating the dijet to the diphoton
system, are applied: the difference
between the average pseudorapidity of the two jets and the pseudorapidity
of the diphoton system is required to be less than 2.5,
and the difference in azimuthal angle between the diphoton system
and the dijet system is required to be greater than 2.6 radians.

A multivariate regression is used to extract the photon energy
and a photon-by-photon estimate of the
uncertainty in that measurement.
The calibration of the photon energy scale uses the Z boson mass as a
reference;
ECAL showers coming from electrons in $\cPZ\to\Pe\Pe$ events are
clustered and reconstructed in exactly the same way as photon showers.
The photon selection efficiency, energy resolution, and associated
systematic uncertainties are estimated from data,
using $\cPZ\to\Pe\Pe$ events to derive data/simulation correction factors.
The jet reconstruction efficiency, the efficiency to correctly locate
the vertex position, and the trigger
efficiency, together with the corresponding systematic
uncertainties, are also evaluated from data.

For the multivariate analysis, a boosted decision tree (BDT)~\cite{Yang2005370,Hocker:2007ht}
is trained to give a high output value (score) for signal-like events and for events with
good diphoton invariant mass resolution, based on
the following observables:
(i) the photon quality determined from
electromagnetic shower shape and isolation variables;
(ii) the expected mass resolution;
(iii) the per-event estimate of the probability of
locating the diphoton vertex within 10\mm of its true location along
the beam direction;
and (iv) kinematic characteristics of the photons and the diphoton system.
The kinematic variables are constructed so as to contain no
information about the invariant mass of the diphoton system.
The diphoton events not satisfying the dijet selection are classified into five categories based on the output of the BDT,
with category boundaries optimized for sensitivity to a SM Higgs boson.
Events in the category with smallest expected signal-to-background ratio are rejected,
leaving four categories of events.
Dijet-tagged events with BDT scores smaller than the threshold for the fourth
category are also rejected.
Simulation studies indicate that the background
in the selected event categories is dominated by the irreducible background from QCD production of two photons
and that fewer than 30\% of the diphoton events used in the analysis contain one or more
misidentified photons (predominantly from $\Pgg$+jet production).

Table~\ref{tab:ClassFracs} shows the expected number of signal events
in each event category for a SM Higgs boson
(of $\mH=125$\GeV),
and the background at $\mgg = 125$\GeV, estimated from the fit described below.
The estimated mass resolution is also shown, measured both by
$\sigma_\text{eff}$, half the minimum width
containing 68\% of the signal events, and by the full width
at half maximum (FWHM).
A large variation in the expected signal-to-background ratio between
the categories can be seen, although as a consequence of the
optimization of the category boundaries
the expected signal significances in each category are rather similar.
The differences in the relative signal-to-background ratio between the categories are almost independent
of $\mH$.

\begin{table*}[htbp]
\begin{center}
\topcaption{Expected numbers of SM Higgs boson events ($\mH=125$\GeV) and
estimated background (at $\mgg=125$\GeV) for all event categories of the 7
and 8\TeV data sets.
There are two dijet-tagged categories for the 8\TeV data as
described in the text, and for both data sets the remaining untagged events are
separated into four categories labelled here BDT 0--3, BDT 0 having
the largest expected signal-to-background ratio.
The composition of the SM Higgs boson signal in terms of the production
processes, and its mass resolution, are also given.}

\begin{tabular}{>{\small}c<{\small}|>{\small}r<{\small}||r|>{\small}r<{\small}>{\small}r<{\small}>{\small}r<{\small}>{\small}r<{\small}|>{\centering}b{1.3cm}<{\centering}|>{\centering}b{2.0cm}<{\centering}||r@{\,$\pm$\,}l}
\hline
\multicolumn{2}{c||}{\multirow{2}{*}{\begin{minipage}[t]{2cm}\begin{center}Event categories\end{center}\end{minipage}}} & \multicolumn{7}{c||}{SM Higgs boson expected signal ($\mH=125$\GeV)} & \multicolumn{2}{c}{\multirow{2}{*}{\begin{minipage}[t]{2.5cm}\begin{center}Background \small{$\mgg=125\GeV$ (events/\GeVns)}\end{center}\end{minipage}}}\tabularnewline
\cline{3-9}
\multicolumn{2}{c||}{} & Events & ggH & VBF & VH & ttH & $\sigma_\text{eff}$ \small{(\GeVns)} & \small{FWHM/2.35} \small{(\GeVns)} & \multicolumn{2}{c}{} \tabularnewline
\hline\hline
\multirow{5}{*}{\begin{sideways}7\TeV, 5.1\fbinv\end{sideways}}
& BDT 0       &  3.2 & 61\% & 17\% & 19\% & 3\% & 1.21 & 1.14 & \rule{6mm}{0mm} 3.3 & 0.4 \tabularnewline
& BDT 1       & 16.3 & 88\% &  6\% &  6\% &  -- & 1.26 & 1.08 &  37.5 & 1.3 \tabularnewline
& BDT 2       & 21.5 & 92\% &  4\% &  4\% &  -- & 1.59 & 1.32 &  74.8 & 1.9 \tabularnewline
& BDT 3       & 32.8 & 92\% &  4\% &  4\% &  -- & 2.47 & 2.07 & 193.6 & 3.0 \tabularnewline
& Dijet tag   &  2.9 & 27\% & 72\% &  1\% &  -- & 1.73 & 1.37 &   1.7 & 0.2 \tabularnewline
\hline
\multirow{6}{*}{\begin{sideways}8\TeV, 5.3\fbinv\end{sideways}}
& BDT 0       &  6.1 & 68\% & 12\% & 16\% & 4\% & 1.38 & 1.23 &   7.4 & 0.6 \tabularnewline
& BDT 1       & 21.0 & 87\% &  6\% &  6\% & 1\% & 1.53 & 1.31 &  54.7 & 1.5 \tabularnewline
& BDT 2       & 30.2 & 92\% &  4\% &  4\% &  -- & 1.94 & 1.55 & 115.2 & 2.3 \tabularnewline
& BDT 3       & 40.0 & 92\% &  4\% &  4\% &  -- & 2.86 & 2.35 & 256.5 & 3.4 \tabularnewline
& Dijet tight &  2.6 & 23\% & 77\% &   -- &  -- & 2.06 & 1.57 &   1.3 & 0.2 \tabularnewline
& Dijet loose &  3.0 & 53\% & 45\% &  2\% &  -- & 1.95 & 1.48 &   3.7 & 0.4 \tabularnewline

\hline
\end{tabular}

\label{tab:ClassFracs}
\end{center}
\end{table*}

The background is estimated from data, without the use of MC
simulation,
by fitting the diphoton invariant mass distribution in each of the categories
in a range (100 $< \mgg <$ 180\GeV) extending slightly above and
below that in which the search is performed.
The choices of the function
used to model the background and of the fit range are made based on a study of the possible
bias in the measured signal strength.
Polynomial functions are used.
The degree is chosen by requiring that the potential bias be
at least a factor of 5 smaller
than the statistical accuracy of the fit prediction.
The required polynomial degree ranges from 3 to 5.

A further independent analysis (referred to as the sideband background model) is
performed using a different approach to the background modelling.
Its sensitivity is very similar to that of the standard analysis.
It employs a fit to the output of an additional BDT that
takes as input the diphoton invariant mass and the diphoton BDT output,
and uses a background model
derived from the sidebands of the invariant-mass distribution.
A fit to the diphoton invariant-mass distribution is used to obtain the background normalisation.
This fit is of a power law and excludes a window of width $\pm$2\%$\times\mH$
around the mass hypothesis. The methodology allows a systematic
uncertainty to be assigned to the fit shape.

The expected 95\% CL upper limit on the signal strength
$\sigma/\sigma_\mathrm{SM}$, in the background-only hypothesis,
for the combined 7~and 8\TeV
data, is less than 1.0 in the range
110 $< \mH <$ 140\GeV, with a value of 0.76 at
$\mH$ = 125\GeV.
The observed limit indicates the presence of a significant excess at $\mH$ =
125\GeV in both the 7~and 8\TeV data.
The features of the observed limit are confirmed
by the independent sideband-background-model and cross-check analyses.
The local $p$-value is shown as a function of $\mH$ in
Fig.~\ref{fig:p-value-gg} for the 7~and 8\TeV data,
and for their combination.
The expected (observed) local $p$-value for a SM Higgs boson of mass 125\GeV corresponds to $2.8\,(4.1)\,\sigma.$ In the sideband-background-model and cross-check analyses, the observed local
$p$-values for $\mH=125\GeV$ correspond to 4.6 and $3.7\,\sigma$, respectively.
The best-fit signal strength for a SM Higgs boson mass hypothesis of
125\GeV is $\sigma/\sigma_\mathrm{SM} = 1.6\pm0.4$.

In order to illustrate, in the \mgg distribution, the significance given by the statistical methods,
it is necessary to take into account the large
differences in the expected signal-to-background ratios of the event
categories shown in Table~\ref{tab:ClassFracs}.
The events are weighted according to the category in which they fall.
A weight proportional to $S/(S+B)$ is used, as
suggested in Ref.~\cite{Barlow:1986ek},
where $S$ and $B$ are the number of signal and background events,
respectively, calculated from the simultaneous signal-plus-background fit to all
categories (with varying overall signal strength) and integrating over a $2\sigma_\text{eff}$ wide window, in each category,
centred on 125\GeV.
Figure~\ref{fig:mass-gg} shows the data, the signal
model, and the background model, all weighted.
The weights are
normalised such that the integral of the weighted signal model matches
the number of signal events given by the best fit.
The unweighted distribution, using the same binning but in a more
restricted mass range, is shown as an inset.
The excess at 125\GeV is evident in both the weighted and unweighted distributions.

\begin{figure}[htbp]
   \begin{center}
     \includegraphics[width=\cmsFigWideWidth]{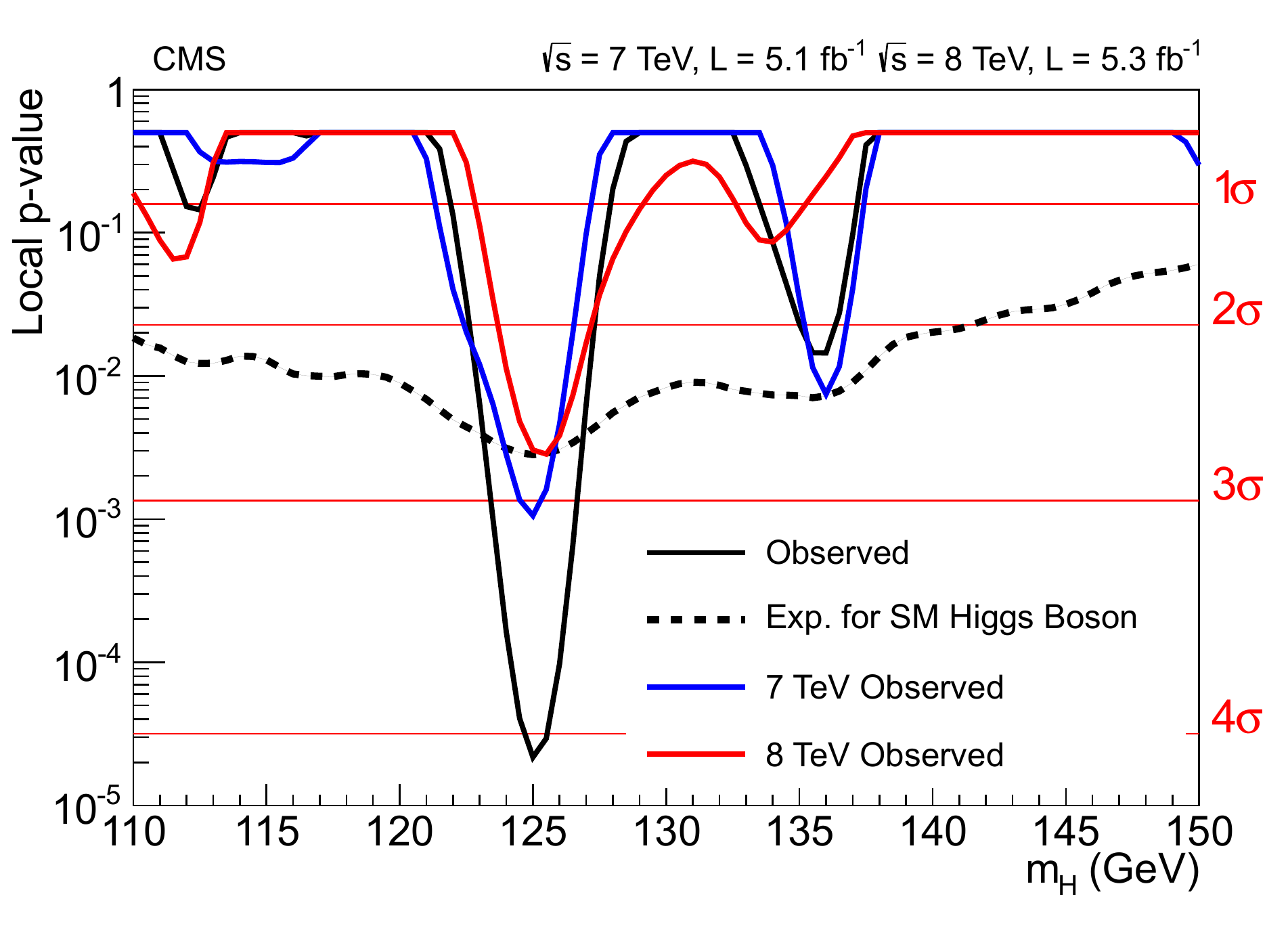}
     \caption{\label{fig:p-value-gg} The local $p$-value as a function
       of $\mH$ in the $\Pgg\Pgg$ decay mode for the combined 7~and 8\TeV data sets. The
       additional lines show the values for the two data sets taken
       individually.
The dashed line shows the expected local $p$-value for the combined
data sets, should a SM Higgs boson exist with mass $\mH$.
}
   \end{center}
\end{figure}

\begin{figure}[htbp]
   \begin{center}
     \includegraphics[width=\cmsFigWideWidth]{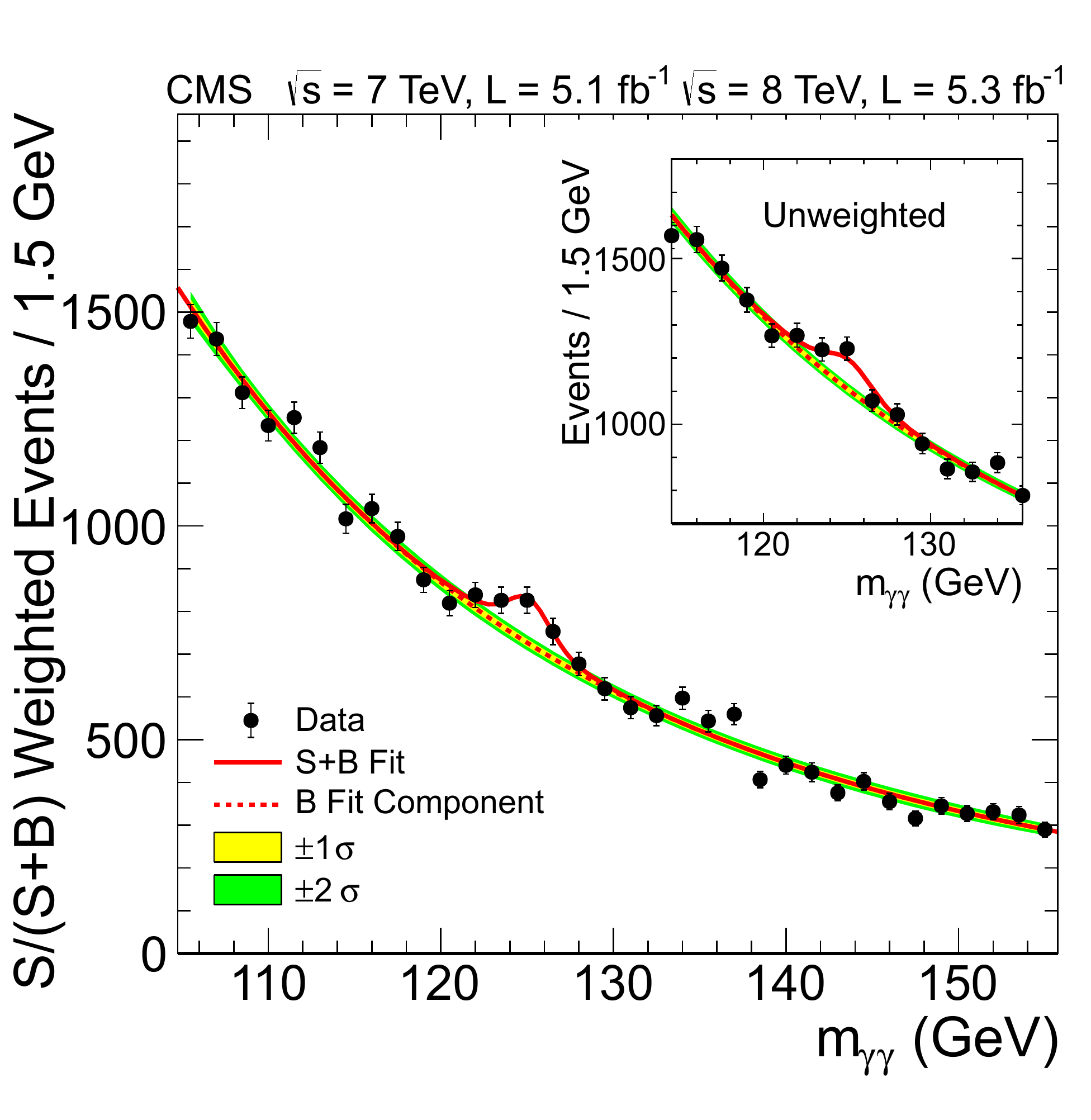}
     \caption{\label{fig:mass-gg} The diphoton invariant mass
       distribution with each event weighted
       by the $S/(S+B)$ value of its category. The lines
       represent the fitted background and signal, and the coloured
       bands represent the $\pm$1 and $\pm$2 standard deviation
       uncertainties in the background estimate.
       The inset shows the central part of the unweighted invariant
       mass distribution.
}
   \end{center}
\end{figure}

\subsection{\texorpdfstring{$\PH\to\cPZ\cPZ$}{H to ZZ}\label{sec:HZZ}}

In the $\PH\to\cPZ\cPZ\to 4\ell$ decay mode a search is made for a narrow four-lepton mass peak
in the presence of a small continuum background.
Early detailed studies outlined the promise of this mode over a
wide range of Higgs boson masses~\cite{DellaNegra1990a}.
Only the search in the range 110--160\GeV is
reported here.
Since there are differences in the reducible background rates and mass resolutions between the subchannels
4e, 4$\mu$, and 2e2$\mu$, they are analysed separately.
The background sources include
an irreducible four-lepton contribution from direct $\cPZ\cPZ$
production via $\cPq\cPaq$ and gluon-gluon processes.
Reducible contributions arise from
$\cPZ+\bbbar$ and $\ttbar$ production where the final states contain
two isolated leptons and two $\cPqb$-quark jets producing secondary
leptons.
Additional background arises from $\cPZ+$jets and $\PW\cPZ+$jets events where jets are misidentified as
leptons.
Compared to the analysis reported in Ref.~\cite{Chatrchyan:2012dg},
the present analysis employs improved muon reconstruction, improved
lepton identification and isolation, and a kinematic discriminant
exploiting the decay kinematics expected for the signal events.
An algorithm to recover final-state radiation (FSR) photons has also been deployed.

Electrons are required to have $\pt > 7$\GeV and $|\eta| < 2.5$.
The corresponding requirements for muons are $\pt > 5$\GeV
and $|\eta| < 2.4$.
Electrons are selected using a multivariate identifier trained using a sample of $\PW+$jets events, and the
working point is optimized using $\cPZ+$jets events.
Both muons and electrons are required to be isolated.
The combined reconstruction and selection efficiency
is measured using electrons and muons in $\cPZ$ boson decays.
Muon reconstruction and identification efficiency for muons with
$\pt < 15$\GeV  is measured using  J/$\psi$ decays.

The electron or muon pairs from $\cPZ$ boson decays are required to originate
from the same primary vertex.
This is ensured by requiring that the significance of the impact
parameter with respect to the event vertex satisfy $| S_\mathrm{IP} | < 4$
for each lepton, where $S_\mathrm{IP} = I/\sigma_{I}$,
$ I$ is the three-dimensional lepton impact parameter
at the point of closest approach to the vertex, and
$\sigma_{I}$ its uncertainty.

Final-state radiation from the leptons is recovered and included
in the computation of the lepton-pair invariant mass.
The FSR recovery is
tuned using simulated samples of $\cPZ\cPZ\to 4\ell$ and tested on data samples
of \cPZ\ boson decays to electrons and muons.
Photons reconstructed within $\abs{\eta} < 2.4$ are
considered as possibly due to FSR.
The photons must satisfy the following requirements. They must be within $\DR < 0.07$ of a muon and have
 $\pt^\gamma > 2$\GeV
(most photon showers within this distance of an electron having already been
automatically clustered with the electron shower);
or if their distance from a lepton is in the range $0.07 < \DR < 0.5$,
they must satisfy $\pt^\gamma > 4$\GeV, and be isolated within $\DR = 0.3$.
Such photon candidates are combined with the lepton if the resulting
three-body invariant mass is less than 100\GeV and closer to the \cPZ\ boson mass
than the mass before the addition of the photon.

The event selection requires two pairs of same-flavour, oppositely charged
leptons.
The pair with invariant mass closest to the Z boson mass is required to have a mass in the
range 40--120\GeV and the other pair is required to have
a mass in the range 12--120\GeV.
The ZZ background is evaluated from MC simulation studies.
Two different approaches are employed to estimate the reducible and instrumental backgrounds from data.
Both start by selecting events in a background control region, well
separated from the signal region,
by relaxing the isolation and identification criteria for two
same-flavour reconstructed leptons.
In the first approach, the additional pair of leptons is required to
have the same charge (to avoid signal contamination) while in the
second, two opposite-charge leptons failing the isolation and identification
criteria are required.
In addition, a control region with three passing leptons and one failing lepton is used to estimate contributions
from backgrounds with three prompt leptons and one misidentified lepton.
The event rates measured in the background control region are extrapolated to
the signal region using the measured probability for a
reconstructed lepton to pass the isolation and identification
requirements.
This probability is measured in an independent sample.
Within uncertainties, comparable background counts in the signal
region are estimated by both methods.

The number of selected $\cPZ\cPZ\rightarrow 4\ell$ candidate events in the mass range $110 < m_{4\ell} < 160$\GeV,
in each of the three final states, is given in
Table~\ref{tab:SelectYieldsLowMass}, where $m_{4\ell}$ is the
four-lepton invariant mass.
The number of predicted background events, in each of the three final states, and their uncertainties are also given,
together with the number of signal events expected from a SM Higgs boson
of $\mH = 125$\GeV.
The $m_{4\ell}$ distribution is shown in Fig.~\ref{fig:ZZmass}.
There is a clear peak at the Z boson mass where the decay
$\cPZ\to 4\ell$ is reconstructed.
This feature of the data is well reproduced by the background estimation.
The figure also shows an excess of events above the expected
background around 125\GeV.
The total background and the numbers of events observed in
the three bins where an excess is seen are also shown in Table~\ref{tab:SelectYieldsLowMass}.
The combined signal reconstruction and selection efficiency, with respect to the $\mH = 125\GeV$ generated signal with  $m_{\ell\ell} > 1\GeV$ as the only cut, is 18\% for the 4\Pe\ channel, 40\% for the 4\Pgm\ channel, and 27\% for the 2\Pe2\Pgm\ channel.

\begin{figure}[htbp]
   \begin{center}
     \includegraphics[width=\cmsFigWideWidth]{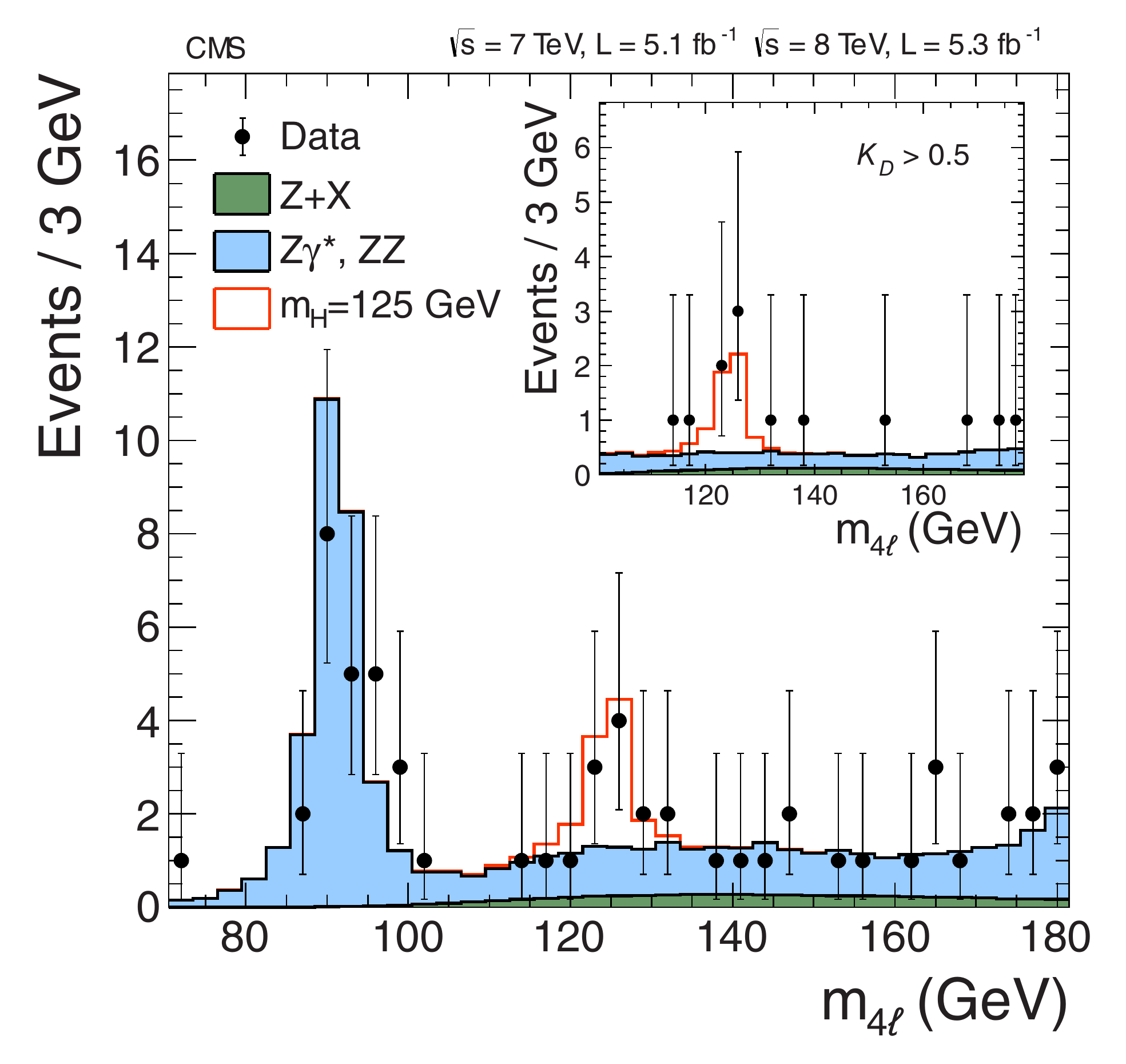}
     \caption{\label{fig:ZZmass} Distribution of the four-lepton invariant mass
       for the $\cPZ\cPZ\to 4\ell$ analysis.
The points represent the data, the filled histograms represent the background,
and the open histogram shows the signal expectation for a Higgs boson
of mass $\mH = 125$\GeV, added to the background expectation.
The inset shows the $m_{4\ell}$ distribution after selection of events with
$K_D > 0.5$, as described in the text.
}
   \end{center}
\end{figure}

\begin{table*}[htbp]
\begin{center}
\topcaption{The number of selected events, compared to the expected
background yields and expected number of signal events ($\mH =
125$\GeV) for each final state in the $\PH\to\cPZ\cPZ$ analysis.
The estimates of the $\cPZ+X$ background are based on data.
These results are given for the mass range from 110 to 160\GeV.
The total background and the observed numbers of events are also shown
for the three bins (``signal region'') of Fig.~\ref{fig:ZZmass} where an excess is seen ($121.5 < m_{4\ell} < 130.5$\GeV).
}
\label{tab:SelectYieldsLowMass}
\begin{tabular}{l|c|c|c||c}
\hline
Channel & $4\Pe$ & $4\Pgm$ & $2\Pe2\Pgm$ & $4\ell$ \\
\hline\hline
$\cPZ\cPZ$ background & 2.7 $\pm$ 0.3 & 5.7 $\pm$ 0.6 & 7.2 $\pm$ 0.8 & 15.6 $\pm$ 1.4 \\
$\cPZ+X$         &  $1.2 ^{ + 1.1}_{ - 0.8 }$ & $0.9 ^{ + 0.7 }_{ - 0.6 }$ & $2.3 ^{ + 1.8 }_{ - 1.4 }$ & $4.4 ^{ + 2.2 \phantom{^0}}_{ - 1.7\phantom{_0} }$ \\ %
\hline
All backgrounds \small{($110 < m_{4\ell} < 160$\,GeV)} &  $4.0 \pm 1.0 $ & $6.6 \pm 0.9 $ & $9.7 \pm 1.8 $ & $20 \pm 3$\\ %
\hline
Observed \small{($110 < m_{4\ell} < 160$\,GeV)} & 6 & 6 & 9 & 21\\
\hline \hline
Signal \small{($\mH = 125$\GeV)} &  1.36  $\pm$  0.22  &  2.74  $\pm$  0.32  &  3.44  $\pm$  0.44 & 7.54 $\pm$ 0.78 \\
\hline \hline
All backgrounds \small{(signal region)} &  $0.7 \pm 0.2 $  &  $1.3 \pm 0.1$  &  $1.9\pm 0.3$ & $3.8 \pm 0.5$\\
\hline
Observed \small{(signal region)} & 1 & 3 & 5 & 9\\
\hline
\end{tabular}
\end{center}
\end{table*}

The kinematics of the $\PH\rightarrow\cPZ\cPZ\rightarrow 4\ell$ process in its centre-of-mass frame,
for a given invariant mass of the four-lepton system,
is fully described by five angles and the invariant masses of the two lepton pairs~\cite{Cabibbo:1965zz,Gao:2010qx,DeRujula:2010ys}.
These seven variables provide
significant discriminating power between signal and background.
The momentum of the $\cPZ\cPZ$ system may further differentiate signal from background, but would introduce
dependence on the production mechanism, and on the modelling of the QCD effects, and is therefore not considered here.
A kinematic discriminant is constructed based on the probability ratio of the signal and background hypotheses,
$K_{D}=\mathcal{P}_\text{sig}/(\mathcal{P}_\text{sig}+\mathcal{P}_\text{bkg})$,
as described in Ref.~\cite{Chatrchyan:2012sn}.
The likelihood ratio is defined for each value of $m_{4\ell}$.
For the signal, the phase-space and \cPZ\ propagator
terms~\cite{Choi:2002jk} are
included in a fully analytic parameterization~\cite{Gao:2010qx},
while the background probability is tabulated using a simulation of the
$\cPq\cPaq\to\cPZ\cPZ/\cPZ\gamma$ process.
The statistical analysis only includes events with $m_{4\ell} >100$\GeV.

Figure~\ref{fig:ZZmela} (upper) shows the distribution of $K_D$ versus $m_{4\ell}$
for events selected in the $4\ell$ subchannels.
The colour-coded regions show the expected background.
Figure~\ref{fig:ZZmela} (lower) shows the same two-dimensional distribution of
events, but this time superimposed on the expected event density from
a SM Higgs boson ($\mH$ = 125\GeV).
A clustering of events is observed around 125\GeV with a large value of $K_D$, where the background expectation is low and
the signal expectation is high, corresponding to
the excess seen in the one-dimensional mass distribution.
The $m_{4\ell}$ distribution of events satisfying $K_D > 0.5$ is shown
in the inset in Fig.~\ref{fig:ZZmass}.

\begin{figure}[htbp]
   \begin{center}
     \includegraphics[width=\cmsFigWideWidth]{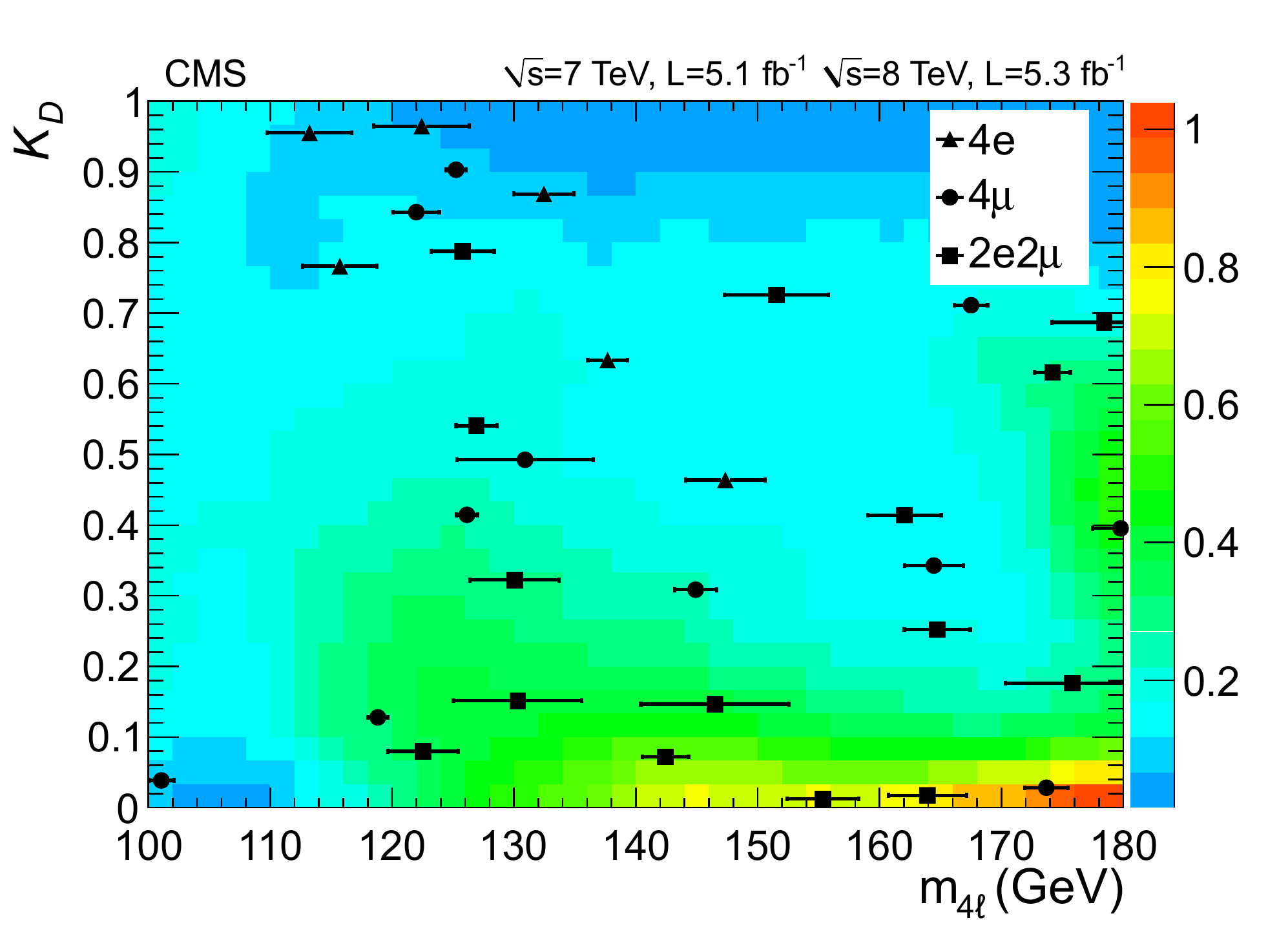}
     \includegraphics[width=\cmsFigWideWidth]{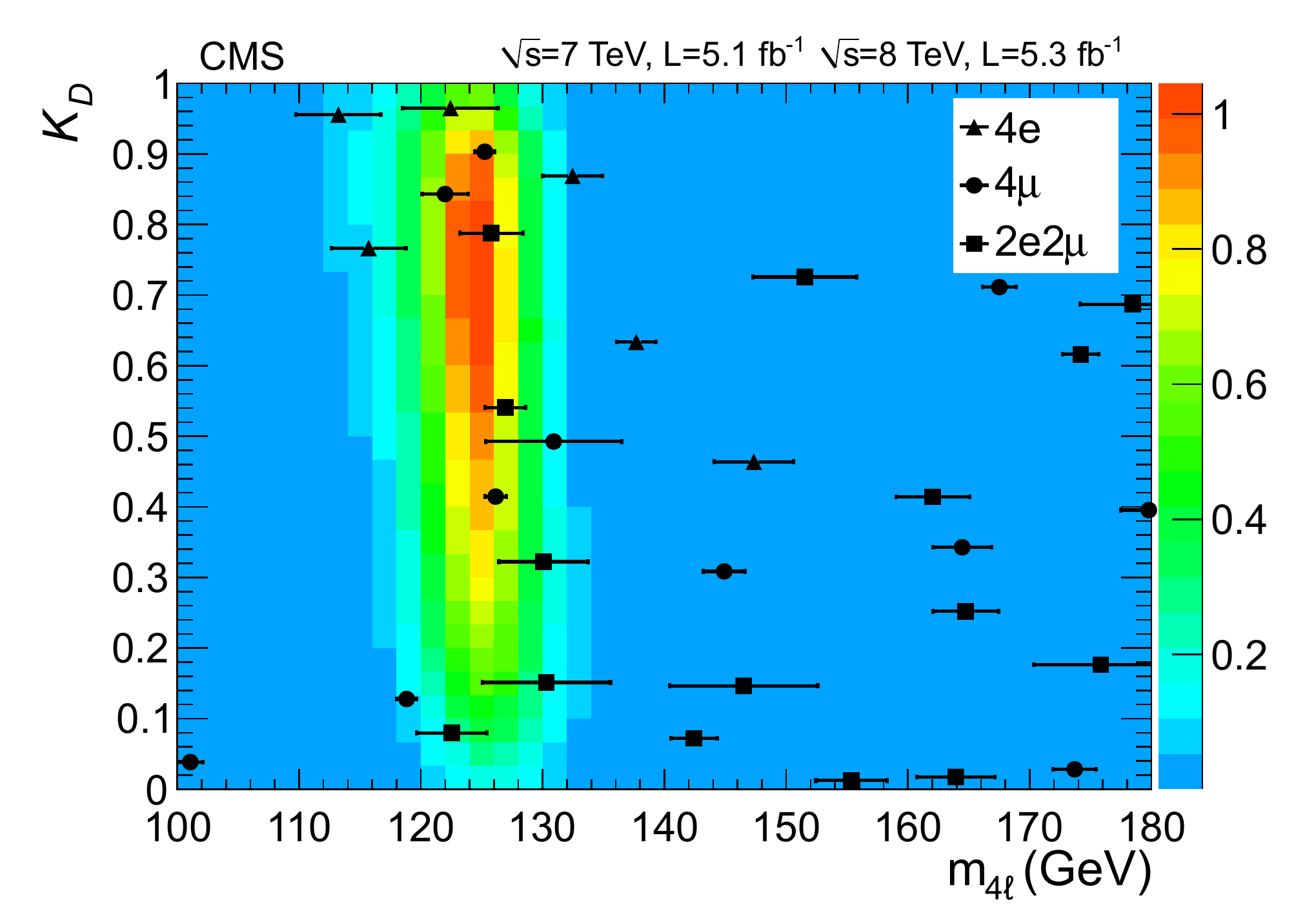}
    \caption{\label{fig:ZZmela} The distribution of events
selected in the $4\ell$ subchannels for the kinematic discriminant $K_{D}$ versus $m_{4\ell}$.
Events in the three final states are marked by filled symbols (defined in the legend).
The horizontal error bars indicate the estimated mass resolution.
In the upper plot the colour-coded regions show the background expectation;
in the lower plot the colour-coded regions show the event density expected from a SM Higgs boson ($\mH$ = 125\GeV)
(both in arbitrary units).
}
   \end{center}
\end{figure}

\begin{figure} [htbp]
\begin{center}
\includegraphics[width=\cmsFigWideWidth]{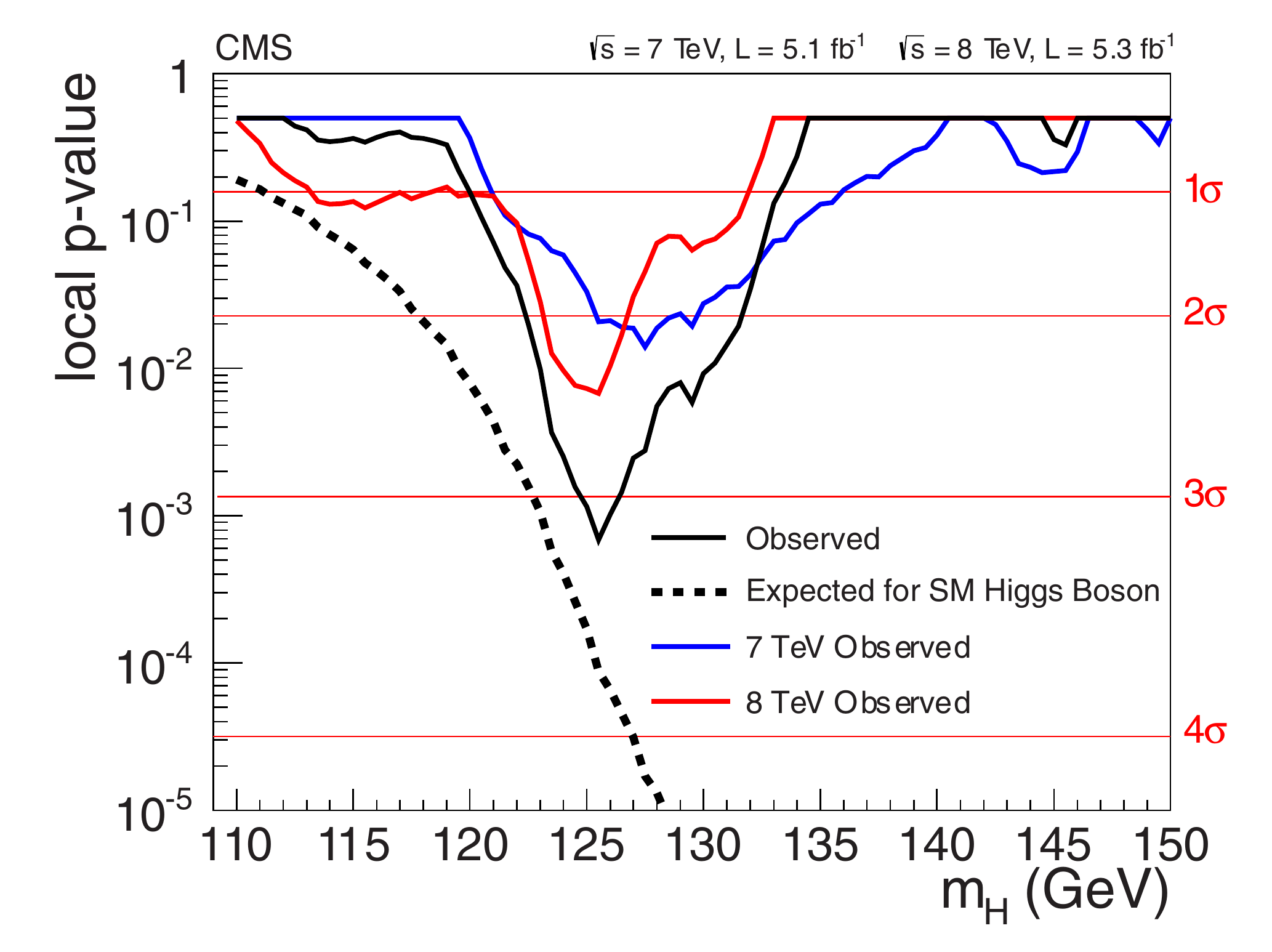}
\caption{
The observed local $p$-value for the $\cPZ\cPZ$ decay mode as a function of the SM Higgs
boson mass.
The dashed line shows the expected local $p$-values
for a SM Higgs boson with a mass $\mH$.
}
\label{fig:pvalue_ZZ}
\end{center}
\end{figure}

There are three final states and two data sets (7~and
8\TeV), and thus the statistical treatment requires
six simultaneous two-dimensional maximum-likelihood fits for each value
of $\mH$, in the variables ${m}_{4\ell}$ and $K_{D}$.
Systematic uncertainties are evaluated from data for the trigger efficiency
and for the combined lepton reconstruction, identification, and isolation efficiencies, as described in~\cite{CMS:2011aa}.
Systematic uncertainties in the energy/momentum calibration
and in the energy resolution are estimated from data.
Additional systematic uncertainties arise from limited statistical precision in the reducible
background control regions.

The expected 95\% CL upper limit on the signal strength
$\sigma/\sigma_\mathrm{SM}$,
in the background-only hypothesis,
for the combined 7~and 8\TeV
data, falls steeply between 110 and 140\GeV, and has a value
of 0.6 at $\mH = 125$\GeV.
The observed upper limit indicates the presence of a significant excess
in the range $120 < \mH < 130$\GeV. The local $p$-value is shown as a function of $\mH$ in Fig.~\ref{fig:pvalue_ZZ} for the 7 and 8\TeV data, and for their combination.
The minimum local $p$-value in the data occurs at $\mH = 125.6$\GeV
and has a significance of 3.2$\,\sigma$ (expected 3.8$\,\sigma$).
The combined best-fit signal strength for a SM Higgs
boson mass hypothesis of 125.6\GeV is $\sigma/\sigma_\mathrm{SM} = 0.7^{+0.4}_{-0.3}$.
\section{Decay modes with low mass resolution}

\subsection{\texorpdfstring{$\PH\to\PW\PW$}{H to WW}\label{sec:HWW}}

The decay mode $\PH\to\PW\PW$ is highly sensitive to a SM Higgs boson
in the mass range around the $\PW\PW$ threshold of 160\GeV.
With the development of tools for lepton identification and $\ETmiss$
reconstruction optimized for LHC pileup conditions, it is
possible to extend the sensitivity down to 120\GeV.
This decay mode is analysed by selecting events in which both $\PW$ bosons decay leptonically, resulting in a
signature with two isolated, oppositely charged leptons (electrons
or muons) and large $\ETmiss$ due to the undetected neutrinos~\cite{Barger:1990mn,Dittmar:1996ss}.
A $\pt$ threshold of 20 (10)\GeV is applied to
the lepton leading (subleading) in $\pt$.
The analysis of the
7\TeV data is described in Ref.~\cite{Chatrchyan:2012ty} and remains
unchanged, while the 8\TeV analysis was modified to cope with more
difficult conditions induced by the higher pileup of the 2012 data
taking.

Events are classified according to the number of jets (0, 1, or 2)
with $\pt>30\GeV$ and within $|\eta|<4.7$ ($|\eta|<5.0$ for the 7\TeV
data set), and further separated into same-flavour ($\Pe\Pe$ and $\Pgm\Pgm$)
or different-flavour ($\Pe\Pgm$) categories.  Events with more than
two jets are rejected.  To improve the sensitivity of the analysis,
the selection criteria are optimized separately for the different
event categories since they are characterised by different dominating
backgrounds.  The zero-jet $\Pe\Pgm$ category has the best signal
sensitivity.  Its main backgrounds are irreducible nonresonant $\PW\PW$
production and reducible W+jets processes, where a jet is
misidentified as a lepton.
The one-jet $\Pe\Pgm$ and zero-jet same-flavour categories
only contribute to the signal sensitivity at the 10\% level because
of larger backgrounds, from top-quark decays and
Drell--Yan production, respectively.  Event selection in the two-jet
category is optimized for the VBF production mechanism.  This
category has the highest expected signal-to-background ratio, but its
contribution to the overall sensitivity is small owing to the
lower cross section relative to inclusive production.

The projected $\ETmiss$ variable~\cite{Chatrchyan:2012ty} is
used to reduce the Drell--Yan background arising from events where the
$\ETmiss$ vector is aligned with the lepton $\pt$, as well as events with
mismeasured $\ETmiss$ associated with poorly reconstructed leptons and jets.
The projected $\ETmiss$ is defined as the transverse component of the $\ETmiss$
vector with respect to the closest lepton direction, if it is closer than
$\pi/2$ in azimuthal angle, or the full $\ETmiss$ otherwise.  Since pileup degrades the
projected $\ETmiss$ resolution, the minimum of
two different projected $\ETmiss$ definitions is used: the first includes
all particle candidates in the event, while the second uses only the charged
particle candidates associated with the primary vertex.  In the $8\TeV$
analysis, the minimum projected $\ETmiss$ defined in this way is then
required to be above a threshold that varies by category.  For $\mH>140\GeV$,
projected $\ETmiss$ is required to be greater than $20\GeV$ in the $\Pe\Pgm$
channel, and greater than 45\GeV in the same-flavour channels.  For
$\mH \le 140\GeV$ in the same-flavour channels, where it is more difficult
to separate the signal from the Drell--Yan background, a
multivariate selection is used, combining kinematic and topological variables.
In the two-jet category, a simple selection of $\ETmiss>45\GeV$ is applied.
To further reduce the Drell--Yan background in the same-flavour final states,
events with a dilepton mass within $15\GeV$ of the $\cPZ$ boson mass are
rejected. The background from low-mass resonances is rejected by requiring a dilepton
invariant mass greater than 12\GeV.

To suppress the top-quark background, a ``top tagging'' technique
based on soft-muon and b-jet tagging is
applied.  The first method is designed to veto events containing muons
from $\cPqb$ jets coming from decays of top quarks.  The second method
uses a $\cPqb$-jet tagging algorithm, which looks within jets for tracks
with large impact parameters.  The algorithm is applied also in the case
of zero-jet events, which may contain low-$\pt$ jets below the
selection threshold.  To reduce the background from $\PW\cPZ$ production,
events with a third lepton passing the identification and isolation
requirements are rejected.

Yields for the dominant backgrounds are estimated using control
regions in the data.  The W+jets contribution is derived
from data using a ``tight-loose'' sample in which one lepton passes the
standard criteria and the other does not, but instead
satisfies a ``loose'' set of requirements.  The efficiency
$\epsilon_{\text{loose}}$ for a jet that satisfies the loose selection
to pass the tight selection is determined using data from an independent
loose lepton-trigger sample dominated by jets.  The background contamination
is then estimated using the events of the ``tight-loose'' sample weighted
by $\epsilon_{\text{loose}}/(1-\epsilon_{\text{loose}})$.  The
normalisation of the top-quark background
is estimated by counting the number of top-tagged events and applying the
corresponding top-tagging efficiency.  The nonresonant $\PW\PW$ contribution
is normalised by using events with a dilepton mass larger than $100\GeV$,
where the Higgs boson signal contamination is negligible, extrapolated
to the signal region using simulated samples.  The same-flavour Drell--Yan background
is normalised using the number of events observed with a dilepton
mass within 7.5\GeV of the $\cPZ$ boson mass, after subtracting the non-Drell--Yan contribution.
Other minor backgrounds from $\PW\cPZ$, $\cPZ\cPZ$, and
$\PW\gamma$ are estimated from simulation.

The $7\TeV$ data are
analysed by training a BDT for each
Higgs boson mass hypothesis in the zero-jet and one-jet event categories,
while a simple selection strategy is employed in the VBF
category~\cite{Chatrchyan:2012ty}.  In the BDT analysis, the Higgs boson
signal is separated
from the background by using a binned maximum-likelihood fit
to the classifier distribution.  The $8\TeV$ analysis is based on a
simple selection strategy optimized for each mass hypothesis, where
additional kinematic and topological
requirements are applied to improve the signal-to-background ratio.
One of the most sensitive variables to discriminate between
$\PH \to \PW\PW$ decays and nonresonant $\PW\PW$ production
is the dilepton invariant mass $m_{\ell\ell}$.
This quantity is shown in Fig.~\ref{fig:WWdilep}
for the zero-jet $\Pe\Pgm$ category after the full selection
for $\mH=125$\GeV, except for the
selection on $m_{\ell\ell}$ itself.
Table~\ref{tab:hwwselection} shows for the $8\TeV$ analysis
the number of events selected in data, background estimates, and signal
predictions for $\mH$ = 125\GeV in each
analysis category after applying all the selection requirements.
About 97\% of the signal events selected in the zero-jet
$\Pe\Pgm$ category are expected to be produced by the gluon-gluon fusion process,
whereas 83\% of the signal in the two-jet $\Pe\Pgm$ category is expected
to be produced by the VBF process.
The 95\% CL expected and
observed limits for the combination of the 7 and 8\TeV
analyses are shown in Fig.~\ref{fig:WWlimit}.
A broad excess is observed that is consistent with a SM Higgs boson of mass 125\GeV.
This is illustrated by the dotted curve in Fig.~\ref{fig:WWlimit} showing the median expected limit
in the presence of a SM Higgs boson with $\mH=125\GeV.$
The expected significance for a SM Higgs of mass
125\GeV is 2.4$\,\sigma$ and the observed significance is 1.6$\,\sigma$.

\begin{table*}[htbp]
  \begin{center}
  \topcaption{Observed number of events, background estimates, and signal
  predictions for $\mH = 125$\GeV in each category
  of the $\PW\PW$ analysis of the 8\TeV data set.  All the selection requirements
  have been applied.  The combined experimental and theoretical, systematic
  and statistical uncertainties are shown.
  The Z$\gamma$ process includes the dimuon, dielectron, and $\Pgt\Pgt\to\ell\ell$ final states.}
   \label{tab:hwwselection}
 {
\begin{tabular} {l|r@{$\,\pm\,$}l|r@{$\,\pm\,$ }l|r@{$\,\pm\,$}l|r@{$\,\pm\,$}l|r@{$\,\pm\,$}l|r@{$\,\pm\,$}l}
\hline
Category: & \multicolumn{2}{c|}{0-jet $\Pe\mu$} & \multicolumn{2}{c|}{0-jet $\ell\ell$} & \multicolumn{2}{c|}{1-jet $\Pe\mu$} & \multicolumn{2}{c|}{1-jet $\ell\ell$} & \multicolumn{2}{c|}{2-jet $\Pe\mu$} & \multicolumn{2}{c}{2-jet $\ell\ell ^{\phantom{^0} }_{ \phantom{_0} }$ }\\
\hline \hline
$\PW\PW$ & 87.6&9.5 & 60.4&6.7 & 19.5&3.7 & 9.7&1.9 & 0.4&0.1 & 0.3&0.1 \\
$\PW\cPZ +\cPZ\cPZ +\cPZ\Pgg$ & 2.2&0.2 & 37.7&12.5 & 2.4&0.3 & 8.7&4.9 & 0.1&0.0 & 3.1&1.8 \\
Top  & 9.3&2.7 & 1.9&0.5 & 22.3&2.0 & 9.5&1.1 & 3.4&1.9 & 2.0&1.2 \\
$\PW+\text{jets}$ & 19.1&7.2 & 10.8&4.3 & 11.7&4.6 & 3.9&1.7 & 0.3&0.3 & 0.0&0.0 \\
$\PW\gamma^{(*)}$ & 6.0&2.3 & 4.6&2.5 & 5.9&3.2 & 1.3&1.2 & 0.0&0.0 & 0.0&0.0 \\
\hline
All backgrounds & 124.2&12.4 & 115.5&15.0 & 61.7&7.0 & 33.1&5.7 & 4.1&1.9 & 5.4&2.2\\
\hline
Signal \small{$(\mH = 125\GeV)$} & 23.9&5.2 & 14.9&3.3 & 10.3&3.0 & 4.4&1.3 & 1.5&0.2 & 0.8&0.1 \\
\hline
Data & \multicolumn{2}{c|}{158} & \multicolumn{2}{c|}{123} & \multicolumn{2}{c|}{54} & \multicolumn{2}{c|}{43} & \multicolumn{2}{c|}{6} & \multicolumn{2}{c}{7}\\
\hline
\end{tabular}
  }
  \end{center}
\end{table*}

\begin{figure}[htbp]
  \begin{center}
    \includegraphics[width=\cmsFigWideWidth]{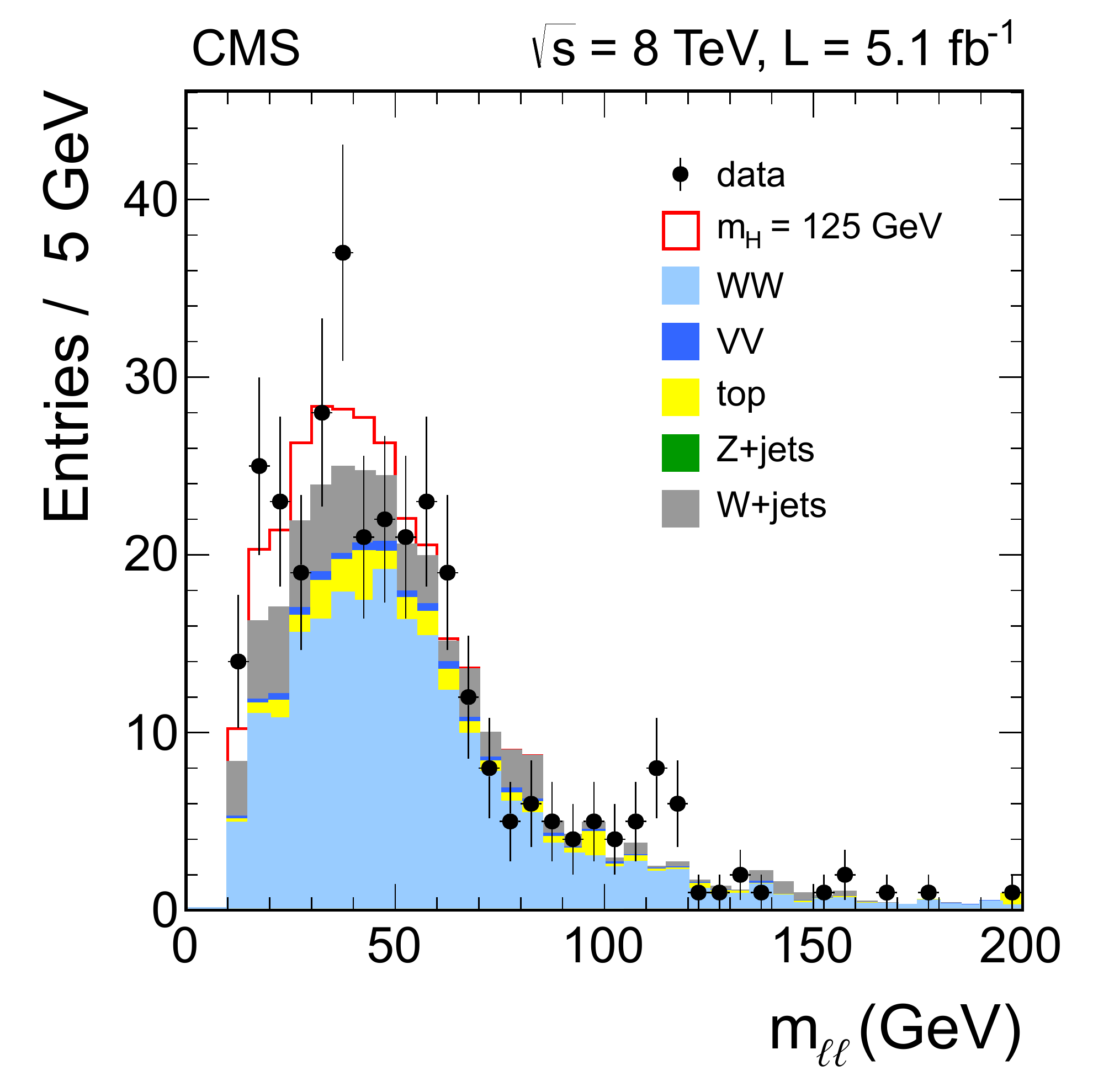}
    \caption{Distribution of $m_{\ell\ell}$ for the
    zero-jet $\Pe\Pgm$ category in the $\PH\to\PW\PW$ search
    at 8\TeV. The signal expected from a Higgs boson with a mass $\mH
    = 125$\GeV is shown added to the background.
}
    \label{fig:WWdilep}
  \end{center}
\end{figure}

\begin{figure}[htbp]
  \begin{center}
    \includegraphics[width=\cmsFigWideWidth]{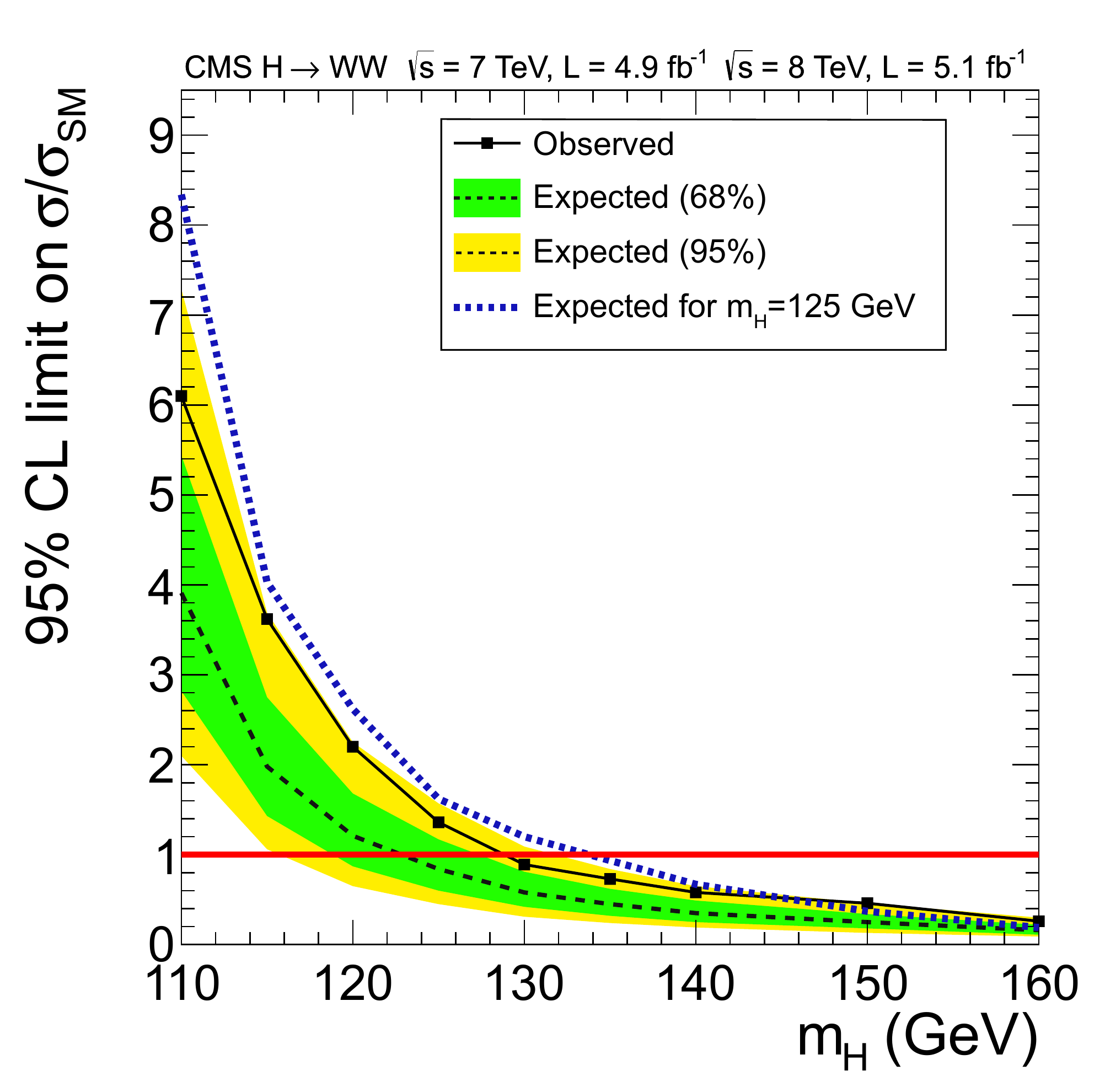}
    \caption{The 95\% CL limit on $\sigma/\sigma_\mathrm{SM}$ for a Higgs boson
      decaying, via a W boson pair, to two leptons and two neutrinos,
       for the combined 7 and 8\TeV
      data sets. The symbol $\sigma/\sigma_\mathrm{SM}$ denotes the
      production cross section times the relevant branching fractions, relative to the SM expectation.
The background-only expectations are represented by their median (dashed line) and by the 68\%  and 95\% CL bands. The dotted curve shows the median expected limit for a SM Higgs boson with $\mH=125\GeV$.
}
    \label{fig:WWlimit}
  \end{center}
\end{figure}

\subsection{\texorpdfstring{$\PH\to\Pgt\Pgt$}{H to tau tau}\label{sec:Htt}}

The decay mode $\PH\to\Pgt\Pgt$ is searched for in four
exclusive subchannels, corresponding to different decays of the $\Pgt$ pair:
$\Pe\Pgm$, $\Pgm\Pgm$, $\Pe \Pgt_{\mathrm{h}}$, and
$\Pgm \Pgt_{\mathrm{h}}$, where electrons and muons arise from
leptonic $\Pgt$ decays, and $\Pgt_{\mathrm{h}}$ denotes hadronic
$\Pgt$ decays.  The latter are reconstructed by selecting
$\Pgt$ decays consistent with the hypothesis of
three charged pions, or one charged pion and up to
two neutral pions~\cite{CMS-PAPERS-TAU-11-001}.
The search is made in the mass range 110--145\GeV, and a
signal should appear as a broad excess in the distribution of the $\Pgt$-pair
invariant mass $m_{\Pgt\Pgt}$.

The sensitivity of the search is improved by classifying the
events according to jet multiplicity and the transverse
momentum of the reconstructed $\Pgt$.  The multiplicity of jets
with $\pt>30\GeV$ reflects the production mechanism:
events with zero or one jet are likely to come from the
gluon-gluon fusion process, while events with two jets are candidates for VBF
production. Events including $\cPqb$ jets with $\pt>20\GeV$ are removed from zero- and one-jet categories.
The signal purities in the zero- and one-jet
categories are increased, and the $m_{\Pgt\Pgt}$ resolution is
improved, by separating events into low- and high-$\pt$ subchannels.
The high-$\pt$ subchannels are defined by
$\pt^{\Pgt_{\mathrm{h}}}>40\GeV$ in channels with a hadronic
$\Pgt$ decay, and $\pt^{\mu}>35\,(30)\GeV$ in the
$\Pe\Pgm\,(\Pgm\Pgm)$ channel.
The mass $m_{\Pgt\Pgt}$ is reconstructed with an algorithm~\cite{Chatrchyan:2011nx} combining the
visible $\Pgt$ decay products and the missing transverse energy, achieving a resolution of about 20\% on $m_{\Pgt\Pgt}$.
Figure~\ref{fig:HttMass} shows
as an example the reconstructed $m_{\Pgt\Pgt}$ distribution in
the $\mu\Pgt_{\mathrm h}$ VBF category for the combined $7$ and
$8\TeV$ data samples.

\begin{figure}[htbp]
  \begin{center}
    \includegraphics[width=\cmsFigWideWidth]{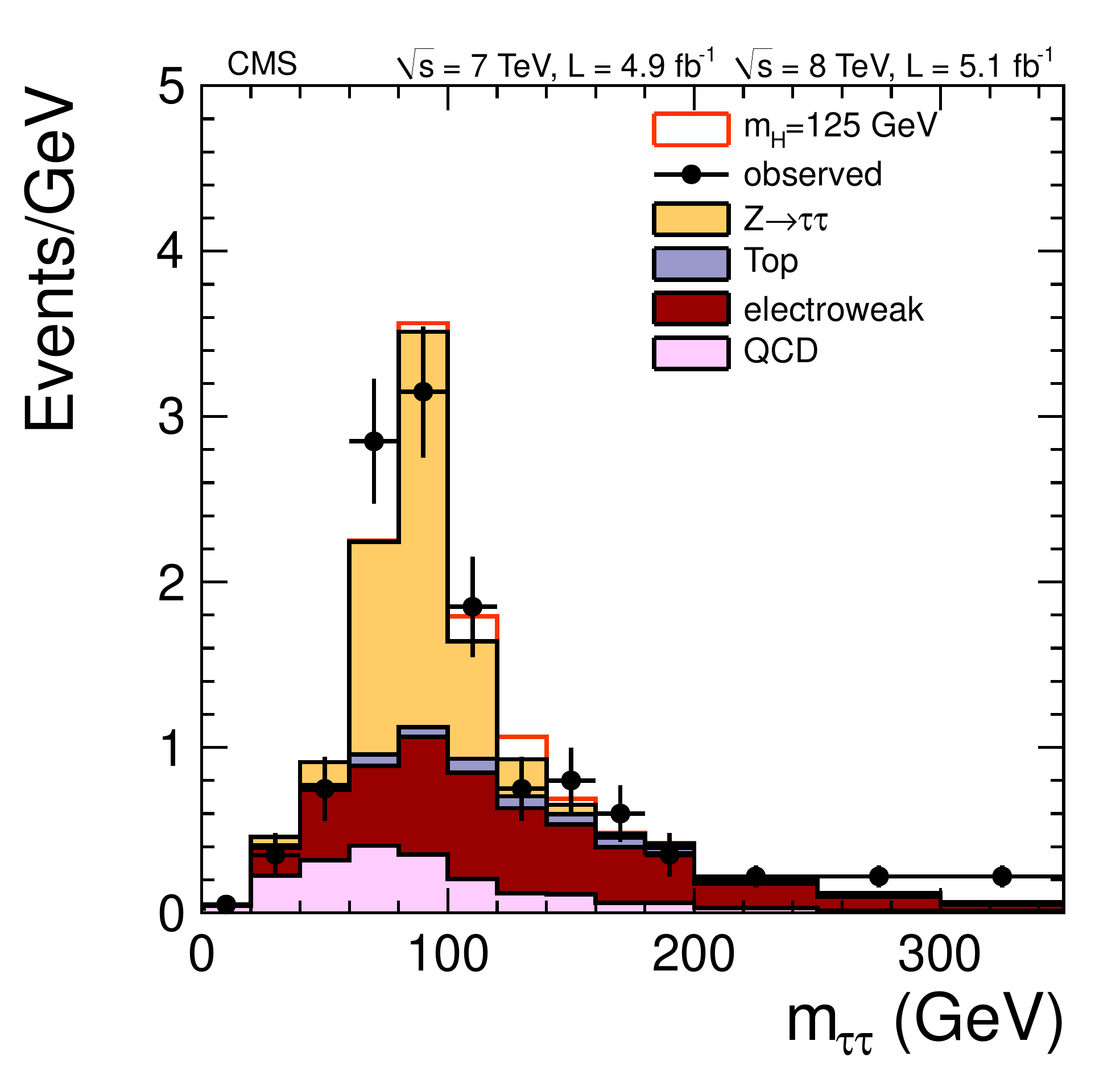}
    \caption{Distribution of $m_{\Pgt\Pgt}$ in the combined 7~and 8\TeV data sets for the $\mu\Pgt_{\mathrm h}$ VBF category of the
    $\PH\to\Pgt\Pgt$ search. The signal expected from a SM Higgs boson
    ($\mH = 125$\GeV) is added to the background.}
    \label{fig:HttMass}
  \end{center}
\end{figure}

Backgrounds in the $\Pe\Pgm$ and $\Pgm\Pgm$ channels arise
from $\ttbar$ and Drell--Yan production, while $\PW$ and $\cPZ$
production with a misidentified $\Pgt_{\mathrm{h}}$ candidate
from an electron, muon, or jet dominates in the hadronic channels.
Backgrounds from
$\cPZ\to\Pgt\Pgt$ decays are modelled with
$\cPZ\to\Pgm\Pgm$ events in data where each muon is replaced with particles
from simulated decays of a $\Pgt$ with the same momentum as the
muon.  Reducible backgrounds, comprising $\PW+$jets, QCD multijet production, and residual
$\cPZ\to\Pe\Pe$ events, are estimated from the data~\cite{Chatrchyan:2012vp}.
An improved signal-to-background ratio is achieved by including
explicitly in the event selection for the
VBF production mechanism the pseudorapidity
separation between forward jets and the large invariant mass of the dijet
system.
Table~\ref{tab:TAUVBF7+8TeV} shows the numbers of expected and observed
events in the most sensitive event categories (VBF) for the 7 and 8\TeV data sets.
The expected signal yields for a SM Higgs boson with $\mH = 125$\GeV
are also shown.

\begin{table*}[htbp]
\begin{center}
\topcaption{
  Numbers of expected and observed events in the most sensitive event
  categories (VBF) in the $\PH\to\Pgt\Pgt$ analysis for the
  7~and~8\TeV data sets. The expected signal yields for a SM Higgs boson with
  $\mH = 125$\GeV are also shown.  Combined statistical and systematic
  uncertainties in each estimate are reported.
}
\begin{tabular}{l|r@{$ \,\,\pm\,\, $}l|r@{$\,\,\pm\,\,$}l|r@{$\,\,\pm\,\,$}l|r@{$\,\,\pm\,\,$}l}
\hline
Subchannel & \multicolumn{2}{c|}{$\Pe\Pgt_\mathrm{h}$} & \multicolumn{2}{c|}{$\Pgm\Pgt_\mathrm{h}$} & \multicolumn{2}{c|}{$\Pe\Pgm$} & \multicolumn{2}{c}{$\Pgm\Pgm$}\\
\hline\hline
$\cPZ\to\Pgt\Pgt$ & 53 & 5 & 100 & 9 & 56 & 12  & 5.3 & 0.4\\
QCD & 35 & 7 & 41 & 9 & 7.4 & 1.4 & \multicolumn{2}{c}{$-$} \\
$\PW+$jets & 46 & 10 & 72 & 15 & \multicolumn{2}{c|}{$-$}& \multicolumn{2}{c}{$-$} \\
$\cPZ+$jets & 13 & 2 & 2.5 & 0.6 & \multicolumn{2}{c|}{$-$} & \multicolumn{2}{c}{$-$}\\
$\cPZ\to\Pgm\Pgm$ & \multicolumn{2}{c|}{$-$} & \multicolumn{2}{c|}{$-$}& \multicolumn{2}{c|}{$-$}& 70 & 8 \\
$\ttbar$ &  7.0 & 1.7 & 14 & 3 & 24 & 2 & 6.7 & 1.5\\
Dibosons & 1.2 & 0.9 & 2.9 & 2.1 & 11 & 2 & 2.4 & 0.9\\
\hline
All backgrounds & 156 & 13 & 233 & 20 & 99 & 13 & 85 & 9\\
\hline
Signal \small{($\mH = 125$\GeV)} & 4.3 & 0.6 & 7.7 & 1.1  & 3.5 & 0.4 & 0.8 & 0.1\\
\hline
Data & \multicolumn{2}{c|}{142} & \multicolumn{2}{c|}{263}
&\multicolumn{2}{c|}{110} &\multicolumn{2}{c}{83} \\
\hline
\end{tabular}
\label{tab:TAUVBF7+8TeV}
\end{center}
\end{table*}

\begin{figure}[htbp]
  \begin{center}
    \includegraphics[width=\cmsFigWideWidth]{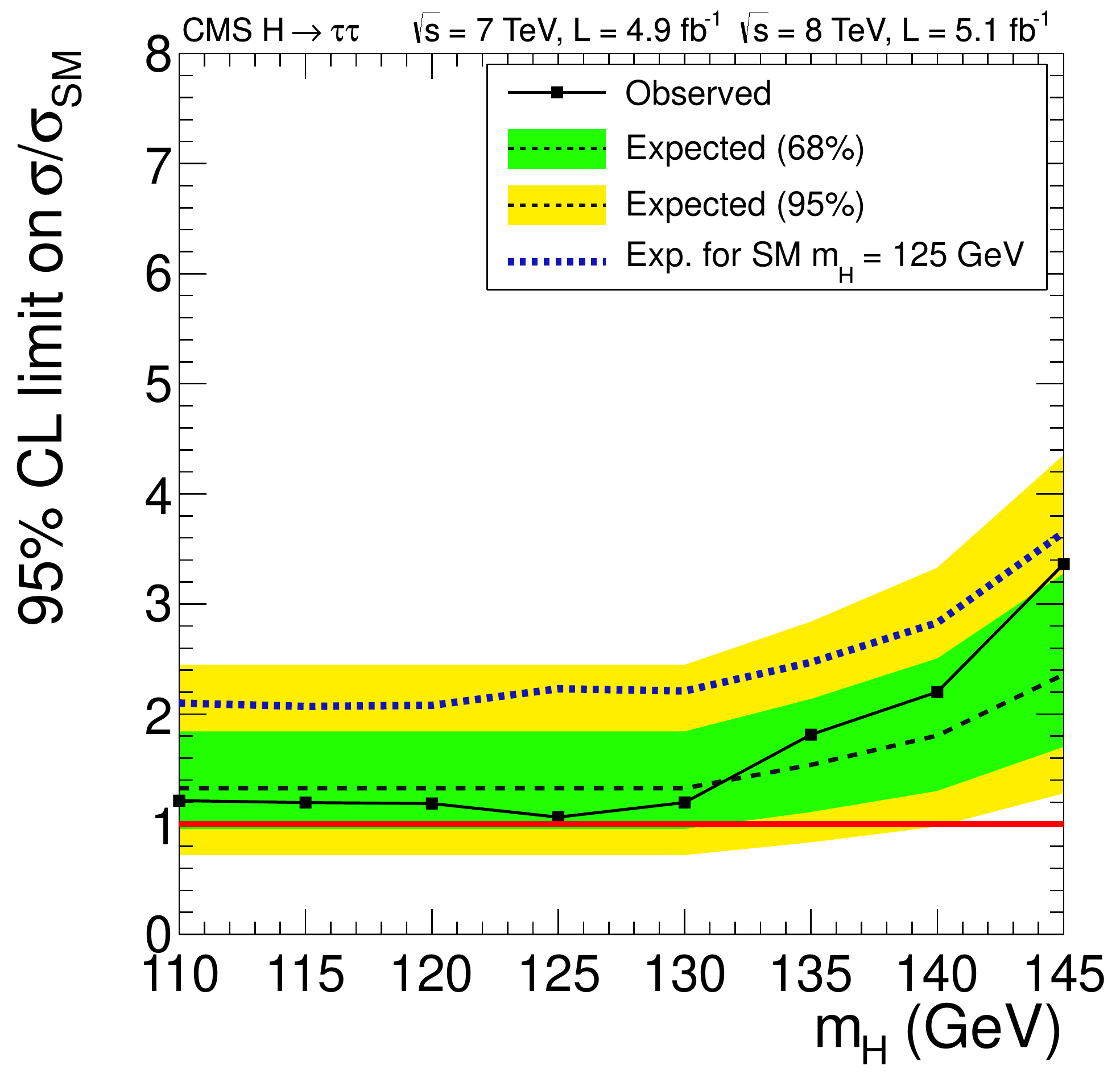}
    \caption{The 95\% CL limit on the signal strength $\sigma/\sigma_\mathrm{SM}$ for a Higgs boson
      decaying to $\Pgt$ pairs, for the combined 7 and 8\TeV data sets.
The symbol $\sigma/\sigma_\mathrm{SM}$ denotes the
production cross section times the relevant branching fractions,
relative to the SM expectation.
The background-only expectations are represented by their median (dashed line) and by the 68\%  and 95\% CL bands. The dotted curve shows the median expected limit for a SM Higgs boson with $\mH = 125\GeV.$
}
    \label{fig:HttLimit}
  \end{center}
\end{figure}

To search for the presence of a Higgs boson signal in the selected events,
a binned maximum-likelihood fit to $m_{\Pgt\Pgt}$ is performed jointly
across the four final states, each with five event categories.  Systematic
uncertainties are represented by nuisance parameters in the fitting process.
The expected
and observed 95\% CL limits on the signal strength for the combination of all categories are shown in Fig.~\ref{fig:HttLimit}.
The expected and observed limits are 1.3 and 1.1 times the
SM Higgs boson cross section at mass 125\GeV, respectively.
The expected significance for a SM Higgs boson of mass
125\GeV is 1.4$\,\sigma$, and the observed value is zero.

\subsection{\texorpdfstring{$\PH\to\cPqb\cPqb$}{H to bb}\label{sec:Hbb}}

For $\mH\leq 135$\GeV, the decay $\PH\to\cPqb\cPqb$ has the
largest branching fraction of the five search modes, but the
inclusive signal is overwhelmed by QCD production of bottom quarks.  The analysis
is therefore designed to search for the associated production of the Higgs
boson in events where a dijet resonance is produced at high $\pt$ in association
with a $\PW$ or $\cPZ$ boson; this largely suppresses the QCD background.
Five independent search channels are explored corresponding
to different decays of the vector boson: $\cPZ(\ell\ell)\PH$,
$\cPZ(\cPgn\cPgn)\PH$, and $\PW(\ell\cPgn)\PH$. Events are further
separated into two categories based on the $\pt$ of the vector boson,
ranging from $50$--$100\GeV$ for the lowest bin in the $\cPZ(\ell\ell)$
search, to greater than 170\GeV for the highest bin in the $\PW(\ell\cPgn)$
search.  For the $\cPZ(\cPgn\cPgn)$ search, two subchannels
are defined as $120<\ETmiss<160\GeV$ and $\ETmiss>160\GeV$.
The two jets comprising the candidate Higgs
boson decay are required to be identified as $\cPqb$ jets, and the
dijet system must satisfy a $\pt$ threshold that is
optimized within each channel: greater than 120\GeV for $\PW\PH$,
160\GeV for $\cPZ(\cPgn\cPgn)\PH$, and no explicit threshold for $\cPZ(\ell\ell)\PH.$

Dominant backgrounds arise from production of vector bosons in
association with jets, pair- or single-production of top quarks, and
diboson production ($\PW\PW$, $\PW\cPZ$, $\cPZ\cPZ$) with one of
the bosons decaying hadronically.  Significant background rejection
is achieved in general by requiring large $\pt$ for the dijet, while
also requiring that there be minimal additional jet activity and
that the vector boson and dijet be back to back in azimuth.
The effect on the signal efficiency of this selection due to
higher-order electroweak~\cite{Denner:2011id} and
QCD~\cite{Ferrera:2011bk}
corrections are accounted for in the systematic uncertainties.
Further signal discrimination is obtained from the dijet invariant mass,
which is expected to peak near $\mH$.  A multivariate regression
algorithm to better estimate $\cPqb$-jet $\pt$
is trained on jets in simulated signal events and achieves a final
dijet mass resolution of 8--9\% for $\mH = 125\GeV$.
The performance of the regression algorithm is checked in data
using $\PW/\cPZ+$jets and $\ttbar$ events.

A search for the signal is made in the distribution of scores of a BDT trained at
discrete mass points.  Input variables to the BDT algorithm exploit
kinematic and topological information about the vector boson and dijet
systems, and the colour-singlet nature of the Higgs boson~\cite{Gallicchio:2010sw}.  The
distribution of scores in simulated background events is checked using control regions in the data designed
to enrich individual background contributions.
Figure~\ref{fig:HbbBDT} shows as an example the BDT scores for the high-$\pt$ subchannel of the $\cPZ(\cPgn\cPgn)\PH$ channel in the $8\TeV$ data set, after all selection criteria have been applied.

\begin{figure}[htbp]
  \begin{center}
    \includegraphics[width=\cmsFigWideWidth]{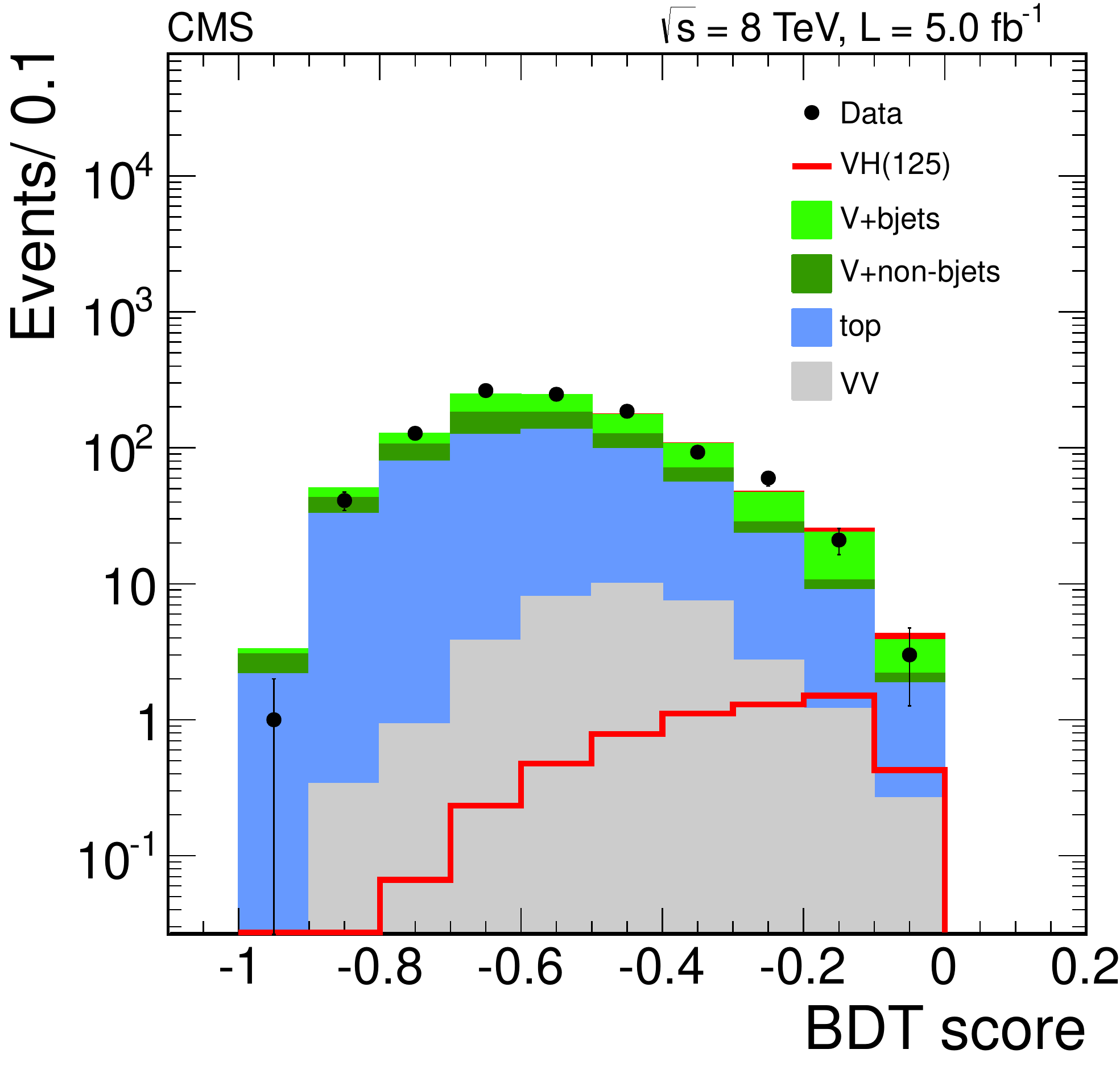}
    \caption{Distribution of BDT scores for the high-$\pt$ subchannel of the  $\cPZ(\cPgn\cPgn)\PH(\cPqb\cPqb)$ search in the $8\TeV$ data set after all selection  criteria have been applied.
The signal expected from a Higgs boson ($\mH = 125\GeV$), including $\PW(\ell\cPgn)\PH$ events where the charged lepton is not  reconstructed, is shown added to the background and also overlaid for comparison with the diboson background.}
    \label{fig:HbbBDT}
  \end{center}
\end{figure}

The rates for the dominant backgrounds arising from production of
$\PW/\cPZ+$jets and top-quark pairs are estimated in data~\cite{Chatrchyan:2012ww}, while
contributions from single-top and diboson production are estimated from
simulation studies.
The signal is then searched for as an excess in
the BDT score distribution using
the predicted shapes for signal and background events, for Higgs
boson masses in the range 110--135\GeV.

Combined results for expected and observed $95\%$ CL limits
obtained from the 7 and 8\TeV data sets are displayed in
Fig.~\ref{fig:HbbLimit}. The expected and observed limits are 1.6 and 2.1 times
the SM Higgs boson cross section at mass 125\GeV.
The  expected local $p$-value for a SM Higgs of mass
125\GeV corresponds to 1.9$\,\sigma$, while the observed value corresponds to 0.7$\,\sigma$.

\begin{figure}[htbp]
  \begin{center}
    \includegraphics[width=\cmsFigWideWidth]{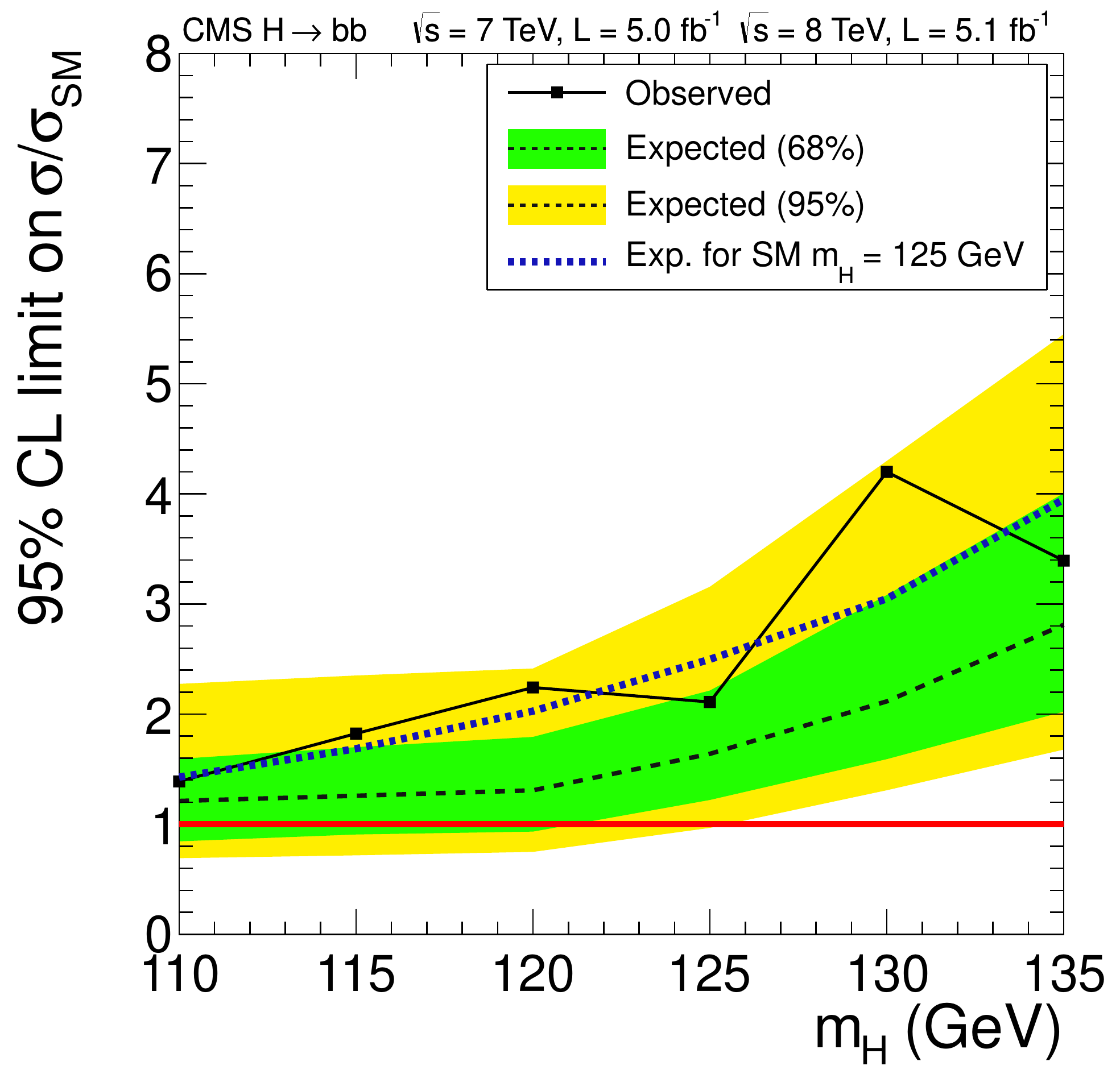}
    \caption{The 95\% CL limit on the signal strength $\sigma/\sigma_\mathrm{SM}$ for a Higgs boson
      decaying to two b quarks, for the combined 7 and 8\TeV data sets.
The symbol $\sigma/\sigma_\mathrm{SM}$ denotes the production cross section times the relevant branching fractions,
relative to the SM expectation.
The background-only expectations are represented by their median
(dashed line) and by the 68\%  and 95\% CL bands. The dotted curve shows the median expected limit for a SM Higgs boson with $\mH = 125\GeV.$
}
    \label{fig:HbbLimit}
  \end{center}
\end{figure}

\section{Combined results}\label{sec:Results}

The individual results for the channels analysed for the five decay
modes, summarised in Table~\ref{tab:chans}, are combined using the
methods outlined in Section~\ref{sec:Strategy}.
The combination assumes the relative branching fractions predicted by the SM and takes into account the experimental statistical and
systematic uncertainties as well as the theoretical uncertainties, which are
dominated by the imperfect knowledge of the QCD scale and parton distribution functions.
The $\CLs$ is shown in Fig.~\ref{fig:CLs} as a function of the Higgs boson mass hypothesis.
The observed values
are shown by the solid points.
The dashed  line indicates the median of the expected results for
the background-only hypothesis,
with the green (dark) and yellow (light) bands indicating the ranges in which
the $\CLs$ values are expected to lie in 68\% and 95\%
of the experiments under the background-only hypothesis.
The probabilities for an observation, in the absence of a signal,  to lie above or below the 68\% (95\%) band are 16\% (2.5\%) each.
The thick  horizontal lines indicate $\CLs$ values of  0.05, 0.01, and 0.001.
The mass regions where the observed $\CLs$ values are below these lines are excluded
with the corresponding ($1-\CLs$) confidence levels.
Our previously published results exclude the SM Higgs boson from 127 to 600\GeV~\cite{Chatrchyan:2012tx}.
In the search described here, the SM Higgs boson is excluded at 95\%
CL in the range $110 < \mH < \ObsNFL$\GeV.
In the range $\ObsNFL < \mH < \ObsNFH$\GeV  a significant excess is seen and the SM Higgs boson
cannot be excluded at 95\% CL.

\begin{figure} [htbp]
\begin{center}
\includegraphics[width=\cmsFigWideWidth]{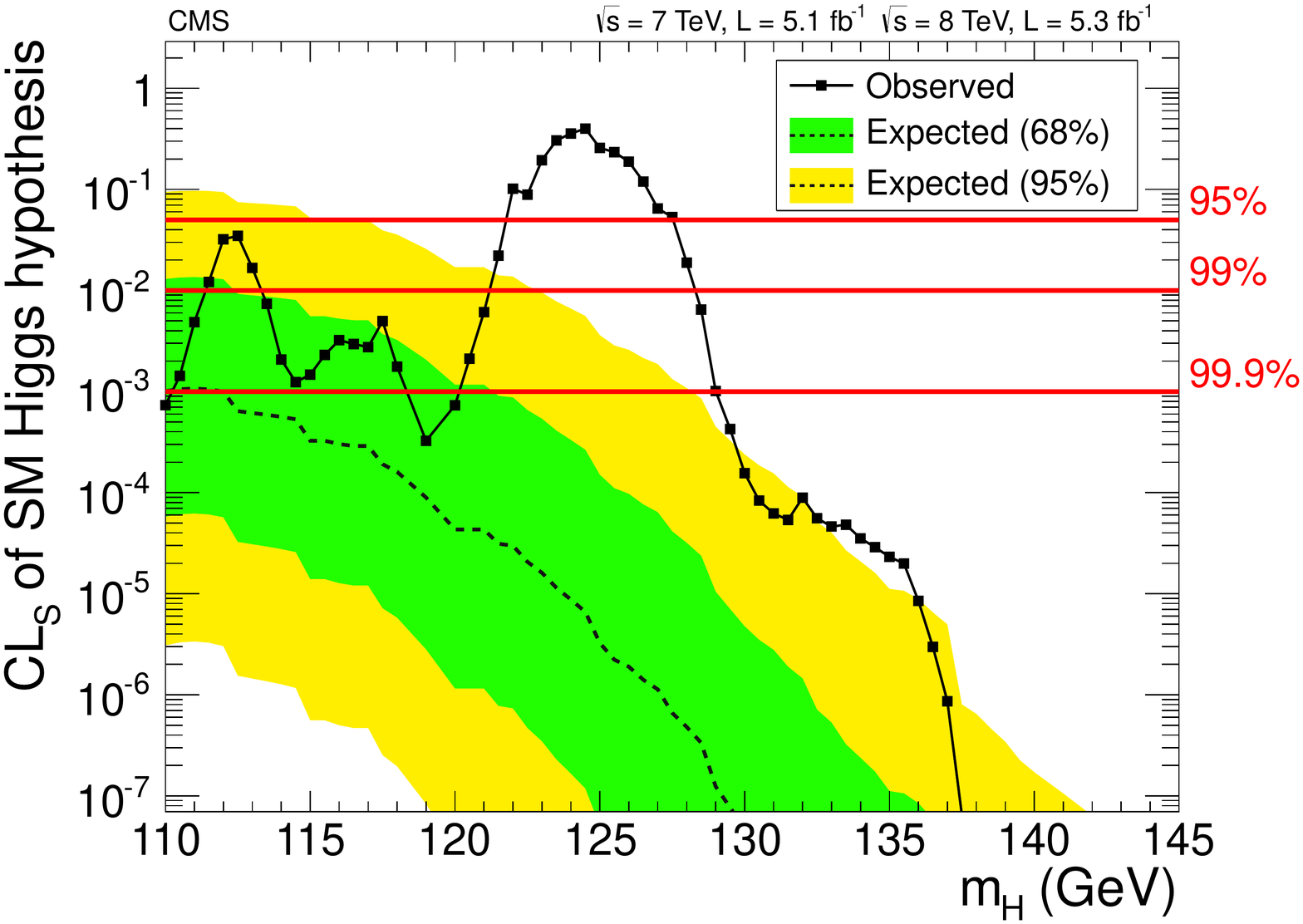}
\caption{ The $\CLs$ values for the SM Higgs boson hypothesis
as a function of the Higgs boson mass in the range 110--145\GeV.
The background-only expectations are represented by their median (dashed line) and by the 68\%  and 95\% CL bands.
    }
\label{fig:CLs}
\end{center}
\end{figure}

\subsection{Significance of the observed excess}

The consistency of the observed excess with the background-only hypothesis may be judged from
Fig.~\ref{fig:pvalue}, which shows a scan of the
local $p$-value for the 7~and 8\TeV data sets
and their combination.
The 7~and 8\TeV data sets
exhibit an excess of $\MaxLocalZseven \,\sigma$ and $\MaxLocalZeight \,\sigma$ significance, respectively,
for a Higgs boson mass of approximately 125\GeV.
In the overall combination the significance is 5.0\,$\sigma$ for $\mH=125.5\GeV$.
Figure~\ref{fig:pvalue_chans} gives the local $p$-value for the five
decay modes individually and displays the expected overall $p$-value.

The largest contributors to the overall excess in the combination
are the $\Pgg\Pgg$ and $\cPZ\cPZ$ decay modes. They both have very good
mass resolution, allowing good localization of the invariant mass of
a putative resonance responsible for the excess. Their combined
significance reaches 5.0\,$\sigma$ (Fig.~\ref{fig:pvalue_subcomb}).
The $\PW\PW$ decay mode has an exclusion sensitivity  comparable to the
$\gamma\gamma$ and $\cPZ\cPZ$ decay modes
but does not have a good mass resolution.
It has an excess with
local significance 1.6\,$\sigma$ for $\mH\sim 125$\GeV. When added to
the $\gamma\gamma$ and $\cPZ\cPZ$ decay modes, the combined significance
becomes 5.1\,$\sigma$.
Adding the $\Pgt\Pgt$ and $\cPqb\cPqb$ channels  in the combination, the final
significance becomes 5.0\,$\sigma$.
Table~\ref{tab:Signif} summarises the expected and observed local $p$-values for a SM Higgs boson mass
hypothesis of 125.5\GeV for the various combinations of channels.

\begin{table}[htbp]
\begin{center}
\topcaption{
The expected and observed local $p$-values, expressed
as the corresponding number of standard deviations of the observed excess from the
background-only hypothesis,
for $\mH = 125.5$\GeV, for various combinations of decay modes.
}
\label{tab:Signif}
\begin{tabular}{l|c|c}
\hline
Decay mode/combination & Expected ($\sigma$) & Observed ($\sigma$) \\
\hline\hline
$\Pgg\Pgg$ & 2.8 & 4.1 \\ %
$\cPZ\cPZ$  &  3.8 & 3.2 \\ %
\hline
$\Pgt\Pgt$ + $\cPqb\cPqb$ & 2.4 & 0.5 \\
$\Pgg\Pgg$ + $\cPZ\cPZ$ & 4.7 & 5.0 \\
$\Pgg\Pgg$ + $\cPZ\cPZ$ + $\PW\PW$ & 5.2 & 5.1 \\ %
$\Pgg\Pgg$ + $\cPZ\cPZ$ + $\PW\PW$ + $\Pgt\Pgt$ + $\cPqb\cPqb$ & 5.8 & 5.0 \\%
\hline
\end{tabular}
\end{center}
\end{table}

\begin{figure} [htbp]
\begin{center}
\includegraphics[width=\cmsFigWideWidth]{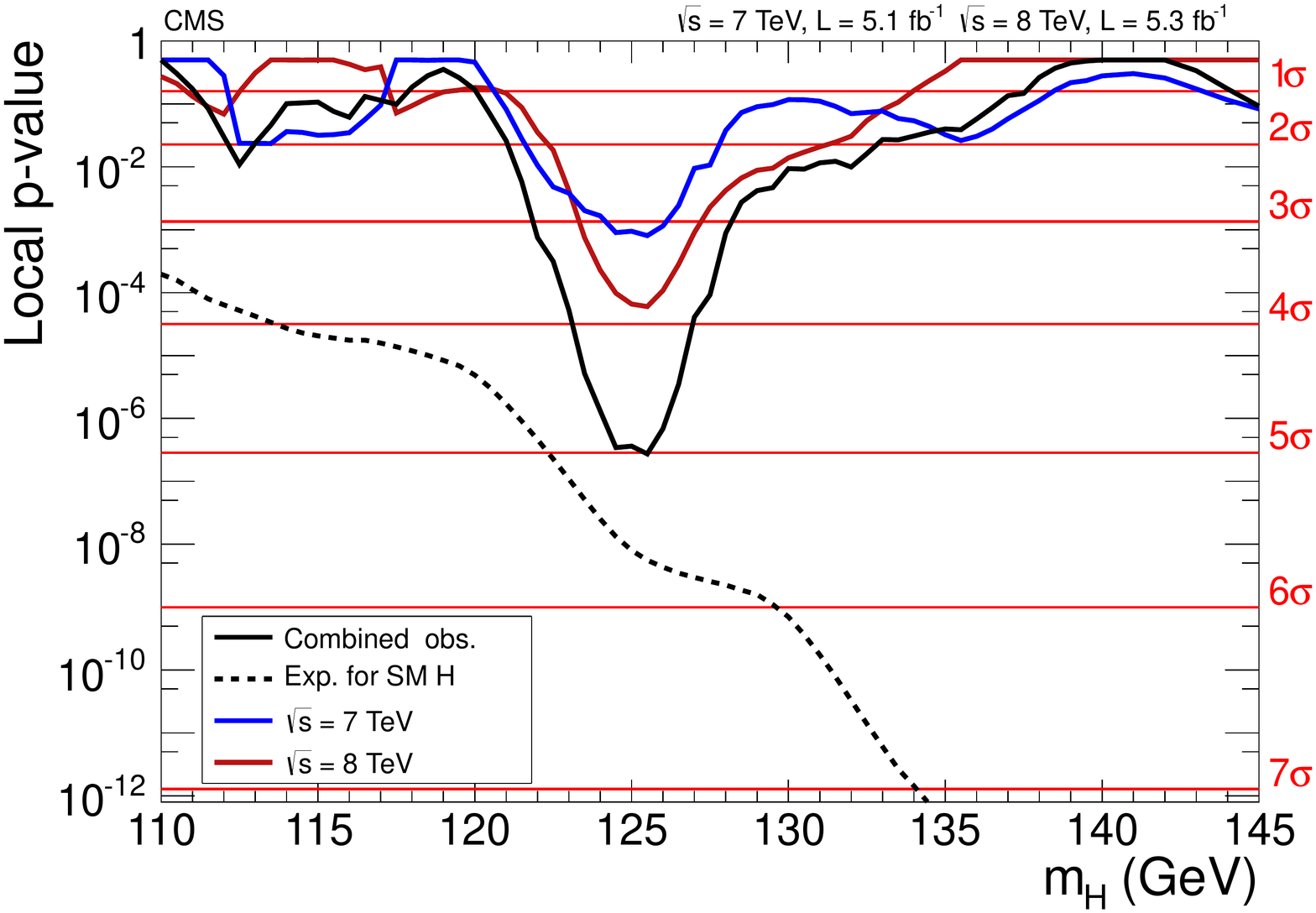}
\caption{
The observed local $p$-value for 7\TeV~and 8\TeV data, and their combination
as a function of the SM Higgs boson mass. The dashed line shows the expected local $p$-values
for a SM Higgs boson with a mass $\mH$.
    }
\label{fig:pvalue}
\end{center}
\end{figure}

\begin{figure} [htbp]
\begin{center}
\includegraphics[width=\cmsFigWideWidth]{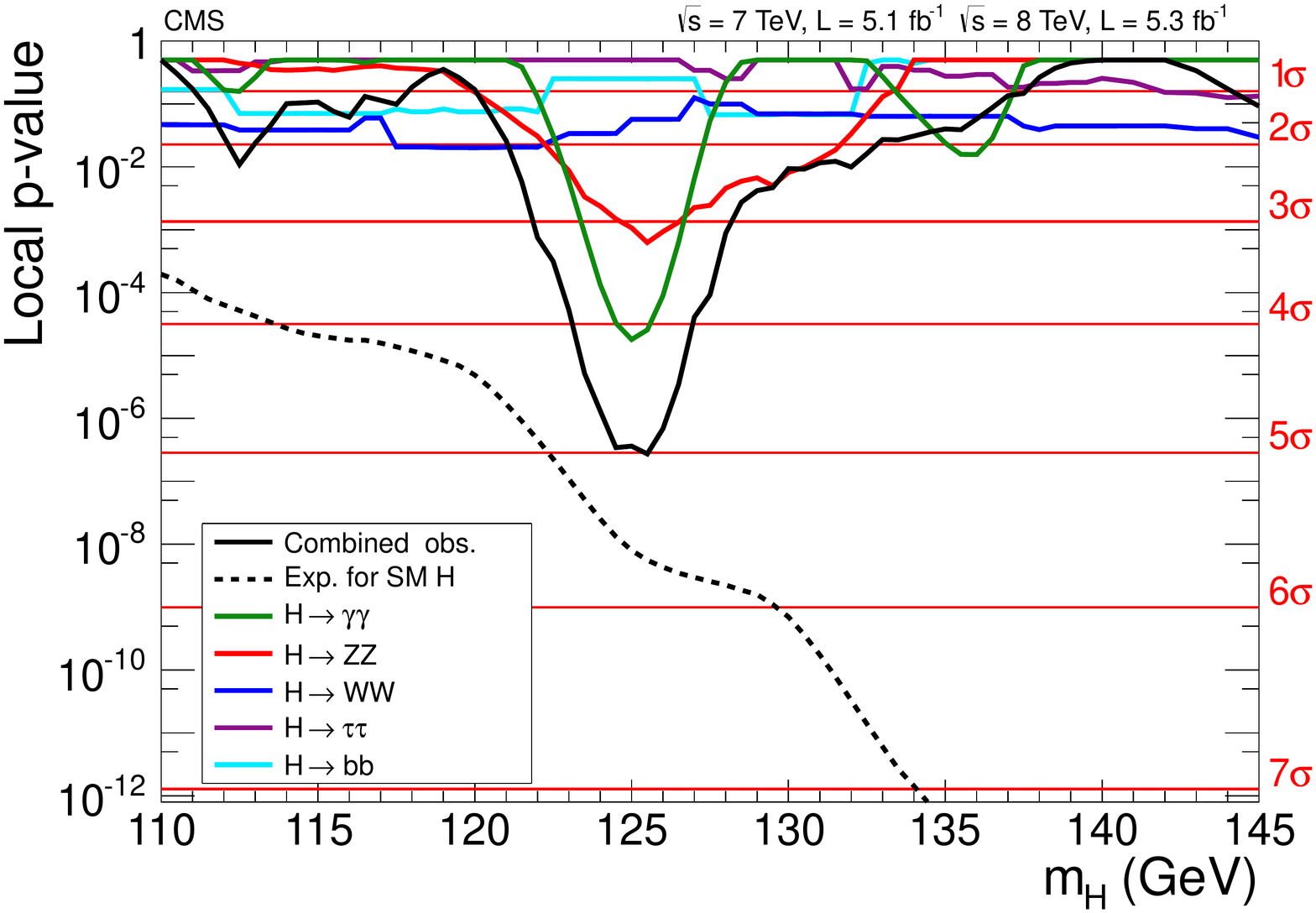}
\caption{
The observed local $p$-value for the five decay modes
and the overall combination as a function of the SM Higgs boson mass.
The dashed line shows the expected local $p$-values
for a SM Higgs boson with a mass $\mH$.
}
\label{fig:pvalue_chans}
\end{center}
\end{figure}

\begin{figure} [htbp]
\begin{center}
\includegraphics[width=\cmsFigWideWidth]{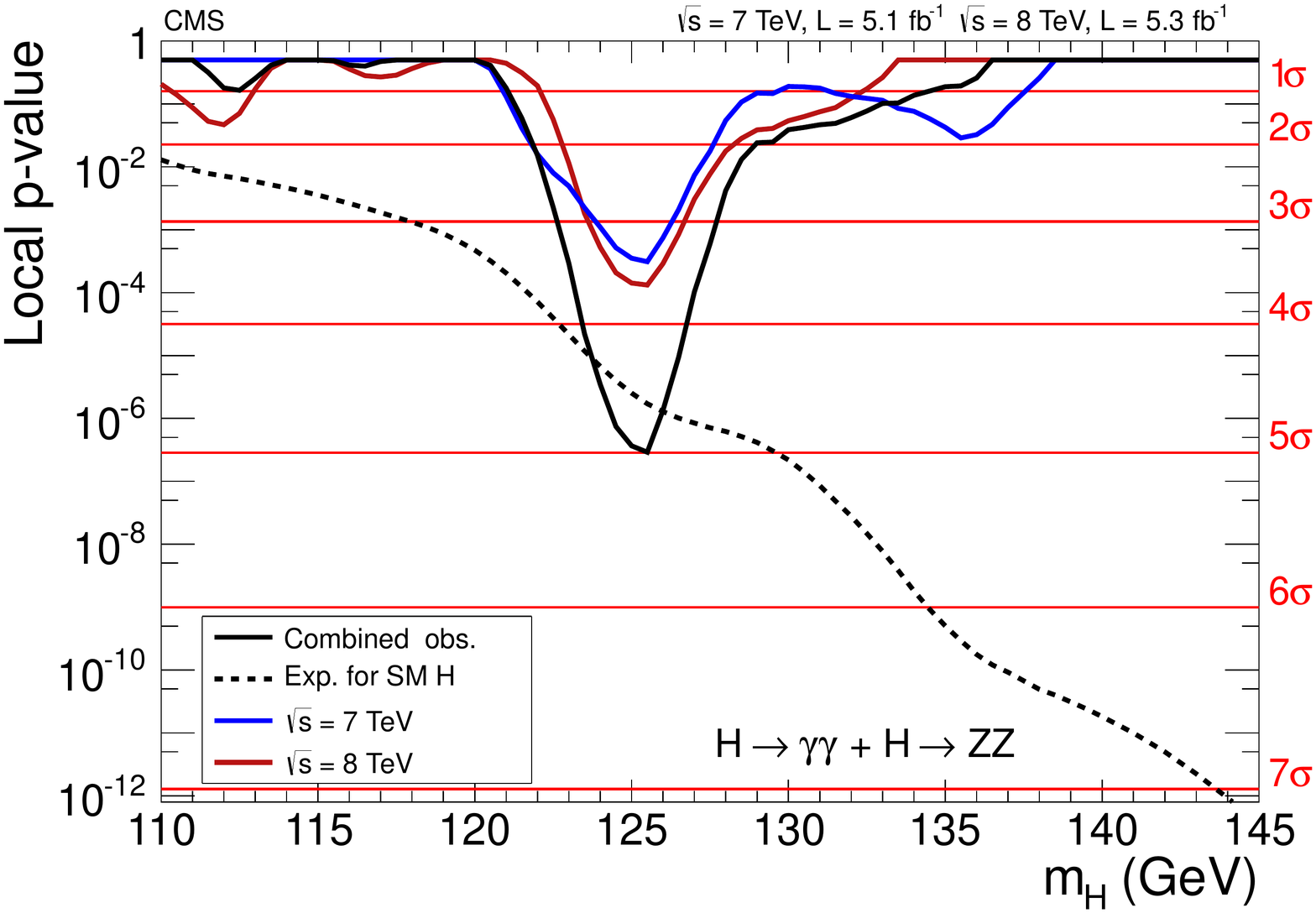}
\caption{
The observed local $p$-value for decay modes with high mass-resolution channels,
$\gamma\gamma$ and $\cPZ\cPZ$, as a function of the SM Higgs
boson mass.
The dashed line shows the expected local $p$-values
for a SM Higgs boson with a mass $\mH$.
}
\label{fig:pvalue_subcomb}
\end{center}
\end{figure}

The global $p$-value for the search range 115--130 (110--145)\GeV
is calculated using the method suggested in Ref.~\cite{LEE}, and corresponds to
$\GlobalZsmall \,\sigma$ ($\GlobalZmedium \,\sigma$).
These results confirm the very low probability for an excess as large as or larger than that observed
to arise from a statistical fluctuation of the background.
The excess constitutes the observation of a new
particle with a mass near 125\GeV, manifesting itself in decays to two photons or to $\cPZ\cPZ$.
These two decay modes indicate that the new particle is a boson; the two-photon decay implies that its spin is different from one~\cite{Landau,Yang}.

\subsection{Mass of the observed boson}

The mass $m_{\mathrm{X}}$ of the observed boson is determined using
the $\gamma\gamma$ and $\cPZ\cPZ$ decay modes, with the former dominating the precision of the measurement.
The calibration of the energy scale in the $\gamma\gamma$ decay mode is achieved with reference to the known \cPZ\ boson mass, as described in Section~\ref{sec:Hgg}.
There are two main sources of systematic uncertainty: (i) imperfect simulation
of the differences between electrons and photons and (ii)
the need to extrapolate from $m_\mathrm{Z}$ to $m_{\mathrm{X}}\approx 125$\GeV.
The systematic uncertainties are evaluated by making comparisons between data
and simulated samples of $\cPZ\to\Pe\Pe$ and $\PH\to\gamma\gamma$
($\mH=90$\GeV).
The two uncertainties, which together amount to 0.5\%, are
assumed to be fully correlated between all the $\gamma\gamma$ event categories in the 7~and
8\TeV data.
For the $\cPZ\cPZ \to 4\ell$  decay mode the energy scale (for electrons)
and momentum scale (for muons) are calibrated using the leptonic decays
of the $\cPZ$ boson, with an assigned uncertainty of 0.4\%.

Figure~\ref{fig:fit_mass} shows the
two-dimensional~68\% CL regions for the signal strength $\sigma/\sigma_\text{SM}$ versus $m_{\mathrm{X}}$ for the three channels
(untagged $\gamma\gamma$, dijet-tagged $\gamma\gamma$, and $\cPZ\cPZ\to 4\ell$).
The combined 68\%~CL contour shown  in Fig.~\ref{fig:fit_mass}
assumes that the relative event yields among the three channels are those expected from  the standard model, while the overall signal strength
is a free parameter.

To extract the value of $m_{\mathrm{X}}$ in a model-independent way,
the signal yields of
the three
channels are allowed to vary independently.
Thus the expected event yields in these channels are scaled by independent factors, while
the signal is assumed to be due to a particle with a unique mass $m_{\mathrm{X}}$.
The combined best-fit mass is $m_{\mathrm{X}} = \MASS$\GeV.

\subsection{Compatibility with the SM Higgs boson hypothesis}

A first test of the compatibility of the observed boson with the SM
Higgs boson is provided by examination of
the best-fit value for the common signal strength $\sigma/\sigma_{\mathrm{SM}}$,
obtained in a combination of all search channels.
Figure~\ref{fig:muhat} shows a scan
of the overall $\sigma/\sigma_{\mathrm{SM}}$ obtained in the combination of all channels versus a hypothesised
Higgs boson mass $\mH$. The band corresponds to the ${\pm}1\,\sigma$ uncertainty (statistical and systematic).
The excesses seen in the 7\TeV and 8\TeV data, and in their combination, around 125\GeV are consistent with unity
within the ${\pm}1\,\sigma$ uncertainties. The observed $\sigma/\sigma_\text{SM}$ value for an excess at 125.5\GeV
in a combination of all data is \MUHAT.
The different decay channels and data sets have been
examined for self-consistency.
Figure~\ref{fig:SelfConsistencyDecayProd}
shows the measured values of  $\sigma/\sigma_\text{SM}$ results
obtained for the different decay modes.
These results are consistent, within uncertainties, with the
expectations for a SM Higgs boson.

\begin{figure} [htbp]
\begin{center}
\includegraphics[width=\cmsFigWideWidth]{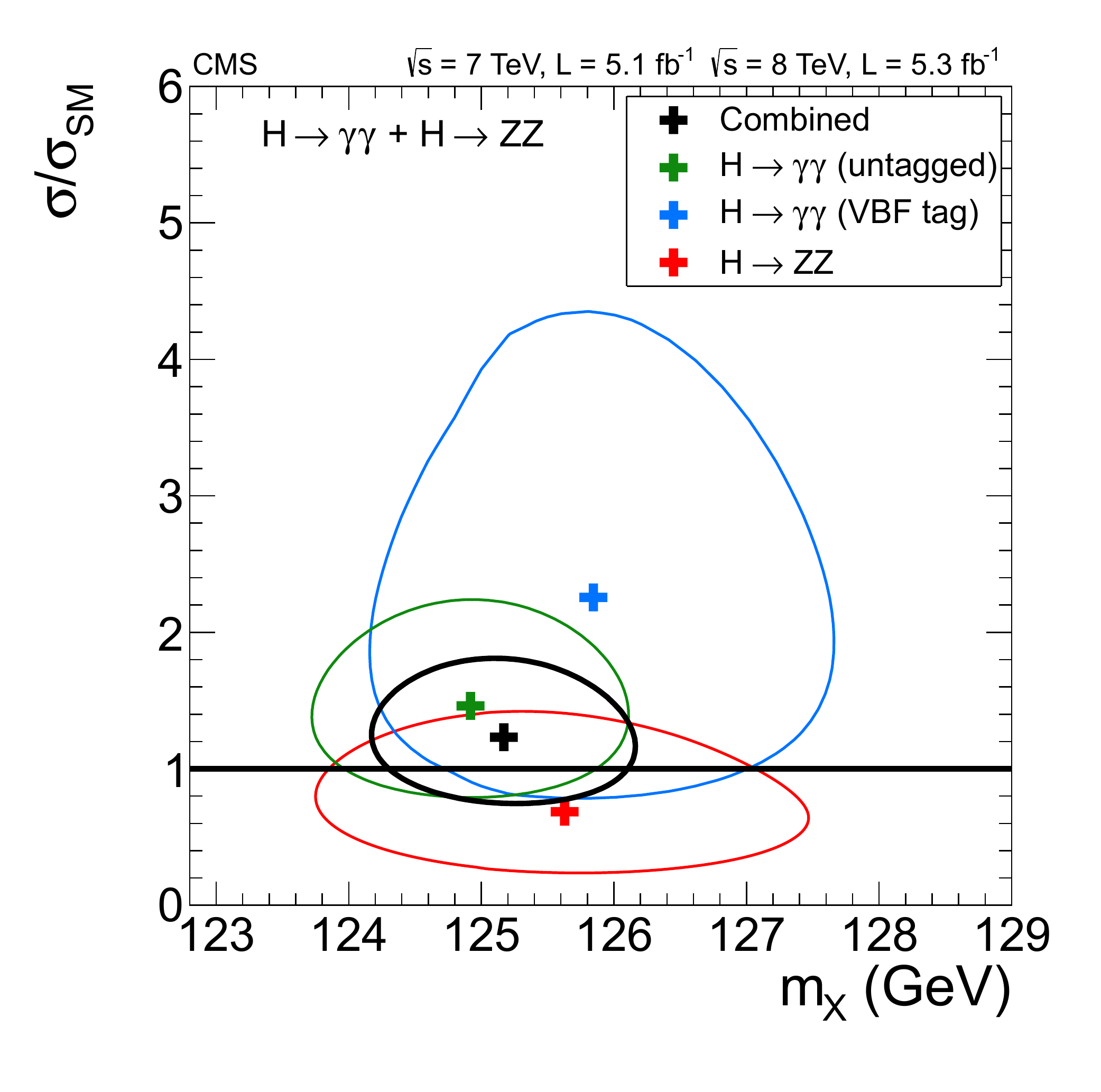}
\caption{
The 68\% CL contours for the signal strength $\sigma/\sigma_\text{SM}$ versus the boson mass $m_\mathrm{X}$ for the untagged $\gamma \gamma$,
   $\gamma \gamma$ with VBF-like dijet, 4$\ell$, and their
   combination.
The symbol $\sigma/\sigma_\mathrm{SM}$ denotes the
      production cross section times the relevant branching fractions,
relative to the SM expectation.   In this combination, the relative signal strengths for the three
   decay modes
   are constrained by the expectations for the SM Higgs boson.
}
\label{fig:fit_mass}
\end{center}
\end{figure}

\begin{figure} [htbp]
\begin{center}
\includegraphics[width=\cmsFigWideWidth]{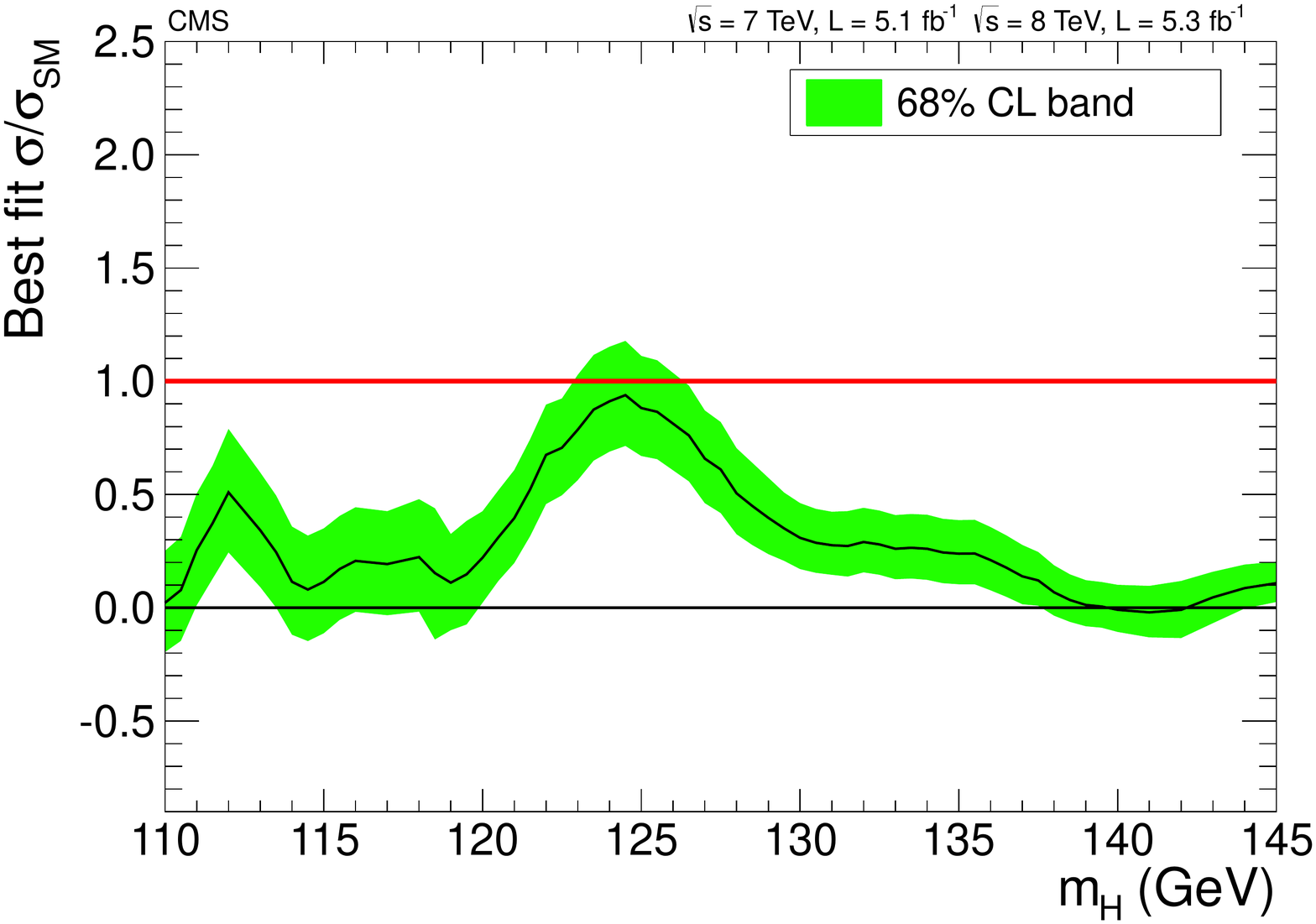}
\caption{The observed best-fit signal strength $\sigma/\sigma_\mathrm{SM}$
as a function of the SM Higgs boson mass in the range 110--145\GeV
for the combined 7~and 8\TeV data sets.
The symbol $\sigma/\sigma_\mathrm{SM}$ denotes the
      production cross section times the relevant branching fractions,
relative to the SM expectation. The band corresponds to the $\pm 1$ standard deviation uncertainty
in $\sigma/\sigma_\mathrm{SM}$.
    }
\label{fig:muhat}
\end{center}
\end{figure}

\begin{figure} [htbp]
\begin{center}
\includegraphics[width=\cmsFigWideWidth]{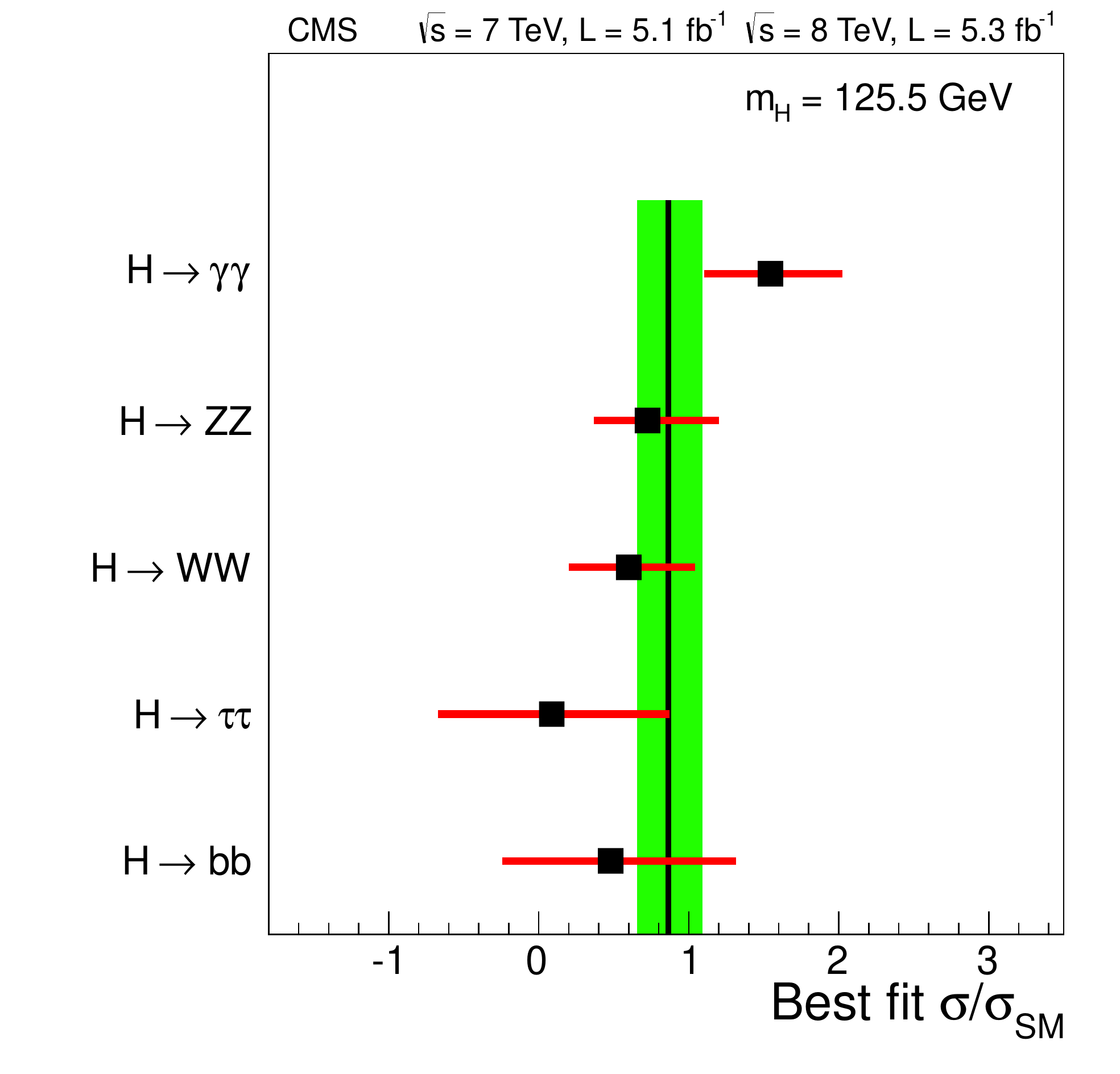}
\caption{Values of $\sigma/\sigma_\text{SM}$ for the combination (solid vertical line)
and for individual decay modes (points).
The vertical band shows the overall $\sigma/\sigma_\text{SM}$ value
\MUHAT.
The symbol $\sigma/\sigma_\mathrm{SM}$ denotes the
      production cross section times the relevant branching fractions,
relative to the SM expectation.
The horizontal bars indicate the $\pm 1$ standard deviation uncertainties
in the $\sigma/\sigma_\text{SM}$ values for individual modes; they include both statistical and systematic uncertainties.
    }
\label{fig:SelfConsistencyDecayProd}
\end{center}
\end{figure}

\section{Conclusions}\label{sec:Conclusion}

Results are presented from searches for the standard model Higgs boson
in proton-proton collisions at $\sqrt{s}=7$~and~8\TeV
in the CMS experiment at the LHC, using data samples corresponding
to integrated luminosities of
up to 5.1\fbinv at 7\TeV and 5.3\fbinv at 8\TeV.
The search is performed in five decay modes:
$\Pgg\Pgg$, $\cPZ\cPZ$, $\PWp\PWm$, $\Pgt^+\Pgt^-$, and $\bbbar$.
An excess of events  is observed above the expected background, with
a local significance of 5.0$\,\sigma$, at a mass near
125\GeV, signalling the production of a new particle.
The expected local significance for a standard model Higgs boson of that mass is 5.8$\,\sigma$.
The global $p$-value in the search range of 115--130 (110--145)\GeV corresponds to 4.6\,$\sigma$ (4.5\,$\sigma$).
The excess is most significant in the two decay modes with
the best mass resolution, $\Pgg\Pgg$ and $\cPZ\cPZ$, and a
fit to these signals gives a mass of
$125.3\pm 0.4\,(\text{stat.})\pm 0.5\,(\text{syst.})\GeV$.
The decay to two photons indicates that the new particle is a boson with spin different from one. The results presented here are consistent,
within uncertainties, with expectations for a standard model Higgs boson.
The collection of further data will enable a more rigorous test of this conclusion and an investigation
of whether the properties of the new particle imply physics beyond the standard model.

\section*{Acknowledgements}

We congratulate our colleagues in the CERN accelerator departments for the excellent performance of the LHC machine. We thank the computing centres in the Worldwide LHC computing Grid for the provisioning and excellent performance of computing
infrastructure essential to our analyses. We gratefully acknowledge the contributions of the technical staff at CERN and other CMS institutes. We also thank the administrative staff at CERN and the other CMS institutes and acknowledge support from BMWF and FWF (Austria); FNRS and FWO (Belgium); CNPq, CAPES, FAPERJ, and FAPESP (Brazil); MES (Bulgaria); CERN; CAS, MoST, and NSFC (China); COLCIENCIAS (Colombia); MSES (Croatia); RPF (Cyprus); MEYS (Czech Republic); MoER, SF0690030s09 and ERDF (Estonia); Academy of Finland, MEC, and HIP (Finland); CEA and CNRS/IN2P3 (France); BMBF, DFG, and HGF (Germany); GSRT (Greece); OTKA and NKTH (Hungary); DAE and DST (India); IPM (Iran); SFI (Ireland); INFN (Italy); NRF and WCU (Korea); LAS (Lithuania); CINVESTAV, CONACYT, SEP, and UASLP-FAI (Mexico); MSI (New Zealand); PAEC (Pakistan); MSHE and NSC (Poland); FCT (Portugal); JINR (Armenia, Belarus, Georgia, Ukraine, Uzbekistan); MON, RosAtom, RAS and RFBR (Russia); MSTD (Serbia); SEIDI and CPAN (Spain); Swiss Funding Agencies (Switzerland); NSC (Taipei); TUBITAK and TAEK (Turkey); NASU (Ukraine); STFC (United Kingdom); DOE and NSF (USA). Individuals have received support from the Marie-Curie programme and the European Research Council (European Union); the Leventis Foundation; the A. P. Sloan Foundation; the Alexander von Humboldt Foundation; the Austrian Science Fund (FWF); the Belgian Federal Science Policy Office; the Fonds pour la Formation \`a la Recherche dans l'Industrie et dans l'Agriculture (FRIA-Belgium); the Agentschap voor Innovatie door Wetenschap en Technologie (IWT-Belgium); the Council of Science and Industrial Research, India; the Compagnia di San Paolo (Torino); and the HOMING PLUS programme of Foundation for Polish Science, cofinanced from European Union, Regional Development Fund.
\bibliography{auto_generated}   

\providecommand{\href}[2]{#2}\begingroup\raggedright\begin{thebibliography}{100}%
\makeatletter
\providecommand{\hrefCMSnoop }[0]{\@secondoftwo}%
\makeatother
\providecommand{\doi}{\texttt{doi:}\begingroup \urlstyle{tt}\Url}

\bibitem{Englert:1964et}
\hrefCMSnoop {} {F.~Englert and R.~Brout, ``{Broken symmetry and the mass of
  gauge vector mesons}'',} \textit{ Phys. Rev. Lett.} \textbf{ 13} (1964) 321,
  \href{http://dx.doi.org/10.1103/PhysRevLett.13.321}{\doi{10.1103/PhysRevLett.13.321}}.

\bibitem{Higgs:1964ia}
\hrefCMSnoop {} {P.~W. Higgs, ``{Broken symmetries, massless particles and
  gauge fields}'',} \textit{ Phys. Lett.} \textbf{ 12} (1964) 132,
  \href{http://dx.doi.org/10.1016/0031-9163(64)91136-9}{\doi{10.1016/0031-9163(64)91136-9}}.

\bibitem{Higgs:1964pj}
\hrefCMSnoop {} {P.~W. Higgs, ``{Broken symmetries and the masses of gauge
  bosons}'',} \textit{ Phys. Rev. Lett.} \textbf{ 13} (1964) 508,
  \href{http://dx.doi.org/10.1103/PhysRevLett.13.508}{\doi{10.1103/PhysRevLett.13.508}}.

\bibitem{Guralnik:1964eu}
\hrefCMSnoop {} {G.~S. Guralnik, C.~R. Hagen, and T.~W.~B. Kibble, ``{Global
  conservation laws and massless particles}'',} \textit{ Phys. Rev. Lett.}
  \textbf{ 13} (1964) 585,
  \href{http://dx.doi.org/10.1103/PhysRevLett.13.585}{\doi{10.1103/PhysRevLett.13.585}}.

\bibitem{Higgs:1966ev}
\hrefCMSnoop {} {P.~W. Higgs, ``{Spontaneous symmetry breakdown without
  massless bosons}'',} \textit{ Phys. Rev.} \textbf{ 145} (1966) 1156,
  \href{http://dx.doi.org/10.1103/PhysRev.145.1156}{\doi{10.1103/PhysRev.145.1156}}.

\bibitem{Kibble:1967sv}
\hrefCMSnoop {} {T.~W.~B. Kibble, ``{Symmetry breaking in non-Abelian gauge
  theories}'',} \textit{ Phys. Rev.} \textbf{ 155} (1967) 1554,
  \href{http://dx.doi.org/10.1103/PhysRev.155.1554}{\doi{10.1103/PhysRev.155.1554}}.

\bibitem{Glashow:1961tr}
\hrefCMSnoop {} {S.~L. Glashow, ``{Partial-symmetries of weak interactions}'',}
  \textit{ Nucl. Phys.} \textbf{ 22} (1961) 579,
\href{http://dx.doi.org/10.1016/0029-5582(61)90469-2}{\doi{10.1016/0029-5582(61)90469-2}}.

\bibitem{Weinberg:1967tq}
\hrefCMSnoop {} {S.~Weinberg, ``{A Model of Leptons}'',} \textit{ Phys. Rev.
  Lett.} \textbf{ 19} (1967) 1264,
\href{http://dx.doi.org/10.1103/PhysRevLett.19.1264}{\doi{10.1103/PhysRevLett.19.1264}}.

\bibitem{sm_salam}
\hrefCMSnoop {} {A.~Salam, ``Weak and electromagnetic interactions'',} in
  \textit{ Elementary particle physics: relativistic groups and analyticity},
  N.~Svartholm, ed., p.~367.
\newblock Almqvist \& Wiskell, 1968.
\newblock Proceedings of the eighth Nobel symposium.

\bibitem{Cornwall:1973tb}
\hrefCMSnoop {} {J.~M. Cornwall, D.~N. Levin, and G.~Tiktopoulos, ``{Uniqueness
  of spontaneously broken gauge theories}'',} \textit{ Phys. Rev. Lett.}
  \textbf{ 30} (1973) 1268,
\href{http://dx.doi.org/10.1103/PhysRevLett.30.1268}{\doi{10.1103/PhysRevLett.30.1268}}.

\bibitem{Cornwall:1974km}
\hrefCMSnoop {} {J.~M. Cornwall, D.~N. Levin, and G.~Tiktopoulos, ``{Derivation
  of Gauge Invariance from High-Energy Unitarity Bounds on the s Matrix}'',}
  \textit{ Phys. Rev. D} \textbf{ 10} (1974) 1145,
  \href{http://dx.doi.org/10.1103/PhysRevD.10.1145}{\doi{10.1103/PhysRevD.10.1145}}.
Also Erratum, \doi{10.1103/PhysRevD.11.972}.

\bibitem{LlewellynSmith:1973ey}
\hrefCMSnoop {} {C.~H. Llewellyn~Smith, ``{High-Energy Behavior and Gauge
  Symmetry}'',} \textit{ Phys. Lett. B} \textbf{ 46} (1973) 233,
\href{http://dx.doi.org/10.1016/0370-2693(73)90692-8}{\doi{10.1016/0370-2693(73)90692-8}}.

\bibitem{Lee:1977eg}
\hrefCMSnoop {} {B.~W. Lee, C.~Quigg, and H.~B. Thacker, ``{Weak interactions
  at very high energies: The role of the Higgs-boson mass}'',} \textit{ Phys.
  Rev. D} \textbf{ 16} (1977) 1519,
\href{http://dx.doi.org/10.1103/PhysRevD.16.1519}{\doi{10.1103/PhysRevD.16.1519}}.

\bibitem{EWKlimits}
\href {http://cdsweb.cern.ch/record/1313716} {{ALEPH, CDF, D0, DELPHI, L3,
  OPAL, SLD Collaborations, the LEP Electroweak Working Group, the Tevatron
  Electroweak Working Group, and the SLD Electroweak and Heavy Flavour Groups},
  ``Precision Electroweak Measurements and Constraints on the Standard
  Model'',} CERN PH-EP-2010-095, (2010).
\newblock At this time, the most up-to-date Higgs boson mass constraints come
  from http://lepewwg.web.cern.ch/LEPEWWG/plots/winter2012/.

\bibitem{LEPlimits}
\hrefCMSnoop {} {{ALEPH, DELPHI, L3, OPAL Collaborations, and LEP Working Group
  for Higgs Boson Searches}, ``{Search for the standard model Higgs boson at
  LEP}'',} \textit{ Phys. Lett. B} \textbf{ 565} (2003) 61,
  \href{http://dx.doi.org/10.1016/S0370-2693(03)00614-2}{\doi{10.1016/S0370-2693(03)00614-2}},
\href{http://www.arXiv.org/abs/hep-ex/0306033}{\texttt{ arXiv:hep-ex/0306033}}.

\bibitem{TEVHIGGS_2010}
\hrefCMSnoop {} {{CDF and D0 Collaborations}, ``Combination of {T}evatron
  Searches for the Standard Model {H}iggs Boson in the {$W^+W^-$} Decay
  Mode'',} \textit{ Phys. Rev. Lett.} \textbf{ 104} (2010) 061802,
  \href{http://dx.doi.org/10.1103/PhysRevLett.104.061802}{\doi{10.1103/PhysRevLett.104.061802}}.

\bibitem{CDF:Hbb}
\hrefCMSnoop {} {{CDF Collaboration}, ``Combined search for the standard model
  {H}iggs boson decaying to a $b\bar{b}$ pair using the full {CDF} data set'',}
  (2012). \href{http://www.arXiv.org/abs/1207.1707}{\texttt{ arXiv:1207.1707}}.
  Submitted to Phys. Rev. Lett.

\bibitem{CDFD0:HbbCombined}
\hrefCMSnoop {} {{CDF and D0 Collaborations}, ``Evidence for a particle
  produced in association with weak bosons and decaying to a bottom-antibottom
  quark pair in {H}iggs boson search at the {Tevatron}'',} (2012).
  \href{http://www.arXiv.org/abs/1207.6436}{\texttt{ arXiv:1207.6436}}.
  Submitted to Phys. Rev. Lett.

\bibitem{D0CombHbb}
\hrefCMSnoop {} {{ D0} Collaboration, ``Combined search for the standard model
  {H}iggs boson decaying to $b \bar{b}$ using the {D0 Run II} data set'',}
  (2012). \href{http://www.arXiv.org/abs/1207.6631}{\texttt{ arXiv:1207.6631}}.
  Submitted to Phys. Rev. Lett.

\bibitem{lhc}
\hrefCMSnoop {} {L.~Evans and P.~{Bryant (editors)}, ``{LHC Machine}'',}
  \textit{ JINST} \textbf{ 3} (2008) S08001,
\href{http://dx.doi.org/10.1088/1748-0221/3/08/S08001}{\doi{10.1088/1748-0221/3/08/S08001}}.

\bibitem{Chatrchyan:2012tx}
\hrefCMSnoop {} {{ CMS} Collaboration, ``{Combined results of searches for the
  standard model Higgs boson in pp collisions at $\sqrt{s} = 7$ TeV}'',}
  \textit{ Phys. Lett. B} \textbf{ 710} (2012) 26,
  \href{http://dx.doi.org/10.1016/j.physletb.2012.02.064}{\doi{10.1016/j.physletb.2012.02.064}},
\href{http://www.arXiv.org/abs/1202.1488}{\texttt{ arXiv:1202.1488}}.

\bibitem{ATLAScombJul2012_7TeV}
\hrefCMSnoop {} {{ ATLAS} Collaboration, ``{Combined search for the Standard
  Model Higgs boson in $pp$ collisions at $\sqrt{s} = 7$ TeV with the ATLAS
  detector}'',} \textit{ Phys. Rev. D} \textbf{ 86} (2012) 032003,
  \href{http://dx.doi.org/10.1103/PhysRevD.86.032003}{\doi{10.1103/PhysRevD.86.032003}},
\href{http://www.arXiv.org/abs/1207.0319}{\texttt{ arXiv:1207.0319}}.

\bibitem{LHCHiggsCrossSectionWorkingGroup:2011ti}
{LHC Higgs Cross Section Working Group}\href
  {http://cdsweb.cern.ch/record/1318996} { {et~al.}, ``{Handbook of LHC Higgs
  Cross Sections: 1. Inclusive Observables}'',} (CERN, Geneva, 2011).
  \href{http://www.arXiv.org/abs/1101.0593}{\texttt{ arXiv:1101.0593}}.

\bibitem{Chatrchyan:2012tw}
\hrefCMSnoop {} {{ CMS} Collaboration, ``Search for the standard model {H}iggs
  boson decaying into two photons in {$\Pp\Pp$} collisions at $\sqrt{s}=7$
  {TeV}'',} \textit{ Phys. Lett. B} \textbf{ 710} (2012) 403,
  \href{http://dx.doi.org/10.1016/j.physletb.2012.03.003}{\doi{10.1016/j.physletb.2012.03.003}},
\href{http://www.arXiv.org/abs/1202.1487}{\texttt{ arXiv:1202.1487}}.

\bibitem{Chatrchyan:2012dg}
\hrefCMSnoop {} {{ CMS} Collaboration, ``Search for the standard model {H}iggs
  boson in the decay channel {$\PH\rightarrow\cPZ\cPZ\rightarrow 4\ell$ in
  $\Pp\Pp$ collisions at $\sqrt{s} = 7\TeV$}'',} \textit{ Phys. Rev. Lett.}
  \textbf{ 108} (2012) 111804,
  \href{http://dx.doi.org/10.1103/PhysRevLett.108.111804}{\doi{10.1103/PhysRevLett.108.111804}},
\href{http://www.arXiv.org/abs/1202.1997}{\texttt{ arXiv:1202.1997}}.

\bibitem{Chatrchyan:2012ty}
\hrefCMSnoop {} {{ CMS} Collaboration, ``Search for the standard model {H}iggs
  boson decaying to {\PWp{}\PWm} in the fully leptonic final state in pp
  collisions at {$\sqrt{s} = 7\TeV$}'',} \textit{ Phys. Lett. B} \textbf{ 710}
  (2012) 91,
  \href{http://dx.doi.org/10.1016/j.physletb.2012.02.076}{\doi{10.1016/j.physletb.2012.02.076}},
\href{http://www.arXiv.org/abs/1202.1489}{\texttt{ arXiv:1202.1489}}.

\bibitem{Chatrchyan:2012vp}
\hrefCMSnoop {} {{ CMS} Collaboration, ``Search for neutral {H}iggs bosons
  decaying to tau pairs in {$\Pp\Pp$} collisions at {$\sqrt{s}=7\TeV$}'',}
  \textit{ Phys. Lett. B} \textbf{ 713} (2012) 68,
  \href{http://dx.doi.org/10.1016/j.physletb.2012.05.028}{\doi{10.1016/j.physletb.2012.05.028}},
\href{http://www.arXiv.org/abs/1202.4083}{\texttt{ arXiv:1202.4083}}.

\bibitem{Chatrchyan:2012ww}
\hrefCMSnoop {} {{ CMS} Collaboration, ``Search for the standard model {H}iggs
  boson decaying to bottom quarks in pp collisions at $\sqrt{s}=7$ {TeV}'',}
  \textit{ Phys. Lett. B} \textbf{ 710} (2012) 284,
  \href{http://dx.doi.org/10.1016/j.physletb.2012.02.085}{\doi{10.1016/j.physletb.2012.02.085}},
\href{http://www.arXiv.org/abs/1202.4195}{\texttt{ arXiv:1202.4195}}.

\bibitem{Pimia:1990zy}
\href {http://cdsweb.cern.ch/record/215299/files/CERN-90-10-V-3.pdf}
  {M.~Della~Negra {et~al.}, ``Muon trigger and identification'',} in \textit{
  Proceedings of the Large Hadron Collider Workshop}, G.~Jarlskog and D.~Rein,
  eds., p.~467.
\newblock Aachen, Germany, 1990.
\newblock {CERN 90-10-V-3/ECFA 90-133-V-3}.

\bibitem{DellaNegra:1992hp}
\href {https://cdsweb.cern.ch/record/290808} {{ CMS} Collaboration, ``{Letter
  of intent by the CMS Collaboration for a general purpose detector at the
  LHC}'',} Technical Report CERN-LHCC-92-03, CERN-LHCC-I-1, CERN, (1992).

\bibitem{Ellis:1994sq}
\hrefCMSnoop {} {N.~Ellis and T.~S. Virdee, ``{Experimental challenges in high
  luminosity collider physics}'',} \textit{ Ann. Rev. Nucl. Part. Sci.}
  \textbf{ 44} (1994) 609,
\href{http://dx.doi.org/10.1146/annurev.ns.44.120194.003141}{\doi{10.1146/annurev.ns.44.120194.003141}}.

\bibitem{Chatrchyan:2008zzk}
\hrefCMSnoop {} {{ CMS} Collaboration, ``The {CMS} experiment at the {CERN}
  {LHC}'',} \textit{ JINST} \textbf{ 3} (2008) S08004,
  \href{http://dx.doi.org/10.1088/1748-0221/3/08/S08004}{\doi{10.1088/1748-0221/3/08/S08004}}.

\bibitem{CMS-PAS-BTV-11-004}
\href {https://cdsweb.cern.ch/record/1427247} {{CMS Collaboration}, ``b-Jet
  Identification in the {CMS} Experiment'',} CMS Physics Analysis Summary
  CMS-PAS-BTV-11-004, (2012).

\bibitem{PhysRevD.83.112004}
\hrefCMSnoop {} {{ CMS} Collaboration, ``Upsilon production cross section in
  $pp$ collisions at $\sqrt{s}=7$ {TeV}'',} \textit{ Phys. Rev. D} \textbf{ 83}
  (2011) 112004,
  \href{http://dx.doi.org/10.1103/PhysRevD.83.112004}{\doi{10.1103/PhysRevD.83.112004}},
  \href{http://www.arXiv.org/abs/1012.5545}{\texttt{ arXiv:1012.5545}}.

\bibitem{CMS-PAS-PFT-09-001}
\href {http://cdsweb.cern.ch/record/1194487} {{CMS Collaboration},
  ``Particle--Flow Event Reconstruction in {CMS} and Performance for Jets,
  Taus, and {\MET}'',} CMS Physics Analysis Summary CMS-PAS-PFT-09-001, (2009).

\bibitem{CMS-PAS-PFT-10-001}
\href {http://cdsweb.cern.ch/record/1247373} {{CMS Collaboration},
  ``Commissioning of the Particle-flow Event Reconstruction with the first
  {LHC} collisions recorded in the {CMS} detector'',} CMS Physics Analysis
  Summary CMS-PAS-PFT-10-001, (2010).

\bibitem{Cacciari:2008gp}
\hrefCMSnoop {} {M.~Cacciari, G.~P. Salam, and G.~Soyez, ``{The anti-$k_t$ jet
  clustering algorithm}'',} \textit{ JHEP} \textbf{ 04} (2008) 063,
\href{http://dx.doi.org/10.1088/1126-6708/2008/04/063}{\doi{10.1088/1126-6708/2008/04/063}}.

\bibitem{Cacciari:2007fd}
\hrefCMSnoop {} {M.~Cacciari and G.~P. Salam, ``{Pileup subtraction using jet
  areas}'',} \textit{ Phys. Lett. B} \textbf{ 659} (2008) 119,
\href{http://dx.doi.org/10.1016/j.physletb.2007.09.077}{\doi{10.1016/j.physletb.2007.09.077}}.

\bibitem{Cacciari:2008gn}
\hrefCMSnoop {} {M.~Cacciari, G.~P. Salam, and G.~Soyez, ``The catchment area
  of jets'',} \textit{ JHEP} \textbf{ 04} (2008) 005,
\href{http://dx.doi.org/10.1088/1126-6708/2008/04/005}{\doi{10.1088/1126-6708/2008/04/005}}.

\bibitem{Cacciari:2011ma}
\hrefCMSnoop {} {M.~Cacciari, G.~P. Salam, and G.~Soyez, ``{FastJet user manual
  (for version 3.0.2)}'',} \textit{ Eur. Phys. J. C} \textbf{ 72} (2012) 1896,
  \href{http://dx.doi.org/10.1140/epjc/s10052-012-1896-2}{\doi{10.1140/epjc/s10052-012-1896-2}},
\href{http://www.arXiv.org/abs/1111.6097}{\texttt{ arXiv:1111.6097}}.

\bibitem{CMS-JME-10-011}
\hrefCMSnoop {} {{ CMS} Collaboration, ``Determination of jet energy
  calibration and transverse momentum resolution in {CMS}'',} \textit{ JINST}
  \textbf{ 6} (2011) 11002,
  \href{http://dx.doi.org/10.1088/1748-0221/6/11/P11002}{\doi{10.1088/1748-0221/6/11/P11002}},
  \href{http://www.arXiv.org/abs/1107.4277}{\texttt{ arXiv:1107.4277}}.

\bibitem{CMS-JME-10-009}
\hrefCMSnoop {} {{ CMS} Collaboration, ``Missing transverse energy performance
  of the {CMS} detector'',} \textit{ JINST} \textbf{ 6} (2011) 9001,
  \href{http://dx.doi.org/10.1088/1748-0221/6/09/P09001}{\doi{10.1088/1748-0221/6/09/P09001}},
  \href{http://www.arXiv.org/abs/1106.5048}{\texttt{ arXiv:1106.5048}}.

\bibitem{CMS-PAS-PFT-10-003}
\href {http://cdsweb.cern.ch/record/1279347} {{CMS Collaboration},
  ``Commissioning of the particle-flow event reconstruction with leptons from
  {J}/$\psi$ and {W} decays at 7 {TeV}'',} CMS Physics Analysis Summary
  CMS-PAS-PFT-10-003, (2010).

\bibitem{CMS-PAS-EGM-10-004}
\href {http://cdsweb.cern.ch/record/1299116} {{CMS Collaboration}, ``Electron
  Reconstruction and Identification at $\sqrt{s} = 7$ {TeV}'',} CMS Physics
  Analysis Summary CMS-PAS-EGM-10-004, (2010).

\bibitem{Baffioni:2006cd}
S.~Baffioni\hrefCMSnoop {} { {et~al.}, ``{Electron reconstruction in CMS}'',}
  \textit{ Eur. Phys. J. C} \textbf{ 49} (2007) 1099,
\href{http://dx.doi.org/10.1140/epjc/s10052-006-0175-5}{\doi{10.1140/epjc/s10052-006-0175-5}}.

\bibitem{Adam2005}
W.~Adam\hrefCMSnoop {} { {et~al.}, ``{Reconstruction of electrons with the
  Gaussian-sum filter in the CMS tracker at the LHC}'',} \textit{ J. Phys. G}
  \textbf{ 31} (2005) N9,
  \href{http://dx.doi.org/10.1088/0954-3899/31/9/N01}{\doi{10.1088/0954-3899/31/9/N01}}.

\bibitem{CMS-PAS-SMP-12-008}
\href {https://cdsweb.cern.ch/record/1434360} {{CMS Collaboration}, ``Absolute
  Calibration of the Luminosity Measurement at {CMS}: Winter 2012 Update'',}
  CMS Physics Analysis Summary CMS-PAS-SMP-12-008, (2012).

\bibitem{vanderMeer:296752}
\href {https://cdsweb.cern.ch/record/296752} {S.~van~der Meer, ``Calibration of
  the effective beam height in the {ISR}'',} Technical Report
  CERN-ISR-PO-68-31, CERN, Geneva, (1968).

\bibitem{Ellis:1975ap}
\hrefCMSnoop {} {J.~R. Ellis, M.~K. Gaillard, and D.~V. Nanopoulos, ``{A
  phenomenological profile of the Higgs boson}'',} \textit{ Nucl. Phys. B}
  \textbf{ 106} (1976) 292,
\href{http://dx.doi.org/10.1016/0550-3213(76)90382-5}{\doi{10.1016/0550-3213(76)90382-5}}.

\bibitem{Georgi:1977gs}
H.~M. Georgi\hrefCMSnoop {} { {et~al.}, ``{Higgs Bosons from Two Gluon
  Annihilation in Proton Proton Collisions}'',} \textit{ Phys. Rev. Lett.}
  \textbf{ 40} (1978) 692,
\href{http://dx.doi.org/10.1103/PhysRevLett.40.692}{\doi{10.1103/PhysRevLett.40.692}}.

\bibitem{Glashow:1978ab}
\hrefCMSnoop {} {S.~L. Glashow, D.~V. Nanopoulos, and A.~Yildiz, ``{Associated
  Production of Higgs Bosons and Z Particles}'',} \textit{ Phys. Rev. D}
  \textbf{ 18} (1978) 1724,
\href{http://dx.doi.org/10.1103/PhysRevD.18.1724}{\doi{10.1103/PhysRevD.18.1724}}.

\bibitem{Cahn:1986zv}
R.~N. Cahn\hrefCMSnoop {} { {et~al.}, ``{Transverse Momentum Signatures for
  Heavy Higgs Bosons}'',} \textit{ Phys. Rev. D} \textbf{ 35} (1987) 1626,
\href{http://dx.doi.org/10.1103/PhysRevD.35.1626}{\doi{10.1103/PhysRevD.35.1626}}.

\bibitem{Gunion:1987ke}
\hrefCMSnoop {} {J.~F. Gunion, G.~L. Kane, and J.~Wudka, ``{Search techniques
  for charged and neutral intermediate mass {H}iggs bosons}'',} \textit{ Nucl.
  Phys. B} \textbf{ 299} (1988) 231,
\href{http://dx.doi.org/10.1016/0550-3213(88)90284-2}{\doi{10.1016/0550-3213(88)90284-2}}.

\bibitem{Rainwater:1997dg}
\hrefCMSnoop {} {D.~L. Rainwater and D.~Zeppenfeld, ``{Searching for $\PH\to
  \gamma \gamma$ in weak boson fusion at the LHC}'',} \textit{ JHEP} \textbf{
  12} (1997) 005,
  \href{http://dx.doi.org/10.1088/1126-6708/1997/12/005}{\doi{10.1088/1126-6708/1997/12/005}},
\href{http://www.arXiv.org/abs/hep-ph/9712271}{\texttt{ arXiv:hep-ph/9712271}}.

\bibitem{Rainwater:1998kj}
\hrefCMSnoop {} {D.~L. Rainwater, D.~Zeppenfeld, and K.~Hagiwara, ``{Searching
  for H $\to \tau \tau $ in weak boson fusion at the CERN LHC}'',} \textit{
  Phys. Rev. D} \textbf{ 59} (1998) 014037,
  \href{http://dx.doi.org/10.1103/PhysRevD.59.014037}{\doi{10.1103/PhysRevD.59.014037}},
\href{http://www.arXiv.org/abs/hep-ph/9808468}{\texttt{ arXiv:hep-ph/9808468}}.

\bibitem{Rainwater:1999sd}
\hrefCMSnoop {} {D.~L. Rainwater and D.~Zeppenfeld, ``{Observing $H \to
  W^{(*)}W^{(*)} \to e^{\pm}\mu^{\mp}\;/\!\!\!p_T$ in weak boson fusion with
  dual forward jet tagging at the CERN LHC}'',} \textit{ Phys. Rev. D} \textbf{
  60} (1999) 113004,
  \href{http://dx.doi.org/10.1103/PhysRevD.60.113004}{\doi{10.1103/PhysRevD.60.113004}},
  \href{http://www.arXiv.org/abs/hep-ph/9906218}{\texttt{
  arXiv:hep-ph/9906218}}.
Also Erratum, \doi{10.1103/PhysRevD.61.099901}.

\bibitem{Djouadi:1991tka}
\hrefCMSnoop {} {A.~Djouadi, M.~Spira, and P.~M. Zerwas, ``{Production of Higgs
  bosons in proton colliders: QCD corrections}'',} \textit{ Phys. Lett. B}
  \textbf{ 264} (1991) 440,
\href{http://dx.doi.org/10.1016/0370-2693(91)90375-Z}{\doi{10.1016/0370-2693(91)90375-Z}}.

\bibitem{Dawson:1990zj}
\hrefCMSnoop {} {S.~Dawson, ``{Radiative corrections to Higgs boson
  production}'',} \textit{ Nucl. Phys. B} \textbf{ 359} (1991) 283,
\href{http://dx.doi.org/10.1016/0550-3213(91)90061-2}{\doi{10.1016/0550-3213(91)90061-2}}.

\bibitem{Spira:1995rr}
M.~Spira\hrefCMSnoop {} { {et~al.}, ``{Higgs boson production at the LHC}'',}
  \textit{ Nucl. Phys. B} \textbf{ 453} (1995) 17,
  \href{http://dx.doi.org/10.1016/0550-3213(95)00379-7}{\doi{10.1016/0550-3213(95)00379-7}},
\href{http://www.arXiv.org/abs/hep-ph/9504378}{\texttt{ arXiv:hep-ph/9504378}}.

\bibitem{Harlander:2002wh}
\hrefCMSnoop {} {R.~V. Harlander and W.~B. Kilgore, ``{Next-to-next-to-leading
  order Higgs production at hadron colliders}'',} \textit{ Phys. Rev. Lett.}
  \textbf{ 88} (2002) 201801,
  \href{http://dx.doi.org/10.1103/PhysRevLett.88.201801}{\doi{10.1103/PhysRevLett.88.201801}},
\href{http://www.arXiv.org/abs/hep-ph/0201206}{\texttt{ arXiv:hep-ph/0201206}}.

\bibitem{Anastasiou:2002yz}
\hrefCMSnoop {} {C.~Anastasiou and K.~Melnikov, ``{Higgs boson production at
  hadron colliders in NNLO QCD}'',} \textit{ Nucl. Phys. B} \textbf{ 646}
  (2002) 220,
  \href{http://dx.doi.org/10.1016/S0550-3213(02)00837-4}{\doi{10.1016/S0550-3213(02)00837-4}},
\href{http://www.arXiv.org/abs/hep-ph/0207004}{\texttt{ arXiv:hep-ph/0207004}}.

\bibitem{Ravindran:2003um}
\hrefCMSnoop {} {V.~Ravindran, J.~Smith, and W.~L. van Neerven, ``{NNLO
  corrections to the total cross section for Higgs boson production in hadron
  hadron collisions}'',} \textit{ Nucl. Phys. B} \textbf{ 665} (2003) 325,
  \href{http://dx.doi.org/10.1016/S0550-3213(03)00457-7}{\doi{10.1016/S0550-3213(03)00457-7}},
\href{http://www.arXiv.org/abs/hep-ph/0302135}{\texttt{ arXiv:hep-ph/0302135}}.

\bibitem{Catani:2003zt}
S.~Catani\hrefCMSnoop {} { {et~al.}, ``{Soft-gluon resummation for Higgs boson
  production at hadron colliders}'',} \textit{ JHEP} \textbf{ 07} (2003) 028,
\href{http://dx.doi.org/10.1088/1126-6708/2003/07/028}{\doi{10.1088/1126-6708/2003/07/028}}.

\bibitem{Aglietti:2004nj}
U.~Aglietti\hrefCMSnoop {} { {et~al.}, ``{Two-loop light fermion contribution
  to Higgs production and decays}'',} \textit{ Phys. Lett. B} \textbf{ 595}
  (2004) 432,
  \href{http://dx.doi.org/10.1016/j.physletb.2004.06.063}{\doi{10.1016/j.physletb.2004.06.063}},
\href{http://www.arXiv.org/abs/hep-ph/0404071}{\texttt{ arXiv:hep-ph/0404071}}.

\bibitem{Degrassi:2004mx}
\hrefCMSnoop {} {G.~Degrassi and F.~Maltoni, ``{Two-loop electroweak
  corrections to Higgs production at hadron colliders}'',} \textit{ Phys. Lett.
  B} \textbf{ 600} (2004) 255,
  \href{http://dx.doi.org/10.1016/j.physletb.2004.09.008}{\doi{10.1016/j.physletb.2004.09.008}},
\href{http://www.arXiv.org/abs/hep-ph/0407249}{\texttt{ arXiv:hep-ph/0407249}}.

\bibitem{Actis:2008ug}
S.~Actis\hrefCMSnoop {} { {et~al.}, ``{NLO} Electroweak Corrections to {H}iggs
  Boson Production at Hadron Colliders'',} \textit{ Phys. Lett. B} \textbf{
  670} (2008) 12,
  \href{http://dx.doi.org/10.1016/j.physletb.2008.10.018}{\doi{10.1016/j.physletb.2008.10.018}},
\href{http://www.arXiv.org/abs/0809.1301}{\texttt{ arXiv:0809.1301}}.

\bibitem{Anastasiou:2008tj}
\hrefCMSnoop {} {C.~Anastasiou, R.~Boughezal, and F.~Petriello, ``{Mixed
  QCD-electroweak corrections to Higgs boson production in gluon fusion}'',}
  \textit{ JHEP} \textbf{ 04} (2009) 003,
  \href{http://dx.doi.org/10.1088/1126-6708/2009/04/003}{\doi{10.1088/1126-6708/2009/04/003}},
\href{http://www.arXiv.org/abs/0811.3458}{\texttt{ arXiv:0811.3458}}.

\bibitem{deFlorian:2009hc}
\hrefCMSnoop {} {D.~de~Florian and M.~Grazzini, ``{Higgs production through
  gluon fusion: updated cross sections at the Tevatron and the LHC}'',}
  \textit{ Phys. Lett. B} \textbf{ 674} (2009) 291,
  \href{http://dx.doi.org/10.1016/j.physletb.2009.03.033}{\doi{10.1016/j.physletb.2009.03.033}},
\href{http://www.arXiv.org/abs/0901.2427}{\texttt{ arXiv:0901.2427}}.

\bibitem{Baglio:2010ae}
\hrefCMSnoop {} {J.~Baglio and A.~Djouadi, ``{Higgs production at the lHC}'',}
  \textit{ JHEP} \textbf{ 03} (2011) 055,
  \href{http://dx.doi.org/10.1007/JHEP03(2011)055}{\doi{10.1007/JHEP03(2011)055}},
\href{http://www.arXiv.org/abs/1012.0530}{\texttt{ arXiv:1012.0530}}.

\bibitem{deFlorian:2012yg}
\hrefCMSnoop {} {D.~de~Florian and M.~Grazzini, ``{Higgs production at the LHC:
  updated cross sections at $\sqrt{s}=8$ TeV}'',} (2012).
\href{http://www.arXiv.org/abs/1206.4133}{\texttt{ arXiv:1206.4133}}.

\bibitem{Bozzi:2005wk}
G.~Bozzi\hrefCMSnoop {} { {et~al.}, ``{Transverse-momentum resummation and the
  spectrum of the Higgs boson at the LHC}'',} \textit{ Nucl. Phys. B} \textbf{
  737} (2006) 73,
  \href{http://dx.doi.org/10.1016/j.nuclphysb.2005.12.022}{\doi{10.1016/j.nuclphysb.2005.12.022}},
\href{http://www.arXiv.org/abs/hep-ph/0508068}{\texttt{ arXiv:hep-ph/0508068}}.

\bibitem{deFlorian:2011xf}
D.~de~Florian\hrefCMSnoop {} { {et~al.}, ``Transverse-momentum resummation:
  {H}iggs boson production at the {T}evatron and the {LHC}'',} \textit{ JHEP}
  \textbf{ 11} (2011) 064,
\href{http://dx.doi.org/10.1007/JHEP11(2011)064}{\doi{10.1007/JHEP11(2011)064}}.

\bibitem{Passarino:2010qk}
\hrefCMSnoop {} {G.~Passarino, C.~Sturm, and S.~Uccirati, ``{H}iggs
  Pseudo-Observables, Second {R}iemann Sheet and All That'',} \textit{ Nucl.
  Phys. B} \textbf{ 834} (2010) 77,
  \href{http://dx.doi.org/10.1016/j.nuclphysb.2010.03.013}{\doi{10.1016/j.nuclphysb.2010.03.013}},
\href{http://www.arXiv.org/abs/1001.3360}{\texttt{ arXiv:1001.3360}}.

\bibitem{Stewart:2011cf}
\hrefCMSnoop {} {I.~W. Stewart and F.~J. Tackmann, ``Theory Uncertainties for
  {H}iggs and Other Searches Using Jet Bins'',} \textit{ Phys. Rev. D} \textbf{
  85} (2012) 034011,
  \href{http://dx.doi.org/10.1103/PhysRevD.85.034011}{\doi{10.1103/PhysRevD.85.034011}},
\href{http://www.arXiv.org/abs/1107.2117}{\texttt{ arXiv:1107.2117}}.

\bibitem{Djouadi:1997yw}
\hrefCMSnoop {} {A.~Djouadi, J.~Kalinowski, and M.~Spira, ``{HDECAY: A program
  for Higgs boson decays in the standard model and its supersymmetric
  extension}'',} \textit{ Comput. Phys. Commun.} \textbf{ 108} (1998) 56,
  \href{http://dx.doi.org/10.1016/S0010-4655(97)00123-9}{\doi{10.1016/S0010-4655(97)00123-9}},
\href{http://www.arXiv.org/abs/hep-ph/9704448}{\texttt{ arXiv:hep-ph/9704448}}.

\bibitem{hdecay2}
A.~Djouadi\hrefCMSnoop {} { {et~al.}, ``{An update of the program HDECAY}'',}
  in \textit{ {The Les Houches 2009 workshop on TeV colliders: The tools and
  Monte Carlo working group summary report}}.
\newblock 2010.
\newblock \href{http://www.arXiv.org/abs/1003.1643}{\texttt{ arXiv:1003.1643}}.

\bibitem{Bredenstein:2006rh}
A.~Bredenstein\hrefCMSnoop {} { {et~al.}, ``{Precise predictions for the
  Higgs-boson decay H $\rightarrow$ WW/ZZ $\rightarrow$ 4 leptons}'',} \textit{
  Phys. Rev. D} \textbf{ 74} (2006) 013004,
  \href{http://dx.doi.org/10.1103/PhysRevD.74.013004}{\doi{10.1103/PhysRevD.74.013004}},
\href{http://www.arXiv.org/abs/hep-ph/0604011}{\texttt{ arXiv:hep-ph/0604011}}.

\bibitem{Bredenstein:2006ha}
A.~Bredenstein\hrefCMSnoop {} { {et~al.}, ``{Radiative corrections to the
  semileptonic and hadronic Higgs-boson decays H $\rightarrow
  $WW/ZZ$\rightarrow$ 4 fermions}'',} \textit{ JHEP} \textbf{ 02} (2007) 080,
  \href{http://dx.doi.org/10.1088/1126-6708/2007/02/080}{\doi{10.1088/1126-6708/2007/02/080}},
  \href{http://www.arXiv.org/abs/hep-ph/0611234}{\texttt{
  arXiv:hep-ph/0611234}}.

\bibitem{Actis:2008ts}
S.~Actis\hrefCMSnoop {} { {et~al.}, ``{NNLO} Computational Techniques: the
  Cases {$H \to \gamma \gamma$ and $H \to g g$}'',} \textit{ Nucl. Phys. B}
  \textbf{ 811} (2009) 182,
  \href{http://dx.doi.org/10.1016/j.nuclphysb.2008.11.024}{\doi{10.1016/j.nuclphysb.2008.11.024}},
\href{http://www.arXiv.org/abs/0809.3667}{\texttt{ arXiv:0809.3667}}.

\bibitem{Denner:2011mq}
A.~Denner\hrefCMSnoop {} { {et~al.}, ``Standard Model {H}iggs-Boson Branching
  Ratios with Uncertainties'',} \textit{ Eur. Phys. J. C} \textbf{ 71} (2011)
  1753,
  \href{http://dx.doi.org/10.1140/epjc/s10052-011-1753-8}{\doi{10.1140/epjc/s10052-011-1753-8}},
\href{http://www.arXiv.org/abs/1107.5909}{\texttt{ arXiv:1107.5909}}.

\bibitem{Ciccolini:2007jr}
\hrefCMSnoop {} {M.~Ciccolini, A.~Denner, and S.~Dittmaier, ``Strong and
  Electroweak Corrections to the Production of a {H}iggs Boson+2 Jets via Weak
  Interactions at the {Large Hadron Collider}'',} \textit{ Phys. Rev. Lett.}
  \textbf{ 99} (2007) 161803,
  \href{http://dx.doi.org/10.1103/PhysRevLett.99.161803}{\doi{10.1103/PhysRevLett.99.161803}},
  \href{http://www.arXiv.org/abs/0707.0381}{\texttt{ arXiv:0707.0381}}.

\bibitem{Ciccolini:2007ec}
\hrefCMSnoop {} {M.~Ciccolini, A.~Denner, and S.~Dittmaier, ``{Electroweak and
  QCD corrections to Higgs production via vector-boson fusion at the LHC}'',}
  \textit{ Phys. Rev. D} \textbf{ 77} (2008) 013002,
  \href{http://dx.doi.org/10.1103/PhysRevD.77.013002}{\doi{10.1103/PhysRevD.77.013002}},
\href{http://www.arXiv.org/abs/0710.4749}{\texttt{ arXiv:0710.4749}}.

\bibitem{Figy:2003nv}
\hrefCMSnoop {} {T.~Figy, C.~Oleari, and D.~Zeppenfeld, ``{Next-to-leading
  order jet distributions for Higgs boson production via weak-boson fusion}'',}
  \textit{ Phys. Rev. D} \textbf{ 68} (2003) 073005,
  \href{http://dx.doi.org/10.1103/PhysRevD.68.073005}{\doi{10.1103/PhysRevD.68.073005}},
\href{http://www.arXiv.org/abs/hep-ph/0306109}{\texttt{ arXiv:hep-ph/0306109}}.

\bibitem{Arnold:2008rz}
K.~Arnold\hrefCMSnoop {} { {et~al.}, ``{VBFNLO: A parton level Monte Carlo for
  processes with electroweak bosons}'',} \textit{ Comput. Phys. Commun.}
  \textbf{ 180} (2009) 1661,
  \href{http://dx.doi.org/10.1016/j.cpc.2009.03.006}{\doi{10.1016/j.cpc.2009.03.006}},
\href{http://www.arXiv.org/abs/0811.4559}{\texttt{ arXiv:0811.4559}}.

\bibitem{Bolzoni:2010xr}
P.~Bolzoni\hrefCMSnoop {} { {et~al.}, ``{Higgs production via vector-boson
  fusion at NNLO in QCD}'',} \textit{ Phys. Rev. Lett.} \textbf{ 105} (2010)
  011801,
  \href{http://dx.doi.org/10.1103/PhysRevLett.105.011801}{\doi{10.1103/PhysRevLett.105.011801}},
\href{http://www.arXiv.org/abs/1003.4451}{\texttt{ arXiv:1003.4451}}.

\bibitem{Han:1991ia}
\hrefCMSnoop {} {T.~Han and S.~Willenbrock, ``{QCD correction to the $pp \to
  WH$ and $ZH$ total cross-sections}'',} \textit{ Phys. Lett. B} \textbf{ 273}
  (1991) 167,
\href{http://dx.doi.org/10.1016/0370-2693(91)90572-8}{\doi{10.1016/0370-2693(91)90572-8}}.

\bibitem{Brein:2003wg}
\hrefCMSnoop {} {O.~Brein, A.~Djouadi, and R.~Harlander, ``{NNLO QCD
  corrections to the Higgs-strahlung processes at hadron colliders}'',}
  \textit{ Phys. Lett. B} \textbf{ 579} (2004) 149,
  \href{http://dx.doi.org/10.1016/j.physletb.2003.10.112}{\doi{10.1016/j.physletb.2003.10.112}},
\href{http://www.arXiv.org/abs/hep-ph/0307206}{\texttt{ arXiv:hep-ph/0307206}}.

\bibitem{Ciccolini:2003jy}
\hrefCMSnoop {} {M.~L. Ciccolini, S.~Dittmaier, and M.~{Kr\"amer},
  ``{Electroweak radiative corrections to associated $WH$ and $ZH$ production
  at hadron colliders}'',} \textit{ Phys. Rev. D} \textbf{ 68} (2003) 073003,
  \href{http://dx.doi.org/10.1103/PhysRevD.68.073003}{\doi{10.1103/PhysRevD.68.073003}},
\href{http://www.arXiv.org/abs/hep-ph/0306234}{\texttt{ arXiv:hep-ph/0306234}}.

\bibitem{Hamberg:1990np}
\hrefCMSnoop {} {R.~Hamberg, W.~L. van Neerven, and T.~Matsuura, ``{A complete
  calculation of the order $\alpha_\mathrm{S}^{2}$ correction to the Drell-Yan
  $K$ factor}'',} \textit{ Nucl. Phys. B} \textbf{ 359} (1991) 343,
\href{http://dx.doi.org/10.1016/0550-3213(91)90064-5}{\doi{10.1016/0550-3213(91)90064-5}}.

\bibitem{Denner:2011rn}
A.~Denner\hrefCMSnoop {} { {et~al.}, ``{EW corrections to Higgs strahlung at
  the Tevatron and the LHC with HAWK}'',} (2011).
\href{http://www.arXiv.org/abs/1112.5258}{\texttt{ arXiv:1112.5258}}.

\bibitem{Ferrera:2011bk}
\hrefCMSnoop {} {G.~Ferrera, M.~Grazzini, and F.~Tramontano, ``{Associated WH
  production at hadron colliders: a fully exclusive QCD calculation at
  NNLO}'',} \textit{ Phys. Rev. Lett.} \textbf{ 107} (2011) 152003,
  \href{http://dx.doi.org/10.1103/PhysRevLett.107.152003}{\doi{10.1103/PhysRevLett.107.152003}},
\href{http://www.arXiv.org/abs/1107.1164}{\texttt{ arXiv:1107.1164}}.

\bibitem{Beenakker:2001rj}
W.~Beenakker\hrefCMSnoop {} { {et~al.}, ``{Higgs radiation off top quarks at
  the Tevatron and the LHC}'',} \textit{ Phys. Rev. Lett.} \textbf{ 87} (2001)
  201805,
  \href{http://dx.doi.org/10.1103/PhysRevLett.87.201805}{\doi{10.1103/PhysRevLett.87.201805}},
\href{http://www.arXiv.org/abs/hep-ph/0107081}{\texttt{ arXiv:hep-ph/0107081}}.

\bibitem{Beenakker:2002nc}
W.~Beenakker\hrefCMSnoop {} { {et~al.}, ``{NLO QCD corrections to $\ttbar$ H
  production in hadron collisions.}'',} \textit{ Nucl. Phys. B} \textbf{ 653}
  (2003) 151,
  \href{http://dx.doi.org/10.1016/S0550-3213(03)00044-0}{\doi{10.1016/S0550-3213(03)00044-0}},
\href{http://www.arXiv.org/abs/hep-ph/0211352}{\texttt{ arXiv:hep-ph/0211352}}.

\bibitem{Dawson:2002tg}
S.~Dawson\hrefCMSnoop {} { {et~al.}, ``{Associated top quark Higgs boson
  production at the LHC}'',} \textit{ Phys. Rev. D} \textbf{ 67} (2003) 071503,
  \href{http://dx.doi.org/10.1103/PhysRevD.67.071503}{\doi{10.1103/PhysRevD.67.071503}},
\href{http://www.arXiv.org/abs/hep-ph/0211438}{\texttt{ arXiv:hep-ph/0211438}}.

\bibitem{Dawson:2003zu}
S.~Dawson\hrefCMSnoop {} { {et~al.}, ``{Associated Higgs production with top
  quarks at the Large Hadron Collider: NLO QCD corrections}'',} \textit{ Phys.
  Rev. D} \textbf{ 68} (2003) 034022,
  \href{http://dx.doi.org/10.1103/PhysRevD.68.034022}{\doi{10.1103/PhysRevD.68.034022}},
\href{http://www.arXiv.org/abs/hep-ph/0305087}{\texttt{ arXiv:hep-ph/0305087}}.

\bibitem{Botje:2011sn}
M.~Botje\hrefCMSnoop {} { {et~al.}, ``The {PDF4LHC Working Group} Interim
  Recommendations'',} (2011).
  \href{http://www.arXiv.org/abs/1101.0538}{\texttt{ arXiv:1101.0538}}.

\bibitem{Alekhin:2011sk}
S.~Alekhin\hrefCMSnoop {} { {et~al.}, ``{The PDF4LHC Working Group} Interim
  Report'',} (2011). \href{http://www.arXiv.org/abs/1101.0536}{\texttt{
  arXiv:1101.0536}}.

\bibitem{Lai:2010vv}
H.-L. Lai\hrefCMSnoop {} { {et~al.}, ``{New parton distributions for collider
  physics}'',} \textit{ Phys. Rev. D} \textbf{ 82} (2010) 074024,
  \href{http://dx.doi.org/10.1103/PhysRevD.82.074024}{\doi{10.1103/PhysRevD.82.074024}},
  \href{http://www.arXiv.org/abs/1007.2241}{\texttt{ arXiv:1007.2241}}.

\bibitem{Martin:2009iq}
A.~D. Martin\hrefCMSnoop {} { {et~al.}, ``{Parton distributions for the
  LHC}'',} \textit{ Eur. Phys. J. C} \textbf{ 63} (2009) 189,
  \href{http://dx.doi.org/10.1140/epjc/s10052-009-1072-5}{\doi{10.1140/epjc/s10052-009-1072-5}},
  \href{http://www.arXiv.org/abs/0901.0002}{\texttt{ arXiv:0901.0002}}.

\bibitem{Ball:2011mu}
\hrefCMSnoop {} {{ NNPDF} Collaboration, ``Impact of Heavy Quark Masses on
  Parton Distributions and {LHC} Phenomenology'',} \textit{ Nucl. Phys. B}
  \textbf{ 849} (2011) 296,
  \href{http://dx.doi.org/10.1016/j.nuclphysb.2011.03.021}{\doi{10.1016/j.nuclphysb.2011.03.021}},
  \href{http://www.arXiv.org/abs/1101.1300}{\texttt{ arXiv:1101.1300}}.

\bibitem{Anastasiou:2012hx}
C.~Anastasiou\hrefCMSnoop {} { {et~al.}, ``{Inclusive Higgs boson cross-section
  for the LHC at 8 TeV}'',} \textit{ JHEP} \textbf{ 04} (2012) 004,
  \href{http://dx.doi.org/10.1007/JHEP04(2012)004}{\doi{10.1007/JHEP04(2012)004}},
\href{http://www.arXiv.org/abs/1202.3638}{\texttt{ arXiv:1202.3638}}.

\bibitem{Dittmaier:2012vm}
{LHC Higgs Cross Section Working Group}\hrefCMSnoop {} { {et~al.}, ``{Handbook
  of LHC Higgs Cross Sections: 2. Differential Distributions}'',} (CERN,
  Geneva, 2012).
\href{http://www.arXiv.org/abs/1201.3084}{\texttt{ arXiv:1201.3084}}.

\bibitem{Agostinelli:2002hh}
\hrefCMSnoop {} {{ GEANT4} Collaboration, ``{GEANT4}---a simulation toolkit'',}
  \textit{ Nucl. Instrum. Meth. A} \textbf{ 506} (2003) 250,
  \href{http://dx.doi.org/10.1016/S0168-9002(03)01368-8}{\doi{10.1016/S0168-9002(03)01368-8}}.

\bibitem{powheg1}
S.~Alioli\hrefCMSnoop {} { {et~al.}, ``{NLO Higgs boson production via gluon
  fusion matched with shower in POWHEG}'',} \textit{ JHEP} \textbf{ 04} (2009)
  002,
\href{http://dx.doi.org/10.1088/1126-6708/2009/04/002}{\doi{10.1088/1126-6708/2009/04/002}}.

\bibitem{powheg2}
\hrefCMSnoop {} {P.~Nason and C.~Oleari, ``{NLO Higgs boson production via
  vector-boson fusion matched with shower in POWHEG}'',} \textit{ JHEP}
  \textbf{ 02} (2010) 037,
\href{http://dx.doi.org/10.1007/JHEP02(2010)037}{\doi{10.1007/JHEP02(2010)037}}.

\bibitem{Sjostrand:2006za}
\hrefCMSnoop {} {T.~Sj{\"o}strand, S.~Mrenna, and P.~Z. Skands, ``{PYTHIA 6.4
  physics and manual}'',} \textit{ JHEP} \textbf{ 05} (2006) 026,
\href{http://dx.doi.org/10.1088/1126-6708/2006/05/026}{\doi{10.1088/1126-6708/2006/05/026}}.

\bibitem{HqT1}
G.~Bozzi\hrefCMSnoop {} { {et~al.}, ``{The $q_T$ spectrum of the Higgs boson at
  the LHC in QCD perturbation theory}'',} \textit{ Phys. Lett. B} \textbf{ 564}
  (2003) 65,
\href{http://dx.doi.org/10.1016/S0370-2693(03)00656-7}{\doi{10.1016/S0370-2693(03)00656-7}}.

\bibitem{FeHiPro1}
\hrefCMSnoop {} {C.~Anastasiou, K.~Melnikov, and F.~Petriello, ``Fully
  differential {H}iggs boson production and the di-photon signal through
  next-to-next-to-leading order'',} \textit{ Nucl. Phys. B} \textbf{ 724}
  (2005) 197,
  \href{http://dx.doi.org/10.1016/j.nuclphysb.2005.06.036}{\doi{10.1016/j.nuclphysb.2005.06.036}}.

\bibitem{FeHiPro2}
\hrefCMSnoop {} {C.~Anastasiou, S.~Bucherer, and Z.~Kunszt, ``{HPro}: A {NLO}
  {Monte-Carlo} for {H}iggs production via gluon fusion with finite heavy quark
  masses'',} \textit{ JHEP} \textbf{ 10} (2009) 068,
  \href{http://dx.doi.org/10.1088/1126-6708/2009/10/068}{\doi{10.1088/1126-6708/2009/10/068}}.

\bibitem{Gieseke:2006ga}
S.~Gieseke\hrefCMSnoop {} { {et~al.}, ``{Herwig++ 2.0 Release Note}'',} (2006).
\href{http://www.arXiv.org/abs/hep-ph/0609306}{\texttt{ arXiv:hep-ph/0609306}}.

\bibitem{Alwall:2007st}
J.~Alwall\hrefCMSnoop {} { {et~al.}, ``{MadGraph/MadEvent v4: the new web
  generation}'',} \textit{ JHEP} \textbf{ 09} (2007) 028,
  \href{http://dx.doi.org/10.1088/1126-6708/2007/09/028}{\doi{10.1088/1126-6708/2007/09/028}},
\href{http://www.arXiv.org/abs/0706.2334}{\texttt{ arXiv:0706.2334}}.

\bibitem{LHC-HCG-Report}
\href {http://cdsweb.cern.ch/record/1379837} {{ATLAS and CMS Collaborations,
  LHC Higgs Combination Group}, ``Procedure for the {LHC} {H}iggs boson search
  combination in {S}ummer 2011'',} Technical Report ATL-PHYS-PUB 2011-11, CMS
  NOTE 2011/005, (2011).

\bibitem{Junk:1999kv}
\hrefCMSnoop {} {T.~Junk, ``{Confidence level computation for combining
  searches with small statistics}'',} \textit{ Nucl. Instrum. Meth. A} \textbf{
  434} (1999) 435,
  \href{http://dx.doi.org/10.1016/S0168-9002(99)00498-2}{\doi{10.1016/S0168-9002(99)00498-2}}.

\bibitem{Read1}
\hrefCMSnoop {} {A.~L. Read, ``Presentation of search results: the {CLs}
  technique'',} \textit{ J. Phys. G} \textbf{ 28} (2002) 2693,
  \href{http://dx.doi.org/10.1088/0954-3899/28/10/313}{\doi{10.1088/0954-3899/28/10/313}}.

\bibitem{LEE}
\hrefCMSnoop {} {E.~Gross and O.~Vitells, ``{Trial factors for the look
  elsewhere effect in high energy physics}'',} \textit{ Eur. Phys. J. C}
  \textbf{ 70} (2010) 525,
  \href{http://dx.doi.org/10.1140/epjc/s10052-010-1470-8}{\doi{10.1140/epjc/s10052-010-1470-8}},
\href{http://www.arXiv.org/abs/1005.1891}{\texttt{ arXiv:1005.1891}}.

\bibitem{Cowan:2010st}
G.~Cowan\hrefCMSnoop {} { {et~al.}, ``Asymptotic formulae for likelihood-based
  tests of new physics'',} \textit{ Eur. Phys. J. C} \textbf{ 71} (2011) 1,
  \href{http://dx.doi.org/10.1140/epjc/s10052-011-1554-0}{\doi{10.1140/epjc/s10052-011-1554-0}},
  \href{http://www.arXiv.org/abs/1007.1727}{\texttt{ arXiv:1007.1727}}.

\bibitem{RooStats}
L.~Moneta\href
  {http://pos.sissa.it/archive/conferences/093/057/ACAT2010_057.pdf} {
  {et~al.}, ``The {R}oo{S}tats {P}roject'',} in \textit{ 13$^\text{th}$ Int.
  Workshop on Advanced Computing and Analysis Techniques in Physics Research
  (ACAT2010)}.
\newblock 2010.
\newblock \href{http://www.arXiv.org/abs/1009.1003}{\texttt{ arXiv:1009.1003}}.
\newblock {PoS ACAT:057}.

\bibitem{Seez1990a}
C.~J. Seez\href {http://cdsweb.cern.ch/record/220524} { {et~al.}, ``Photon
  decay modes of the intermediate mass {H}iggs'',} in \textit{ Proceedings of
  the Large Hadron Collider Workshop}, G.~Jarlskog and D.~Rein, eds., p.~474.
\newblock Aachen, Germany, 1990.
\newblock {CERN 90-10-V-2/ECFA 90-133-V-2}.

\bibitem{Yang2005370}
\hrefCMSnoop {} {H.-J. Yang, B.~P. Roe, and J.~Zhu, ``Studies of boosted
  decision trees for MiniBooNE particle identification'',} \textit{ Nucl.
  Instrum. Meth. A} \textbf{ 555} (2005) 370,
  \href{http://dx.doi.org/10.1016/j.nima.2005.09.022}{\doi{10.1016/j.nima.2005.09.022}}.

\bibitem{Hocker:2007ht}
H.~Voss\href {http://pos.sissa.it/archive/conferences/050/040/ACAT_040.pdf} {
  {et~al.}, ``{TMVA: Toolkit for Multivariate Data Analysis with ROOT}'',} in
  \textit{ XI Int. Workshop on Advanced Computing and Analysis Techniques in
  Physics Research}.
\newblock 2007.
\newblock \href{http://www.arXiv.org/abs/physics/0703039}{\texttt{
  arXiv:physics/0703039}}.
\newblock
{PoS ACAT:040}.

\bibitem{Barlow:1986ek}
\hrefCMSnoop {} {R.~J. Barlow, ``{Event classification using weighting
  methods}'',} \textit{ J. Comp. Phys.} \textbf{ 72} (1987) 202,
\href{http://dx.doi.org/10.1016/0021-9991(87)90078-7}{\doi{10.1016/0021-9991(87)90078-7}}.

\bibitem{DellaNegra1990a}
M.~Della~Negra\href {http://cdsweb.cern.ch/record/215298} { {et~al.}, ``Search
  for {$\PH\to\cPZ^*\cPZ^*\to 4$} leptons at the {LHC}'',} in \textit{
  Proceedings of the Large Hadron Collider Workshop}, G.~Jarlskog and D.~Rein,
  eds., p.~509.
\newblock Aachen, Germany, 1990.
\newblock {CERN 90-10-V-2/ECFA 90-133-V-2}.

\bibitem{Cabibbo:1965zz}
\hrefCMSnoop {} {N.~Cabibbo and A.~Maksymowicz, ``{Angular Correlations in
  $K_{e4}$ Decays and Determination of Low-Energy $\pi$-$\pi$ Phase Shifts}'',}
  \textit{ Phys. Rev. B} \textbf{ 137} (1965) 438,
  \href{http://dx.doi.org/10.1103/PhysRev.137.B438}{\doi{10.1103/PhysRev.137.B438}}.
Also Erratum, \doi{10.1103/PhysRev.168.1926}.

\bibitem{Gao:2010qx}
Y.~Gao\hrefCMSnoop {} { {et~al.}, ``{Spin determination of single-produced
  resonances at hadron colliders}'',} \textit{ Phys. Rev. D} \textbf{ 81}
  (2010) 075022,
  \href{http://dx.doi.org/10.1103/PhysRevD.81.075022}{\doi{10.1103/PhysRevD.81.075022}},
\href{http://www.arXiv.org/abs/1001.3396}{\texttt{ arXiv:1001.3396}}.

\bibitem{DeRujula:2010ys}
A.~De~Rujula\hrefCMSnoop {} { {et~al.}, ``{Higgs look-alikes at the LHC}'',}
  \textit{ Phys. Rev. D} \textbf{ 82} (2010) 013003,
  \href{http://dx.doi.org/10.1103/PhysRevD.82.013003}{\doi{10.1103/PhysRevD.82.013003}},
\href{http://www.arXiv.org/abs/1001.5300}{\texttt{ arXiv:1001.5300}}.

\bibitem{Chatrchyan:2012sn}
\hrefCMSnoop {} {{ CMS} Collaboration, ``Search for a {H}iggs boson in the
  decay channel {$\PH \to \cPZ\cPZ^{(*)} \to \cPq\cPaq \ell^-\ell^+$} in pp
  collisions at $\sqrt{s}$ = 7 {TeV}'',} \textit{ JHEP} \textbf{ 04} (2012)
  036,
  \href{http://dx.doi.org/10.1007/JHEP04(2012)036}{\doi{10.1007/JHEP04(2012)036}},
\href{http://www.arXiv.org/abs/1202.1416}{\texttt{ arXiv:1202.1416}}.

\bibitem{Choi:2002jk}
S.~Y. Choi\hrefCMSnoop {} { {et~al.}, ``{Identifying the Higgs spin and parity
  in decays to Z pairs}'',} \textit{ Phys. Lett. B} \textbf{ 553} (2003) 61,
  \href{http://dx.doi.org/10.1016/S0370-2693(02)03191-X}{\doi{10.1016/S0370-2693(02)03191-X}},
\href{http://www.arXiv.org/abs/hep-ph/0210077}{\texttt{ arXiv:hep-ph/0210077}}.

\bibitem{CMS:2011aa}
\hrefCMSnoop {} {{ CMS} Collaboration, ``{Measurement of the inclusive W and Z
  production cross sections in pp collisions at $\sqrt{s} = 7\TeV$ with the CMS
  experiment}'',} \textit{ JHEP} \textbf{ 10} (2011) 132,
  \href{http://dx.doi.org/10.1007/JHEP10(2011)132}{\doi{10.1007/JHEP10(2011)132}}.

\bibitem{Barger:1990mn}
V.~D. Barger\hrefCMSnoop {} { {et~al.}, ``{Intermediate mass Higgs boson at
  hadron supercolliders}'',} \textit{ Phys. Rev. D} \textbf{ 43} (1991) 779,
\href{http://dx.doi.org/10.1103/PhysRevD.43.779}{\doi{10.1103/PhysRevD.43.779}}.

\bibitem{Dittmar:1996ss}
\hrefCMSnoop {} {M.~Dittmar and H.~K. Dreiner, ``{How to find a Higgs boson
  with a mass between 155 GeV and 180 GeV at the CERN LHC}'',} \textit{ Phys.
  Rev. D} \textbf{ 55} (1997) 167,
  \href{http://dx.doi.org/10.1103/PhysRevD.55.167}{\doi{10.1103/PhysRevD.55.167}},
\href{http://www.arXiv.org/abs/hep-ph/9608317}{\texttt{ arXiv:hep-ph/9608317}}.

\bibitem{CMS-PAPERS-TAU-11-001}
\hrefCMSnoop {} {{ CMS} Collaboration, ``Performance of $\tau$-lepton
  reconstruction and identification in {CMS}'',} \textit{ JINST} \textbf{ 7}
  (2011) P01001,
  \href{http://dx.doi.org/10.1088/1748-0221/7/01/P01001}{\doi{10.1088/1748-0221/7/01/P01001}}.

\bibitem{Chatrchyan:2011nx}
\hrefCMSnoop {} {{ CMS} Collaboration, ``{Search for Neutral MSSM Higgs Bosons
  Decaying to Tau Pairs in $pp$ Collisions at $\sqrt{s}=7$ TeV}'',} \textit{
  Phys. Rev. Lett.} \textbf{ 106} (2011) 231801,
  \href{http://dx.doi.org/10.1103/PhysRevLett.106.231801}{\doi{10.1103/PhysRevLett.106.231801}},
\href{http://www.arXiv.org/abs/1104.1619}{\texttt{ arXiv:1104.1619}}.

\bibitem{Denner:2011id}
A.~Denner\hrefCMSnoop {} { {et~al.}, ``{Electroweak corrections to
  Higgs-strahlung off W/Z bosons at the Tevatron and the LHC with HAWK}'',}
  \textit{ JHEP} \textbf{ 03} (2012) 075,
  \href{http://dx.doi.org/10.1007/JHEP03(2012)075}{\doi{10.1007/JHEP03(2012)075}},
\href{http://www.arXiv.org/abs/1112.5142}{\texttt{ arXiv:1112.5142}}.

\bibitem{Gallicchio:2010sw}
\hrefCMSnoop {} {J.~Gallicchio and M.~D. Schwartz, ``{Seeing in Color: Jet
  Superstructure}'',} \textit{ Phys. Rev. Lett.} \textbf{ 105} (2010) 022001,
  \href{http://dx.doi.org/10.1103/PhysRevLett.105.022001}{\doi{10.1103/PhysRevLett.105.022001}},
\href{http://www.arXiv.org/abs/1001.5027}{\texttt{ arXiv:1001.5027}}.

\bibitem{Landau}
\hrefCMSnoop {} {L.~D. Landau, ``On the angular momentum of a two-photon
  system'',} \textit{ Dokl. Akad. Nauk} \textbf{ 60} (1948) 207.

\bibitem{Yang}
\hrefCMSnoop {} {C.~N. Yang, ``Selection Rules for the Dematerialization of a
  Particle into Two Photons'',} \textit{ Phys. Rev.} \textbf{ 77} (1950) 242,
  \href{http://dx.doi.org/10.1103/PhysRev.77.242}{\doi{10.1103/PhysRev.77.242}}.

\end{thebibliography}\endgroup
\end{document}